\documentclass[prd,floatfix,onecolumn,amsmath,amssymb,10pt]{revtex4-2}
\PassOptionsToPackage{table}{xcolor}
\usepackage{newtxtext,newtxmath} 
\usepackage{graphicx,color,dcolumn,booktabs,bm}
\usepackage{subfigure}
\usepackage{amssymb}
\usepackage{longtable}
\usepackage{indentfirst}
\usepackage{epsfig,gensymb,siunitx}
\usepackage{feynmf}   %{feynmp}
\usepackage{epstopdf}   %{feynmp}
\usepackage{slashed}  %for Feynman symbols
\usepackage{cases}
\usepackage{xcolor}
\definecolor{navyblue}{RGB}{0,0,170}
\definecolor{CobaltBlue}{rgb}{0,0.28,.67}
\definecolor{maroon}{RGB}{139,25,150}%burada 0-255 arasi her biri icin numara vererek renk elde et
\usepackage{multirow}
\usepackage{float}
\usepackage{pgf}
\usepackage{graphicx,color,dcolumn,booktabs,bm}
\usepackage[colorlinks, citecolor=purple,anchorcolor=purple,menucolor=purple, linkcolor=purple,filecolor=purple,runcolor=purple,urlcolor=purple,frenchlinks=purple, urlcolor=purple]{hyperref}
\usepackage{physics}
\usepackage{orcidlink}
\usepackage{bbold}

\newcommand{\g}{g_{\alpha\beta}}

\newcommand{\ga}{\gamma_\alpha}

\def\beq{\begin{equation}}
\def\eeq{\end{equation}}
\def\bea{\begin{eqnarray}}
\def\eea{\end{eqnarray}}
\def\beeq{\begin{eqnarray}}
\def\eeeq{\end{eqnarray}}
\newcommand{\nn}{\nonumber}

\def\vel{\left|}
\def\ver{\right|}
\def\nnb{\nonumber}
\def\ga{\left(}
\def\dr{\right)}

\def\rar{\rightarrow}

\def\ba{\begin{array}}
\def\ea{\end{array}}

\def\xis0{{\Xi^{*0}}}

\def\g5{\gamma_5}

\def\es{\!\!\! &=& \!\!\!}

\def\ar{&+& \!\!\!}
\def\ek{&-& \!\!\!}

\begin{document}
	
	\preprint{}
	
\title{\color{navyblue}{Comprehensive analyses of rare $ \Lambda_b \rightarrow \Lambda \ell^+  \ell^-$, $\Sigma_b  \rightarrow \Sigma \ell^+  \ell^-$ 
and $\Xi_b \rightarrow \Xi \ell^+  \ell^-$ decays in the 2HDM}}

	\author{Z.~Tavuko\u glu$^{1}$\orcidlink{0000-0002-7967-2672}}
	\email{zeynep.tavukoglu@okan.edu.tr}
	
	\author{A.~T.~Olgun$^{1}$\orcidlink{0000-0002-2113-4978}}
	\email{tugba.olgun@okan.edu.tr}
	
	\author{K.~Azizi$^{2,3}$\orcidlink{0000-0003-3741-2167}}
	\email{kazem.azizi@ut.ac.ir}
	\thanks{Corresponding author}

\affiliation{
	$^{1}$Vocational School, \href{http://www.okan.edu.tr/en/}{Istanbul Okan University}, Tuzla Campus, 34959 Istanbul, T\"{u}rkiye\\
	$^{2}$Department of Physics, \href{https://ut.ac.ir/en}{University of Tehran}, North Karegar Avenue, Tehran 14395-547, Iran\\
	$^{3}$Department of Physics, \href{https://www.dogus.edu.tr/en}{Dogus University}, Dudullu-\"{U}mraniye, 34775
	Istanbul,  T\"{u}rkiye}	
\date{\today}

	\begin{abstract}
We investigate rare special dileptonic decays of $ \Lambda_b$, $\Sigma_b$ and $\Xi_b $ baryons  in the Standard Model and context of the general Two-Higgs-Doublet Model with Type III. Specifically, we consider the decays $ \Lambda_b \rightarrow \Lambda \ell^+  \ell^-$, $\Sigma_b  \rightarrow \Sigma \ell^+  \ell^-$  and $\Xi_b \rightarrow \Xi \ell^+  \ell^-$, where $\ell$ represents $\mu$ or $\tau$ lepton. By studying these rare decays, we aim to assess the impact of the Two-Higgs-Doublet Model with Type III on various observables, such as the differential decay width, the total decay width, the differential branching ratio, total branching ratio, and lepton forward-backward asymmetries using the decay amplitude and the transition matrix elements in terms of form factors calculated via light cone QCD in full theory.  We compare our results to those of the Standard Model, as well as existing lattice QCD predictions and experimental data, to assess the agreement and viability of the Two-Higgs-Doublet Model with Type III. Furthermore, we highlight the potential for experimental investigations of these decay channels in the near future. The soon-to-be updated LHCb and/or Belle II detectors, renowned for their capabilities in studying rare decays, present excellent opportunities for probing the predicted branching ratios. 
	\end{abstract}
	
	\maketitle
	
	%%%%%%%%%%%%%%%%%%%%%%%%%%%%%%%%%%%%%%%%%%%%%%%%%%%%%%%%%%%%%%%%%%%%%%%%%%%%%%%%%%%
	\section{Introduction}\label{introduction}

The discovery of a scalar particle with a mass of approximately 125 GeV at the ATLAS \cite{ATLAS:2012yve} and CMS \cite{CMS:2012qbp} experiments aligns, within theoretical and experimental uncertainties, with the predictions of the Standard Model (SM) Higgs boson. The properties of the Higgs boson, denoted as h, continue to be extensively measured, including its mass, width, quantum numbers, couplings, and more. The search for additional Higgs bosons remains as a significant focus of the Large Hadron Collider (LHC) physics program. Recent analyses by the CMS Collaboration searching for low-mass scalar resonances at the LHC have conducted searches for scalar di-photon resonances at both 8 TeV and 13 TeV center-of-mass energies. These results revealed a local excess of about 2.9 $\sigma$ at a mass of 95.4 GeV \cite{CMS:2024yhz}. The ATLAS experiment has also performed searches for low-mass scalar resonances in the diphoton channel and reported an excess with a local significance of about 1.7 $\sigma$ near a mass of 95 GeV, consistent with the excess previously observed by CMS \cite{ATLAS:2024bjr}. Following the second long shutdown (LS2) between 2019 and 2021, the LHC Run 3 started in 2022 at a center-of-mass energy of 13.6 TeV and is expected to collect about 250 $ {fb}^{-1}$ of integrated luminosity. Furthermore, during the third long shutdown (LS3), scheduled to begin around 2026, the CMS detector will undergo a major Phase-2 upgrade in preparation for the High-Luminosity LHC (HL-LHC) program \cite{Morovic:2023wgz}.

Several shortcomings and phenomenological indications suggest that the SM may not provide answers to all fundamental questions, reinforcing the motivation for beyond the standard model (BSM) theories.This study emphasizes the importance of examining the Higgs boson's properties and ratios with high precision at the LHC to search for possible hints of BSM physics. While the SM has been highly successful, the quest for a more complete understanding of the Universe necessitates the exploration of extensions to the SM. The measured Higgs boson ratios offer an avenue for discovering new physics phenomena and advancing our knowledge of the fundamental building blocks of nature.

In the search for physics of the BSM, numerous theoretical frameworks propose extensions to the Higgs-boson sector. Therefore, a primary objective of the upcoming LHC Run 3 and High-Luminosity LHC (HL-LHC) is to investigate whether the observed Higgs boson is part of an extended Higgs sector within a BSM model. The Two-Higgs-Doublet Model (2HDM) is regarded as one of the simplest extensions of the SM Higgs-boson sector. As a natural expansion of the SM's Higgs sector, the 2HDM offers a framework to investigate the presence of two distinct Higgs doublets.

The 2HDM \cite{Gunion:1992hs,Branco:2011iw} is important because it provides a theoretical framework to explore physics beyond the SM, our current best theory of particle physics. The 2HDM can help answer fundamental questions that the SM does not fully explain. By introducing an additional Higgs doublet, the 2HDM expands the Higgs sector, leading to the presence of more Higgs particles with different properties. This opens up new possibilities for studying and understanding the Higgs boson, its interactions, and its role in the Universe. The 2HDM makes specific predictions about the properties and interactions of the additional Higgs particles. Experimental tests, such as those conducted at particle colliders like the LHC, can search for these predicted particles and interactions. Confirming or disproving these predictions can guide our understanding of particle physics and the development of new theories. The 2HDM provides a framework to incorporate dark matter, the invisible matter that makes up a significant portion of the Universe. By introducing additional particles, such as a dark matter candidate, the model can help understand the nature of dark matter and its interactions with other particles. The 2HDM builds upon the successful framework of the SM and extends it. It allows scientists to explore new phenomena and study different patterns of particle masses and couplings. This can help us refine and deepen our understanding of the fundamental particles and forces in nature.

The 2HDM introduces five physical Higgs bosons: $h^{0}$ and $H^{0} $(CP-doublet), A (CP-single), and $H^{\pm}$ (charged). In the most general implementation of the this scenario, non-diagonal flavor connections in the fermion sector can lead to flavor-changing neutral current (FCNC) effects that may not align with observed data at the tree level. To accommodate these tree-level FCNCs, both Higgs doublets can couple to leptons and quarks, resulting in a model known as 2HDM Type III, specifically when following the Cheng-Sher ansatz \cite{Cheng:1987rs}.

In 2HDM Type III, the fermion sector requires non-diagonal Yukawa couplings relative to the masses of the corresponding fermions, which are considered dimensionless free parameters. By assuming CP protection and a scenario with suppressed Higgs couplings, the 2HDM can be parameterized by at least seven independent parameters. These parameters include the masses of the Higgs bosons  ($m_{h}$, $m_{H}$,$ m_{H^{\pm}}$,$m_{A}$),
 the ratio of vacuum expectation values of the two Higgs doublets ($tan\beta=\dfrac{\nu^{2}}{\nu_{1}}$),  the mixing angle of the CP-even Higgs states $ \alpha $ and the parameter $ m_{12}^{2}$.
 
 FCNC processes, specifically the $ b \rightarrow  s \ell^+  \ell^-$ transitions, serve as a crucial testing ground for the SM at the one-loop level. In the SM, these transitions are forbidden at the tree level due to the Glashow-Iliopoulos-Maiani (GIM) mechanism, which suppresses FCNC processes. By investigating FCNC decays in both mesonic and baryonic systems, researchers can gain valuable insights into the fundamental interactions and dynamics of particles. These studies contribute to our understanding of the SM, help uncover new physics beyond the SM, and pave the way for further exploration and advancements in particle physics.

The study of baryonic decays, such as $ \Lambda_b$, $\Sigma_b$ and $\Xi_b $ baryon decays, plays a crucial role in understanding the dynamics of quarks within baryons and probing physics  BSM. These decays provide valuable information about the properties of baryons, the interactions between quarks, and the underlying fundamental forces. The dileptonic channel of $\Lambda_b$, $\Sigma_b$, and $\Xi_b$ baryons refers to the decay processes where these baryons decay into a lepton-antilepton pair (dilepton) through weak interactions.  The dileptonic decays of these baryons are of particular interest due to their clean experimental signature and the ability to precisely measure the kinematic properties of the decay products. These decays provide valuable information about the weak interactions and the properties of the baryons involved. 

The CDF Collaboration has reported the first measurement of the $ \Lambda_b \rightarrow \Lambda \ell^+  \ell^-$ decay channel \cite{CDF:2011buy}, identifying $24$ signal events with a statistical significance of $5.8$ Gaussian standard deviations. Using the $p\overline{p}$ collision data corresponding to an integrated luminosity of $6.8 fb^{-1}$ at $\sqrt{s}=1.96$ TeV collected with the CDF II detector, they determined the differential branching ratio for the $ \Lambda_b \rightarrow \Lambda \mu^+  \mu^-$ process as $dBr(\Lambda _b^0 \rightarrow \Lambda \mu^+ \mu^-)/dq^2=[1.73\pm0.42(stat)\pm0.55(syst)]\times10^{-6}$ \cite{Aaltonen:2011qs}. The LHCb Collaboration has also measured the differential branching fraction for the $\Lambda _b^0 \rightarrow \Lambda \mu^+ \mu^-$ decay, obtaining $dBr (\Lambda _b^0 \rightarrow \Lambda \mu^+ \mu^-)/dq^2 = (1.18 \;^{+\,0.09}_{-\,0.08} \pm 0.03 \pm 0.27 )
\times 10^{-7}$ GeV$^2/$c$^4$ in the region $15 $ GeV$^2/$c$^4\leq$   $q^2 \leq 20$  GeV$^2/$c$^4$ \cite{LHCb:2015tgy}. Additionally, LHCb has measured the corresponding lepton forward–backward asymmetry, reporting  $A_{FB}^{\mu}=-0.05 \pm 0.09 (stat)\pm 0.03 (syst)$ for the same $q^2$ interval \cite{LHCb:2015tgy}.
The magnitudes of the branching ratios for $ \Lambda_b \rightarrow \Lambda e^+  e^-$, $ \Lambda_b \rightarrow \Lambda \tau^+  \tau^-$, and for the related decays $\Sigma_b  \rightarrow \Sigma \ell^+  \ell^-$ and $\Xi_b \rightarrow \Xi \ell^+  \ell^-$ (for all lepton species) suggest that these processes are within reach of current LHC experiments. The FCNC transitions in baryonic decays, including $ \Lambda_b \rightarrow \Lambda \ell^+  \ell^-$, $\Sigma_b  \rightarrow \Sigma \ell^+  \ell^-$  and $\Xi_b \rightarrow \Xi \ell^+  \ell^-$ are extremely important for exploring potential contributions from new physics. A substantial literature has systematically analyzed these decay processes, elucidating their complex hadronic behavior and underscoring their capacity to reveal subtle imprints of new physics scenarios \cite{Katirci:2012eh,Azizi:2011mw,Aliev:2010uy,Azizi:2013eta,Gutsche:2013pp,Boer:2014kda,Detmold:2016pkz,Faustov:2016pal,Das:2018iap,Sahoo:2016nvx,Kumar:2015tnz,Wang:2021uzi,Tan:2023opd,Azizi:2016dcj,Azizi:2015hoa,Azizi:2012vy,Azizi:2012yg,Azizi:2011ri,Iguro:2018qzf,Crivellin:2019dun,Iguro:2023jju,Crivellin:2023sig,Coloretti:2025yji}.

The structure of the paper is organized as follows: Section II provides the theoretical framework, including the general overview of the 2HDM and the constraints on its parameters, the effective Hamiltonian and Wilson coefficients, as well as the transition amplitudes and matrix elements. In Section III, we present the numerical analysis and discuss the physical observables, such as the differential decay width, the differential branching ratio, the total decay width, the branching ratio, and the lepton forward-backward asymmetry. Finally, Section IV contains the summary and concluding remarks.

\section{The baryonic $ \Lambda_b \rightarrow \Lambda \ell^+  \ell^-$, $\Sigma_b  \rightarrow \Sigma \ell^+  \ell^-$  and $\Xi_b \rightarrow \Xi \ell^+  \ell^-$ decays in the SM and 2HDM }

\label{sec:Masses}
%%%%%%%%%%%%%%%%%%%%%%%%%%%%%%%%%%%%%%%%%%%%%%%%%%%%%%%%%%%
In this study, a comprehensive analysis of the SM and the 2HDM is presented in the context of the baryonic decays $ \Lambda_b \rightarrow \Lambda \ell^+  \ell^-$, $\Sigma_b  \rightarrow \Sigma \ell^+  \ell^-$  and $\Xi_b \rightarrow \Xi \ell^+  \ell^-$. First, an overview of the theoretical structure of the 2HDM is provided, including its extended scalar sector, the various classifications of the model, and the phenomenological implications arising from the presence of additional Higgs bosons. The study also addresses the theoretical and experimental constraints imposed on the 2HDM parameter space. Subsequently, the effective Hamiltonian governing the $b \rightarrow s \ell^{+}\ell^{-}$ transition is introduced, and the modifications to the Wilson coefficients induced by the extended Higgs sector are discussed. Building upon this framework, the transition amplitudes and matrix elements corresponding to the baryonic decay modes under consideration are derived, incorporating both the short-distance contributions encoded in the Wilson coefficients and the non-perturbative hadronic effects parametrized through appropriate form factors. This approach establishes the necessary foundation for examining the phenomenological implications of the SM and 2HDM in these rare baryonic decay channels. 

%%%%%%%%%%%%%%%%%%%%%%%%%%%%%%%%%%%%%%%%%%%%%%%%%%%%%%%%%%%%%%%%%%%%%%%%%%%%

\subsection{ The general overview of the 2HDM and the constraints on the parameters of the 2HDM }

\label{sec:Decays}
%%%%%%%%%%%%%%%%%%%%%%%%%%%%%%%%%%%%%%%%%%%%%%%%%%%%%%%%%%%
Within the framework of the SM of particle physics, the 2HDM constitutes an extension solely in the Higgs sector, achieved by introducing an additional Higgs doublet while keeping the remaining particle content unchanged. In its most general form, the scalar potential of the 2HDM contains eleven independent parameters; however, the imposition of specific symmetries can reduce the number of free parameters. In many realizations of the Two-Higgs-Doublet Model (2HDM), a discrete $\mathbb{Z}_2$ symmetry is imposed, under which the first scalar doublet transforms as $\Phi_1 \to \Phi_1$ (even under $\mathbb{Z}_2$), while the second transforms as $\Phi_2 \to -\Phi_2$ (odd under $\mathbb{Z}_2$). The presence of this symmetry restricts the scalar potential in the CP-conserving case to eight independent parameters: four physical masses, the mixing angle $\alpha$ in the CP-even sector, the ratio of vacuum expectation values $\tan\beta = v_2/v_1$, and the soft-breaking parameter $m_{12}^2$. A comprehensive discussion of these parameters and their explicit forms can be found in Ref.~\cite{Arhrib:2013oia}.

Analogous to the mechanism of fermion mass generation in the SM, fermions in the 2HDM acquire mass through Yukawa interactions $y_{ij}$ with the scalar doublets. For the first Higgs doublet $\Phi_1$, the Yukawa couplings are constrained to be diagonal both in flavor space and in the mass eigenstate basis. In contrast, for the second doublet $\Phi_2$, the Yukawa couplings are generally non-diagonal and cannot be directly linked to the fermion mass matrices. Since fermion mass eigenstates can be represented as vectors in flavor space, the Yukawa sector of the 2HDM can be reformulated in terms of the physical Higgs mass eigenstates, as discussed in Ref.~\cite{Davidson:2005cw}.

Among the most studied variants, Model I and Model II assign couplings of the two doublets to up- and down-type quarks differently, whereas in Model III both doublets may couple to all quark types \cite{Atwood:1996vj,Cheng:1987rs}. In the general case, the Yukawa Lagrangian for the Type-III 2HDM can be formulated explicitly, allowing both Higgs doublets to couple simultaneously to up- and down-type fermions \cite{Kim:2015zla}:

The fundamental aspects of the 2HDM with Model III can be summarized as follows. Without loss of generality, a basis can be chosen in which only the first scalar doublet is responsible for generating all fermion and gauge boson masses, with its corresponding vacuum expectation values given by
\bea
\left< \phi_1 \right> = \left( \begin{array}{c}
0 \\ \\
\displaystyle{\frac{v}{\sqrt{2}}}	
\end{array} \right)~~~~~,~ \left< \phi_2 \right>=0~. 
\eea
In this basis, the first doublet $\phi_1$ corresponds to the SM Higgs, while all additional Higgs bosons arise from the second doublet $\phi_2$, which can be expressed in the following form.
\bea
\phi_1 = \frac{1}{\sqrt{2}} \left( \begin{array}{c}
\sqrt{2}\, G^+ \\ \\
v + \chi_1^0 + i G^0 
\end{array} \right)~~~~~,
~ \phi_2  = \frac{1}{\sqrt{2}} \left( \begin{array}{c}
\sqrt{2}\, H^+ \\ \\
\chi_2^0 + i A^0 
\end{array} \right)~, 
\eea
where $G^+$ and $G^0$ are the Goldstone bosons. The neutral fields $\chi_1^0$ and $\chi_2^0$ do not correspond to physical mass eigenstates; however, appropriate linear combinations of these fields give rise to the neutral Higgs bosons $H^0$ and $h^0$:
\bea
\chi_1^0 = H^0 \cos \alpha - h^0 \sin \alpha~, \nnb \\
\chi_2^0 = H^0 \sin \alpha + h^0 \cos \alpha~. 
\eea
The general Yukawa Lagrangian can be written as
\bea
{\cal L}_Y = \eta_{ij}^U \bar Q_{iL} \tilde \phi_1 U_{jR} +
\eta_{ij}^{\cal D} \bar Q_{iL} \phi_1 {\cal D}_{jR}
+ \xi_{ij}^U \bar Q_{iL} \tilde \phi_2  U_{jR} +
\xi_{ij}^{\cal D} \bar Q_{iL} \phi_2  {\cal D}_{jR} + h.c.~,
\eea
where $i$, $j$ are the generation indices, $\tilde \phi= i \sigma_2 \phi$,
$\eta_{ij}^{U,{\cal D}}$ and $\xi_{ij}^{U,{\cal D}}$, in general, are the non– diagonal coupling matrices, $L=(1-\gamma_5)/2$ and $R=(1+\gamma_5)/2$ are the left– and right–handed projection operators.
%%%%%%%%%%%%%%%%%%%%%%%%%%%%%%%%%%%%%%%%%%%%%%%%%%%%%%%%%%%%%%%%%%%%%%%%%%%%
\subsection{The effective Hamiltonian and the  Wilson coefficients }

The baryonic processes $ \Lambda_b \rightarrow \Lambda \ell^+ \ell^-$, $\Sigma_b \rightarrow \Sigma \ell^+ \ell^-$, and $\Xi_b \rightarrow \Xi \ell^+ \ell^-$  proceed via the underlying quark-level transition $b \rightarrow s \ell^{+}\ell^{-}$. In both the SM and 2HDM, the $b \rightarrow s \ell^{+}\ell^{-}$ transition is described using the effective Hamiltonian below \cite{Ahmed:2015pua,Alnahdi:2017ogx,Dai:1996vg,Ahmed:2016jgv}
\small{\bea {\cal
H}_{eff}(b\to s \ell^{+}\ell^{-}) &=& -\frac{4 G_F}{\sqrt 2} V_{tb} V^*_{ts} 
 \Bigg\{\sum_{i=1}^{10} C_i( \mu ) O_i( \mu ) + \sum_{i=1}^{10} 
C_{Q_i}( \mu ) Q_i( \mu ) \Bigg\} ~. \eea}
Here, $G_F$  denotes the Fermi coupling constant, while $V_{tb} V^*_{ts}$ represent the relevant elements of the Cabibbo–Kobayashi–Maskawa (CKM) matrix. The initial segment corresponds to the effective Hamiltonian within the SM framework, where the Wilson coefficients are modified by additional contributions stemming from charged Higgs bosons. The latter segment introduces new operators, which originate from the effects of massive neutral Higgs bosons influencing the transition.

The $b\rightarrow s\ell ^{+}\ell ^{-}$ amplitude at quark level is given by the effective Hamiltonian:
\begin{eqnarray}
{\cal
H}_{eff}^{2HDM}(b\rightarrow s\ell^{+}\ell^{-}) &=&-\frac{G_{F}\alpha }{\sqrt{2}\pi }V_{tb}V_{ts}^{\ast }\bigg[\widetilde{C}_{9}^{eff 2HDM}\left( \mu
\right) (\bar{s}\gamma _{\mu }Lb)(\bar{\ell}\gamma ^{\mu }\ell) + \widetilde{C}_{10}^{2HDM}(\bar{s}%
\gamma _{\mu }Lb)(\bar{\ell}\gamma ^{\mu }\gamma ^{5}\ell)  \notag \\
&&-2\widetilde{C}_{7}^{eff 2HDM}\left( \mu \right) \frac{m_{b}}{s}(\bar{s}i\sigma _{\mu
\nu }q^{\nu }Rb)\bar{\ell}\gamma ^{\mu }\ell +C_{Q_1}^{2HDM}\left( \bar{s}Rb\right) \left(\bar{\ell}\ell \right) +C_{Q _2}^{2HDM}\left( \bar{s}Rb\right) \left(\bar{\ell}\gamma^{5}\ell \right)\bigg], 
\end{eqnarray}
where $q$ denotes the momentum transfer.  Including charged Higgs boson contributions modifies the Wilson coefficients $\widetilde{C}_{7}^{eff 2HDM}$, $\widetilde{C}_{9}^{eff 2HDM}$,
$\widetilde{C}_{10}^{2HDM}$, $C_{Q_1}^{2HDM}$ and $C_{Q_2}^{2HDM}$ as follows:
Since charged Higgs exchange diagrams do not generate any new operators, the operator basis for the $b \rightarrow s \ell^+ \ell^-$ transition remains the same as in the SM. The evolution of the Wilson coefficients from the high-energy scale $\mu = m_W$ down to the low-energy scale $\mu = m_b$ is governed by the 
renormalization group equations. A detailed treatment of this subject can be found 
in Ref.~\cite{Buchalla:1995vs}.
Therefore, in model III, the leading-order contributions from the charged Higgs only affect the Wilson coefficients at the electroweak scale $m_W$.
\begin{eqnarray}
C_7^{2HDM}(m_W) &=& C_7^{SM}(m_W) + C_7^{H^\pm}(m_W) \nnb \\
C_9^{2HDM}(m_W) &=& C_9^{SM}(m_W) + C_9^{H^\pm}(m_W) \nnb \\
C_{10}^{2HDM}(m_W) &=& C_{10}^{SM}(m_W) + C_{10}^{H^\pm}(m_W)~.
\end{eqnarray}
The values of the Wilson coefficients $C_7^{SM}(m_W)$, $C_9^{SM}(m_W)$ and $C_{10}^{SM}(m_W)$ in the SM are provided in Refs.  \cite{Buchalla:1995vs,Azizi:2016dcj}.
The coefficient $C_7^{eff}(\mu)$ at the scale 
$\mu=m_b$ in next to leading order (NLO) 
is calculated in \cite{Borzumati:1998tg,Aliev:1999ap}:
\bea
C_7^{eff}(m_b) = C_7^0(m_b) + \frac{\alpha_s(m_b)}{4 \pi} 
C_7^{1,eff}(m_b)~,
\eea
where $C_7^0(m_b)$  is the LO term, and $C_7^{1,eff}(m_b)$  includes the NLO corrections, given explicitly in \cite{Borzumati:1998tg}. For a detailed discussion of the Wilson coefficient $C_7^{2HDM}(m_W) $, see Ref. \cite{Borzumati:1998tg}.

At the scale $m_W$, the leading-order values of the coefficients $C_i^{2HDM}$ are given by:

\begin{eqnarray}
C_9^{2HDM}(m_W) &=& - \frac{1}{\sin^2 \theta_W} \, B(m_W) + 
\frac{1 - 4 \sin^2 \theta_W}{\sin^2 \theta_W} \, C(m_W) + \frac{-19 x^3 + 25 x^2}{36 (x-1)^3}
+\frac{-3 x^4 + 30 x^3 - 54 x^2 + 32 x -8}{18 (x-1)^4} \, \ln x 
+ \frac{4}{9} \nnb \\
&+& \vel \lambda_{tt} \ver^2 \Bigg\{ 
\frac{1 - 4 \sin^2 \theta_W}{\sin^2 \theta_W} \, \frac{x y}{8} \Bigg[ 
\frac{1}{y-1} - \frac{1}{(y-1)^2} \, \ln y \Bigg]- y \Bigg[ \frac{47 y^2 - 79 y + 38}{108 (y-1)^3}
-\frac{3 y^3 - 6 y^2 + 4}{18 (y-1)^4} \, \ln y \Bigg] \Bigg\}~,
\end{eqnarray}
where $\lambda_{tt}$ is the coupling constant associated with the interaction between the top quark and the second Higgs doublet in the 2HDM model. The quantities $B(x)$, $C(x)$, $x$, and $y$ are defined by the following expressions, respectively:
\bea
B(x) &=& - \frac{x}{4 (x-1)} + \frac{x}{4 (x-1)^2} \, \ln x ~, \nnb \\
C(x) &=& - \frac{x}{4} \Bigg[ \frac{x-6}{3 (x-1)} + 
\frac{3 x +2 }{2 (x-1)^2} \ln x \Bigg]~,\nnb \\
x &=& \frac{m_t^2}{m_W^2} ~, \nnb \\
y &=& \frac{m_t^2}{m_{H^\pm}^2}~.
\eea
Here, $\sin^2\theta_W = 0.23$ is taken as the definition of the Weinberg angle.

In Model III, no additional operators arise beyond those present in the SM. Consequently, it is sufficient to implement the substitution $C_9^{SM}(m_W) \rar C_{9}^{2HDM}(m_W)$, as discussed in Refs. \cite{Buchalla:1995vs,Buras:1993xp,Colangelo:1993ux,Buras:1994dj}, in order to determine $C_{9}^{2HDM}$ at the scale $m_b$. After incorporating next-to-leading order QCD corrections, the coefficient $C_9(\mu)$ takes the form:
\bea
\lefteqn{
C_9(\mu) = C_9^{2HDM}(\mu) 
\left[1 + \frac{\alpha_s(\mu)}{\pi} \omega (\hat s) \right]} \nnb \\
&&+ \, g(\hat m_c,\hat s) \Big[ 3 C_1(\mu) + C_2(\mu) + 3 C_3(\mu) + C_4(\mu)
+ 3 C_5(\mu) + C_6(\mu) \Big] \nnb \\
&&- \frac{1}{2} g(0,\hat s) \ga C_3(\mu) + 3 C_4(\mu) \dr
-\,  \frac{1}{2} g \ga 1, \hat s\dr
\ga 4 C_3 + 4 C_4 + 3 C_5 + C_6 \dr \nnb \\
&&- \frac{1}{2} g \ga 0, \hat s\dr \ga C_3 + 3 C_4 \dr
+\, \frac{2}{9} \ga 3 C_3 + C_4 + 3 C_5 + C_6 \dr~, 
\eea
with the definitions $\hat m_c = m_c/m_b~, ~\hat s = p^2/m_b^2$, and
\bea
\lefteqn{
\omega \ga \hat s \dr = - \frac{2}{9} \pi^2 - 
\frac{4}{3} Li_2  \ga \hat s \dr - \frac{2}{3} \ln \ga \hat s\dr 
\,\ln \ga 1 -\hat s \dr} \nnb \\ 
&&- \,\frac{5 + 4 \hat s}{3 \ga 1 + 2 \hat s \dr} \ln \ga 1 -\hat s \dr
-\frac{2 \hat s \ga 1 + \hat s \dr \ga 1 - 2 \hat s \dr}
{3 \ga 1 - \hat s \dr^2 \ga 1 + 2 \hat s \dr} \, \ln \ga \hat s\dr  + \frac{5 + 9 \hat s - 6 {\hat s}^2}
{3 \ga 1 - \hat s \dr \ga 1 + 2 \hat s \dr}~
\eea
The matrix element of $O_9$ receives a ${\cal O}\ga \alpha_s \dr$ correction from one-gluon exchange, in parallel, the function $g \ga \hat m_c, \hat s \dr$ originates from one-loop diagrams involving the four-quark operators $O_1$--$O_6$, whose structure can be written as
\bea 
g \ga \hat m_c, \hat s \dr &=& - \frac{8}{9}  \ln \ga \hat m_i\dr
+ \frac{8}{27} + \frac{4}{9} y_i - \frac{2}{9} \ga 2 + y_i \dr
 \sqrt{\vel 1-y_i \ver} \nnb\\
 &&\Bigg\{ \Theta \ga 1 - y_i \dr 
\Bigg( \ln \frac{1+\sqrt{\vel 1-y_i \ver}}{1-\sqrt{\vel 1-y_i \ver}}
- i \, \pi \Bigg) + \Theta \ga y_i -1 \dr 2 \arctan \frac{1}{\sqrt{y_i - 1}} \Bigg\}~,
\eea
with $y_i = 4 {\hat m_i}^2/{\hat p}^2$.

Long-distance effects also contribute to the Wilson coefficient $C_9$, originating from real $c\bar c$ intermediate states (e.g., $J/\psi$, $\psi^\prime$, etc.). The $J/\psi$ family is included by means of a Breit--Wigner description of the resonances, realized through the replacement (as discussed in Refs. \cite{Grinstein:1989df,e865:1999kah}).
\bea
g \ga \hat m_c, \hat s \dr \rar g \ga \hat m_c, \hat s \dr -
\frac{3\pi}{\alpha^2_{em}}
\, \kappa \sum_{V_i=J/\psi_i,\psi^\prime,\cdots} 
\frac{m_{V_i} \Gamma(V_i \rar \ell^+ \ell^-)}
{(p^2 - m_{V_i}^2) + i m_{V_i} \Gamma_{V_i}}~,
\eea
with the phenomenological parameter assigned the value $\kappa = 2.3$.

After including the contributions from the charged Higgs boson diagrams in addition to the SM results, the Wilson coefficient $C_{10}$ is modified, and the corresponding 2HDM contribution $\widetilde{C}^{2HDM}_{10}(m_W)$ can be expressed as follows:

\begin{eqnarray}
\widetilde{C}^{2HDM}_{10}(m_W)&=&\frac{1}{sin^2\theta_W} ( B(m_W)-C(m_W) ) +\left|\lambda_{tt}\right|^2\frac{1}{sin^2\theta_W}\frac{xy}{8}\left(-\frac{1}{y-1}+\frac{1}{(y-1)^2}\mathrm{ln}y\right).
\end{eqnarray}

At the renormalization scale $\mu = m_W$, the Wilson coefficients $C_i$ appearing in the effective Hamiltonian ${\cal H}_{\text{eff}}(b \to s \ell^+ \ell^-)$ are defined. The corresponding Wilson coefficients, which encode the short-distance contributions of the operators $O_i$, can be found in \cite{Dai:1996vg}, and at the scale $\mu = m_W$, they can be summarized as in \cite{Ahmed:2015pua}.
\bea
C_{Q_1}(m_{W})&=&\frac{m_bm_l}{m^2_{h^0}}\frac{1}{\left|\lambda_{tt}\right|^2}\frac{1}{sin^2\theta_{W}}\frac{x}{4}\bigg[(sin^2\alpha+hcos^2\alpha)f_1(x,y) \nn \\%
  &&+\left[\frac{m^2_{h^0}}{m^2_{W}}+(sin^2\alpha+hcos^2\alpha)(1-z)\right]f_2(x,y) \notag \\
&&+\frac{sin^2 2\alpha}{2m^2_{H^{\pm}}}\left[m^2_{h^0}-\frac{(m^2_{h^0}-m^2_{H^0})^2}{2m^2_{H^0}}\right]f_3(y)\bigg] \label{cq1} \\
C_{Q_2}(m_{W})&=&-\frac{m_bm_l}{m^2_{A^0}}\frac{1}{\left|\lambda_{tt}\right|^2}{f_1(x,y)+\bigg[1+\frac{m^2_{H^\pm}-m^2_{A^0}}{2m^2_{W}}\bigg]f_2(x,y)} \label{cq2} \\
C_{Q_3}(m_{W})&=&\frac{m_{b}e^2}{m_{\ell}g^2}\bigg[C_{Q_1}(m_{W})+C_{Q_2}(m_{W})\bigg] \label{cq3} \\
C_{Q_4}(m_{W})&=&\frac{m_{b}e^2}{m_{\ell}g^2}\bigg[C_{Q_1}(m_{W})-C_{Q_2}(m_{W})\bigg] \label{cq4} \\
C_{Q_i}(m_{W})&=&0 \enskip \enskip  i= 5,...,10 \label{cq5}
\eea
where
\bea
 z=\frac{x}{y}, h=\frac{m^2_{h^0}}{m^2_{H^0}}, \nn \\
f_1(x,y)=\frac{x\mathrm{ln}x}{x-1}-\frac{y\mathrm{ln}y}{y-1},\nn \\
 f_2(x,y)=\frac{x\mathrm{ln}y}{(z-x)(x-1)}-\frac{\mathrm{ln}z}{(z-1)(x-1)}, \nn \\
f_3(y)=\frac{1-y+y\mathrm{ln}y}{(y-1)^2}.
\eea
The scale dependence of the coefficients $C_{Q_1}$ and $C_{Q_2}$ is determined by the anomalous dimensions of $Q_1$ and $Q_2$, respectively:
\beq
C_{Q_i}(m_b)=\eta^{\gamma_Q/\beta_0}C_{Q_i}(m_W), \enskip \enskip i =1, 2
\eeq
with $\gamma_Q = -4$ denoting the anomalous dimension of the operator $\bar{s}_L b_R$ \cite{Dai:1996vg,Ahmed:2015pua}.

% c10 ref arxiv 9906473 
% c7eff arxiv 9802391
%c9  arxiv 9906473 
%cq1 and cq2 1509.08113

\subsection{Transition amplitude and matrix elements}

Typically, the decay amplitude for $ \Lambda_b \rightarrow \Lambda \ell^+  \ell^-$, $\Sigma_b  \rightarrow \Sigma \ell^+  \ell^-$  and $\Xi_b \rightarrow \Xi \ell^+  \ell^-$ baryonic transitions is computed by evaluating the matrix element of the effective weak Hamiltonian between the initial and final baryonic states,
\begin{eqnarray}\label{amplitude}
{\cal M}^{ {\cal B}_{Q}  \rightarrow {\cal B}  \ell^+ \ell^-} = \langle {\cal B} (p) \mid{\cal H}^{eff}\mid 
{\cal B}_{Q} (p+q,s) \rangle~.
\end{eqnarray}
In this framework, ${\cal B}$ denotes the $\Lambda$, $\Sigma$, and $\Xi$ baryons, while $Q$ refers to the $b$ quark.
To evaluate the transition amplitude, we employ the corresponding transition matrix elements, which are parametrized in terms of twelve independent form factors within the full QCD formalism, i.e., without invoking any approximations based on the heavy quark mass expansion or the large energy limit of hadrons.

The computation of the amplitude necessitates knowledge of the following matrix elements, parameterized through twelve form factors in full QCD:
\begin{eqnarray} \langle
B(p_{B}) \mid  \bar s \gamma_\mu (1-\gamma_5) b \mid B_{Q}(p_{B_{Q}})\rangle\es
\bar {u}_B(p_{B}) \Bigg[\gamma_{\mu}f_{1}(q^{2})+{i}
\sigma_{\mu\nu}q^{\nu}f_{2}(q^{2}) + q^{\mu}f_{3}(q^{2}) \nnb \\
\ek \gamma_{\mu}\gamma_5
g_{1}(q^{2})-{i}\sigma_{\mu\nu}\gamma_5q^{\nu}g_{2}(q^{2})
- q^{\mu}\gamma_5 g_{3}(q^{2})
\vphantom{\int_0^{x_2}}\Bigg] u_{B_{Q}}(p_{B_{Q}})~,
\end{eqnarray}
\begin{eqnarray} \langle
B(p_{B}) \mid  \bar s \gamma_\mu (1+\gamma_5) b \mid B_{Q}(p_{B_{Q}})\rangle\es
\bar {u}_B(p_{B}) \Bigg[\gamma_{\mu}f_{1}(q^{2})+{i}
\sigma_{\mu\nu}q^{\nu}f_{2}(q^{2}) + q^{\mu}f_{3}(q^{2}) \nnb \\
&+& \gamma_{\mu}\gamma_5
g_{1}(q^{2})+{i}\sigma_{\mu\nu}\gamma_5q^{\nu}g_{2}(q^{2})
+ q^{\mu}\gamma_5 g_{3}(q^{2})
\vphantom{\int_0^{x_2}}\Bigg] u_{B_{Q}}(p_{B_{Q}})~,
\end{eqnarray}
\begin{eqnarray}
\langle B(p_{B})\mid \bar s i \sigma_{\mu\nu}q^{\nu} (1+ \gamma_5)
b \mid B_{Q}(p_{B_{Q}})\rangle \es\bar{u}_B(p_{B})
\Bigg[\gamma_{\mu}f_{1}^{T}(q^{2})+{i}\sigma_{\mu\nu}q^{\nu}f_{2}^{T}(q^{2})+
q^{\mu}f_{3}^{T}(q^{2}) \nnb \\
\ar \gamma_{\mu}\gamma_5
g_{1}^{T}(q^{2})+{i}\sigma_{\mu\nu}\gamma_5q^{\nu}g_{2}^{T}(q^{2})
+ q^{\mu}\gamma_5 g_{3}^{T}(q^{2})
\vphantom{\int_0^{x_2}}\Bigg] u_{B_{Q}}(p_{B_{Q}})~,
\end{eqnarray}
\begin{eqnarray}
\langle B(p_{B})\mid \bar s i \sigma_{\mu\nu}q^{\nu} (1- \gamma_5)
b \mid B_{Q}(p_{B_{Q}})\rangle \es\bar{u}_B(p_{B})
\Bigg[\gamma_{\mu}f_{1}^{T}(q^{2})+{i}\sigma_{\mu\nu}q^{\nu}f_{2}^{T}(q^{2})+
q^{\mu}f_{3}^{T}(q^{2}) \nnb \\
\ek \gamma_{\mu}\gamma_5
g_{1}^{T}(q^{2})-{i}\sigma_{\mu\nu}\gamma_5q^{\nu}g_{2}^{T}(q^{2})
- q^{\mu}\gamma_5 g_{3}^{T}(q^{2})
\vphantom{\int_0^{x_2}}\Bigg] u_{B_{Q}}(p_{B_{Q}})~,
\end{eqnarray}
\begin{eqnarray}\langle
B(p_{B}) \mid  \bar s (1+\gamma_5) b \mid B_{Q}(p_{B_{Q}})\rangle\es
{1 \over m_b} \bar {u}_B(p_{B}) \Bigg[{\not\!q}f_{1}(q^{2})+{i}
q^{\mu} \sigma_{\mu\nu}q^{\nu}f_{2}(q^{2}) + q^{2}f_{3}(q^{2}) \nnb \\
\ek {\not\!q}\gamma_5
g_{1}(q^{2})-{i}q^{\mu}\sigma_{\mu\nu}\gamma_5q^{\nu}g_{2}(q^{2})
- q^{2}\gamma_5 g_{3}(q^{2})
\vphantom{\int_0^{x_2}}\Bigg] u_{B_{Q}}(p_{B_{Q}})~,
\end{eqnarray}
and
\begin{eqnarray}\langle
B(p_{B}) \mid  \bar s (1-\gamma_5) b \mid B_{Q}(p_{B_{Q}})\rangle\es
{1 \over m_b} \bar {u}_B(p_{B}) \Bigg[{\not\!q} f_{1}(q^{2})+{i}
q^{\mu} \sigma_{\mu\nu}q^{\nu}f_{2}(q^{2}) + q^{2}f_{3}(q^{2}) \nnb \\
\ar {\not\!q}\gamma_5
g_{1}(q^{2})+{i}q^{\mu} \sigma_{\mu\nu}\gamma_5q^{\nu}g_{2}(q^{2})
+ q^{2}\gamma_5 g_{3}(q^{2})
\vphantom{\int_0^{x_2}}\Bigg] u_{B_{Q}}(p_{B_{Q}})~,
\end{eqnarray}
Here, $f^{(T)}i$ and $g^{(T)}i$ (with $i=1,2,3$) denote the transition form factors, while $u_{{\cal B}_{Q}}$ and ${u}_{\cal B}$ represent the spinors of the baryons under consideration. The form factors relevant for the $ \Lambda_b \rightarrow \Lambda \ell^+ \ell^-$, $\Sigma_b \rightarrow \Sigma \ell^+ \ell^-$ and $\Xi_b \rightarrow \Xi \ell^+ \ell^-$ transitions, obtained within the full theory using light-cone QCD sum rules, are adopted from \cite{Aliev:2010uy,Azizi:2011if,Azizi:2011mw}. For the $ \Lambda_b $ channel, results derived from alternative phenomenological approaches can be found in, for example, \cite{Feldmann:2011xf,Boer:2014kda,Wang:2015ndk}. In addition, lattice QCD calculations for the $ \Lambda $ channel have been reported in \cite{Detmold:2016pkz}. There is also a lattice QCD study within the SM framework that presents a calculation of the form factors for the $\Xi_b^- \to \Xi^-$ transition, providing nonperturbative inputs relevant to rare decay channels \cite{ Farrell:2026swf}. 

By employing the transition matrix elements expressed through the form factors, we derive the corresponding transition amplitudes within various scenarios. In the case of the SM, the result is obtained as follows:
\begin{eqnarray}\label{amplitude1}
{\cal M}_{SM}^{ B_{Q} \rightarrow B \ell^+ \ell^-} &=& {G_F \alpha_{em} V_{tb}V_{ts}^\ast \over 2\sqrt{2} \pi} 
\Bigg\{\Big[{\bar u}_B ({p}_{B}) ( \gamma_{\mu}[{\cal A}_1^{SM} R + {\cal B}_1^{SM} L] + 
{i}\sigma_{\mu\nu} q^{\nu}[{\cal A}_2^{SM} R + {\cal B}_2^{SM} L] \nnb \\
&+& q^{\mu} [{\cal A}_3^{SM} R + {\cal B}_3^{SM} L]) 
u_{B_{Q}}(p_{B_{Q}}) \Big] \, (\bar{\ell} 
\gamma^\mu \ell)\nnb \\
&+& \Big[{\bar u}_B ({p}_{B})( \gamma_{\mu}[{\cal D}_1^{SM} R + {\cal E}_1^{SM} L]+ {i}\sigma_{\mu\nu} q^{\nu}[{\cal D}_2^{SM} R 
+{\cal E}_2^{SM} L] \nnb \\
&+& q^{\mu} [{\cal D}_3^{SM} R + {\cal E}_3^{SM} L]) u_{B_{Q}}(p_{B_{Q}}) \Big] \,(\bar{\ell} 
\gamma^\mu \gamma_5 \ell) \Bigg\} ~.\nnb \\
\end{eqnarray}
When the computation is carried out in the context of the 2HDM, the transition amplitude takes the form:

\begin{eqnarray}\label{amplitude1}
{\cal M}^{\mathrm{2HDM}}_{B_{Q} \rightarrow B  \ell^+ \ell^-} &=& 
\frac{G_F \, \alpha_{\mathrm{em}} \, V_{tb} V_{ts}^\ast}{2\sqrt{2}\pi} 
\Bigg\{ \Big[ \bar{u}_B ({p}_{B}) 
\left( \gamma_\mu[{\cal A}_1^{\mathrm{2HDM}} R + {\cal B}_1^{\mathrm{2HDM}} L] 
+ i\sigma_{\mu\nu} q^\nu [{\cal A}_2^{\mathrm{2HDM}} R + {\cal B}_2^{\mathrm{2HDM}} L] \right. \nonumber \\
&+& \left. q^\mu [{\cal A}_3^{\mathrm{2HDM}} R + {\cal B}_3^{\mathrm{2HDM}} L] \right) 
u_{B_{Q}}(p_{B_{Q}}) \Big] (\bar{\ell} \gamma^\mu \ell) \nonumber \\
&+& \Big[ \bar{u}_B ({p}_{B}) 
\left( \gamma_\mu[{\cal D}_1^{\mathrm{2HDM}} R + {\cal E}_1^{\mathrm{2HDM}} L] 
+ i\sigma_{\mu\nu} q^\nu [{\cal D}_2^{\mathrm{2HDM}} R + {\cal E}_2^{\mathrm{2HDM}} L] \right. \nonumber \\
&+& \left. q^\mu [{\cal D}_3^{\mathrm{2HDM}} R + {\cal E}_3^{\mathrm{2HDM}} L] \right) 
u_{B_{Q}}(p_{B_{Q}}) \Big] (\bar{\ell} \gamma^\mu \gamma_5 \ell) \nonumber \\
&+& \Big[ \bar{u}_B ({p}_{B})
\left( \slashed{q}[{\cal G}_1^{\mathrm{2HDM}} R + {\cal H}_1^{\mathrm{2HDM}} L] 
+ i q^\mu \sigma_{\mu\nu} q^\nu [{\cal G}_2^{\mathrm{2HDM}} R + {\cal H}_2^{\mathrm{2HDM}} L] \right. \nonumber \\
&+& \left. q^2 [{\cal G}_3^{\mathrm{2HDM}} R + {\cal H}_3^{\mathrm{2HDM}} L] \right) 
u_{B_{Q}}(p_{B_{Q}}) \Big] (\bar{\ell} \ell) \nonumber \\
&+& \Big[ \bar{u}_B ({p}_{B})
\left( \slashed{q}[{\cal K}_1^{\mathrm{2HDM}} R + {\cal S}_1^{\mathrm{2HDM}} L] 
+ i q^\mu \sigma_{\mu\nu} q^\nu [{\cal K}_2^{\mathrm{2HDM}} R + {\cal S}_2^{\mathrm{2HDM}} L] \right. \nonumber \\
&+& \left. q^2 [{\cal K}_3^{\mathrm{2HDM}} R + {\cal S}_3^{\mathrm{2HDM}} L] \right) 
u_{B_{Q}}(p_{B_{Q}}) \Big] (\bar{\ell} \gamma_5 \ell) \Bigg\} ~. 
\end{eqnarray}
where as previously said, $R=(1+\gamma_5)/2$ denotes the right-handed projector and $L=(1-\gamma_5)/2$ the left-handed projector. 
In the above expressions, the calligraphic coefficients are introduced in 2HDM as
\bea \label{coef-decay-rate} {\cal A}_{1} \es (f_1 - g_1) C_{9}^{eff} - 2 m_b {1\over q^2} (f_1^T + 
g_1^T) C_{7}^{eff} ,~ {\cal A}_{2} = {\cal A}_1 ( 1 \rar 2 ),~ {\cal A}_{3} = {\cal A}_1 \ga 1 \rar 3 \dr ~,
\eea
\bea
{\cal B}_{1} \es (f_1 + g_1 ) C_{9}^{eff} - 2 m_b {1\over q^2} ( f_1^T - 
g_1^T) C_{7}^{eff},~ {\cal B}_{2} = {\cal B}_1 \ga 1 \rar 2 \dr, ~{\cal B}_{3} = {\cal B}_1 \ga 1 \rar 3 \dr ~,
\eea
\bea
{\cal D}_{1} = (f_1 - g_1) C_{10},~~~~~~~~~~~~~~~~ {\cal D}_{2} =  {\cal D}_1 \ga 1 \rar 2 \dr,~
{\cal D}_{3} = {\cal D}_1 \ga 1 \rar 3 \dr,~
\eea
\bea
{\cal E}_{1} = (f_1  + g_1) C_{10},~~~~~~~~~~~~~~~~~ {\cal E}_{2} = {\cal E}_1 \ga 1 \rar 2 \dr,~ 
{\cal E}_{3} = {\cal E}_1 \ga 1 \rar 3 \dr,~
\eea
\bea
{\cal G}_{1} = {1\over m_b}  (f_1 - g_1 ) C_{Q_1} ,~~~~~~~~~ {\cal G}_{2} = {\cal G}_1 \ga 1 \rar 2 \dr,~ 
{\cal G}_{3} = {\cal G}_1 \ga 1 \rar 3 \dr,~
\eea
\bea
{\cal H}_{1} = {1\over m_b} ( f_1  + g_1 )C_{Q_1}  ,~~~~~~~~ {\cal H}_{2} = {\cal H}_1 \ga 1 \rar 2 \dr,~
{\cal H}_{3} =  {\cal H}_1 \ga 1 \rar 3 \dr,~
\eea
\bea
{\cal K}_{1} = {1\over m_b} (f_1  - g_1) C_{Q_2},~~~~~~~~~ {\cal K}_{2} =  {\cal K}_1 \ga 1 \rar 2 \dr,~
{\cal K}_{3} =  {\cal K}_1 \ga 1 \rar 3 \dr,~
\eea
\bea
{\cal S}_{1} = {1\over m_b} ( f_1 + g_1) C_{Q_2} ,~~~~~~~~~~ {\cal S}_{2} = {\cal S}_1 \ga 1 \rar 2 \dr,~
{\cal S}_{3} = {\cal S}_1 \ga 1 \rar 3 \dr,~
\eea
\section{Numerical analysis and physical observables}
In this section, the key physical observables for the $ \Lambda_b \rightarrow \Lambda \ell^+  \ell^-$, $\Sigma_b  \rightarrow \Sigma \ell^+  \ell^-$  and $\Xi_b \rightarrow \Xi \ell^+  \ell^-$ decay channels are systematically analyzed. For each channel, the differential decay widths and differential branching ratios are calculated, providing detailed information on the $q^2$-dependence of the transitions. In addition, the total branching ratios are evaluated, offering a global measure of the decay probabilities. Moreover, the lepton forward–backward asymmetries, which characterize the angular distributions of the final-state leptons, are investigated. The results for all three channels are presented with comprehensive graphical representations, allowing for a direct comparison between the predictions of the SM and the  2HDM with Type-III. The accompanying discussion interprets these findings, highlighting the effects of the extended Higgs sector on the various observables and identifying the kinematic regions where deviations from the SM expectations are most pronounced.
\subsection{The differential decay width}
In the present subsection, we proceed to derive the differential decay width corresponding to the decay channels under consideration.  
By employing the decay amplitude together with the transition matrix elements expressed in terms of form factors,  
the differential decay rate in the 2HDM — which constitutes the most general case among the model examined — is obtained as
\bea\label{DDR} \frac{d^2\Gamma_{2HDM}}{d\hat
sdz}(z,\hat s) = \frac{G_F^2\alpha^2_{em} m_{B_Q}}{16384
\pi^5}| V_{tb}V_{ts}^*|^2 v \sqrt{B(1,r,\hat s)} \, \Bigg[{\cal
T}_{0}^{2HDM}(\hat s)+{\cal T}_{1}^{2HDM}(\hat s) z +{\cal T}_{2}^{2HDM}(\hat s)
z^2\Bigg]~,  \label{dif-decay}
\eea
where $z=\cos\theta$ with $\theta$ being the angle between the momenta of the lepton $l^+$ and the  $B_Q$ in the center 
of mass of leptons, $v=\sqrt{1-\frac{4 m_\ell^2}{q^2}}$ is the lepton velocity, $\lambda=\lambda(1,r,\hat s)=(1-r-\hat s)^2-4r\hat s$ 
is the usual triangle function, $\hat s= q^2/m^2_{B_Q}$ and $r= m^2_{B}/m^2_{B_Q}$. The functions 
${\cal T}_{0}^{2HDM}(\hat s)$, ${\cal T}_{1}^{2HDM}(\hat s)$ and ${\cal T}_{2}^{2HDM}(\hat s)$ are obtained  as
\bea {\cal T}_{0}^{2HDM}(\hat s) \es 32 m_\ell^2
m_{B_Q}^4 \hat s (1+r-\hat s) \Big( \vel {\cal D}_{3} \ver^2 +
\vel {\cal E}_{3} \ver^2 \Big) \nnb \\
\ar 64 m_\ell^2 m_{B_Q}^3 (1-r-\hat s) \, \mbox{\rm Re} \Big[{\cal D}_{1}^\ast
{\cal E}_{3} + {\cal D}_{3}
{\cal E}_1^\ast \Big] \nnb \\
\ar 64 m_{B_Q}^2 \sqrt{r} (6 m_\ell^2 - m_{B_Q}^2 \hat s)
{\rm Re} \Big[{\cal D}_{1}^\ast {\cal E}_{1}\Big] \nnb\\ 
\ar 64 m_\ell^2 m_{B_Q}^3 \sqrt{r} \Bigg\{ 2 m_{B_Q} \hat s
{\rm Re} \Big[{\cal D}_{3}^\ast {\cal E}_{3}\Big] + (1 - r + \hat s)
{\rm Re} \Big[{\cal D}_{1}^\ast {\cal D}_{3} + {\cal E}_{1}^\ast {\cal E}_{3}\Big]\Bigg\} \nnb \\
\ar 32 m_{B_Q}^2 (2 m_\ell^2 + m_{B_Q}^2 \hat s) \Bigg\{ (1
- r + \hat s) m_{B_Q} \sqrt{r} \,
\mbox{\rm Re} \Big[{\cal A}_{1}^\ast {\cal A}_{2} + {\cal B}_{1}^\ast {\cal B}_{2}\Big] \nnb \\
\ek m_{B_Q} (1 - r - \hat s) \, \mbox{\rm Re} \Big[{\cal A}_{1}^\ast {\cal B}_{2} +
{\cal A}_{2}^\ast {\cal B}_{1}\Big] - 2 \sqrt{r} \Big( \mbox{\rm Re} \Big[{\cal A}_{1}^\ast {\cal B}_{1}\Big] +
m_{B_Q}^2 \hat s \,
\mbox{\rm Re} \Big[{\cal A}_{2}^\ast {\cal B}_{2}\Big] \Big) \Bigg\} \nnb \\
\ar 8 m_{B_Q}^2 \Bigg\{ 4 m_\ell^2 (1 + r - \hat s) +
m_{B_Q}^2 \Big[(1-r)^2 - \hat s^2 \Big]
\Bigg\} \Big( \vel {\cal A}_{1} \ver^2 +  \vel {\cal B}_{1} \ver^2 \Big) \nnb \\
\ar 8 m_{B_Q}^4 \Bigg\{ 4 m_\ell^2 \Big[ \lambda + (1 + r -
\hat s) \hat s \Big] + m_{B_Q}^2 \hat s \Big[(1-r)^2 - \hat s^2 \Big]
\Bigg\} \Big( \vel {\cal A}_{2} \ver^2 +  \vel {\cal B}_{2} \ver^2 \Big) \nnb \\
\ek 8 m_{B_Q}^2 \Bigg\{ 4 m_\ell^2 (1 + r - \hat s) -
m_{B_Q}^2 \Big[(1-r)^2 - \hat s^2 \Big]
\Bigg\} \Big( \vel {\cal D}_{1} \ver^2 +  \vel {\cal E}_{1} \ver^2 \Big) \nnb\\
\ar 8 m_{B_Q}^5 \hat s v^2 \Bigg\{ - 8 m_{B_Q} \hat s \sqrt{r}\,
\mbox{\rm Re} \Big[{\cal D}_{2}^\ast {\cal E}_{2}\Big] +
4 (1 - r + \hat s) \sqrt{r} \, \mbox{\rm Re}\Big[{\cal D}_{1}^\ast {\cal D}_{2}+{\cal E}_{1}^\ast {\cal E}_{2}\Big]\nnb \\
\ek 4 (1 - r - \hat s) \, \mbox{\rm Re}\Big[{\cal D}_{1}^\ast {\cal E}_{2}+{\cal D}_{2}^\ast {\cal E}_{1}\Big] +
m_{B_Q} \Big[(1-r)^2 -\hat s^2\Big] \Big( \vel {\cal D}_{2} \ver^2 + \vel
{\cal E}_{2} \ver^2 \Big) \Bigg\} \nnb \\
\ek 8 m_{B_Q}^4 \Bigg\{ 4 m_\ell \Big[(1-r)^2 -\hat s(1+r) \Big]\, \mbox{\rm Re} \Big[{\cal D}_{1}^\ast {\cal K}_{1}
+{\cal E}_{1}^\ast {\cal S}_{1}\Big] \nnb \\ 
\ar (4 m_\ell^2 - m_{B_Q}^2 \hat s) \Big[(1-r)^2 -\hat s(1+r) \Big]\, \Big( \vel {\cal G}_{1} \ver^2 
+ \vel {\cal H}_{1} \ver^2 \Big) \nnb \\
\ar 4 m_{B_Q}^2 \sqrt{r} \hat s^2 (4 m_\ell^2 
- m_{B_Q}^2 \hat s) \, \mbox{\rm Re}\Big[{\cal G}_{3}^\ast {\cal H}_{3}\Big] \Bigg\} \nnb \\
\ek 8 m_{B_Q}^5 \hat s \Bigg\{2 \sqrt{r} (4 m_\ell^2 
- m_{B_Q}^2 \hat s) \,(1 - r + \hat s) \, \mbox{\rm Re}\Big[{\cal G}_{1}^\ast {\cal G}_{3}+{\cal H}_{1}^\ast {\cal H}_{3}\Big] \nnb \\
\ar 4 m_\ell \sqrt{r}(1 - r + \hat s) \mbox{\rm Re}\Big[{\cal D}_{1}^\ast {\cal K}_{3}
+{\cal E}_{1}^\ast {\cal S}_{3}+{\cal D}_{3}^\ast {\cal K}_{1}+{\cal E}_{3}^\ast {\cal S}_{1}\Big] \nnb 
\eea
\bea
\ar 4 m_\ell (1 - r - \hat s) \mbox{\rm Re}\Big[{\cal D}_{1}^\ast {\cal S}_{3}+{\cal E}_{1}^\ast {\cal K}_{3}
+{\cal D}_{3}^\ast {\cal S}_{1}+{\cal E}_{3}^\ast {\cal K}_{1}\Big] \nnb \\
\ar 2 (1 - r - \hat s) (4 m_\ell^2 - m_{B_Q}^2 \hat s) \, \mbox{\rm Re}\Big[{\cal G}_{1}^\ast {\cal H}_{3}
+{\cal H}_{1}^\ast {\cal G}_{3}\Big] \nnb \\
\ek m_{B_Q} \Big[(1-r)^2 -\hat s(1+r) \Big] \Big( \vel {\cal K}_{1} \ver^2 +  \vel {\cal S}_{1} \ver^2 \Big) \Bigg\} \nnb \\
\ek 32 m_{B_Q}^4 \sqrt{r} \hat s \Bigg\{ 2 m_\ell \mbox{\rm Re}\Big[{\cal D}_{1}^\ast {\cal S}_{1}
+{\cal E}_{1}^\ast {\cal K}_{1}\Big]+(4 m_\ell^2 - m_{B_Q}^2 \hat s) \,\mbox{\rm Re}\Big[{\cal G}_{1}^\ast {\cal H}_{1}\Big] \Bigg\} \nnb \\
\ar 8 m_{B_Q}^6 \hat s^2 \Bigg\{ 4 \sqrt{r} \,\mbox{\rm Re}\Big[{\cal K}_{1}^\ast {\cal S}_{1}\Big]
+2 m_{B_Q} \sqrt{r} (1 - r + \hat s) \mbox{\rm Re}\Big[{\cal K}_{1}^\ast {\cal K}_{3}+{\cal S}_{1}^\ast {\cal S}_{3}\Big] \nnb \\
&+& 2 m_{B_Q} (1 - r - \hat s) \mbox{\rm Re}\Big[{\cal K}_{1}^\ast {\cal S}_{3}+{\cal S}_{1}^\ast {\cal K}_{3}\Big] \nnb \\
\ek (4 m_\ell^2 - m_{B_Q}^2 \hat s) (1 + r - \hat s) \Big( \vel {\cal G}_{3} \ver^2 +  \vel {\cal H}_{3} \ver^2 \Big) \nnb \\
\ek 4 m_\ell (1 + r - \hat s) \mbox{\rm Re}\Big[{\cal D}_{3}^\ast {\cal K}_{3}+{\cal E}_{3}^\ast {\cal S}_{3}\Big]
- 8 m_\ell \sqrt{r} \mbox{\rm Re}\Big[{\cal D}_{3}^\ast {\cal S}_{3}+{\cal E}_{3}^\ast {\cal K}_{3}\Big]\Bigg\} \nnb \\
\ar 8 m_{B_Q}^8 \hat s^3 \Bigg\{ (1 + r - \hat s) \Big( \vel {\cal K}_{3} \ver^2 +  \vel {\cal S}_{3} \ver^2 \Big) 
+ 4 \sqrt{r} \mbox{\rm Re}\Big[{\cal K}_{3}^\ast {\cal S}_{3}\Big]\Bigg\}, \nnb \\
\eea
\bea {\cal T}_{1}^{2HDM}(\hat s) &=& -32
m_{B_Q}^4 m_\ell \sqrt{\lambda} v (1 - r)
Re\Big({\cal A}_{1}^* {\cal G}_{1}+{\cal B}_{1}^* {\cal H}_{1}\Big)\nn\\
&-&16 m_{B_Q}^4 \hat s v \sqrt{\lambda} 
\Bigg\{ 2 Re\Big({\cal A}_{1}^* {\cal D}_{1}\Big)-2Re\Big({\cal B}_{1}^* {\cal E}_{1}\Big)\nn\\
&+&2 m_{B_Q} Re\Big({\cal B}_{1}^* {\cal D}_{2}-{\cal B}_{2}^* {\cal D}_{1}+{\cal A}_{2}^* {\cal E}_{1}
-{\cal A}_{1}^*{\cal E}_{2}\Big)\nn\\
&+&2 m_{B_Q} m_\ell Re\Big({\cal A}_{1}^* {\cal H}_{3}+{\cal B}_{1}^* {\cal G}_{3}-{\cal A}_{2}^* {\cal H}_{1}
-{\cal B}_{2}^*{\cal G}_{1}\Big)\Bigg\}\nn\\
&+&32 m_{B_Q}^5 \hat s~ v \sqrt{\lambda} \Bigg\{
m_{B_Q} (1-r)Re\Big({\cal A}_{2}^* {\cal D}_{2} -{\cal B}_{2}^* {\cal E}_{2}\Big)\nn\\
&+& \sqrt{r} Re\Big({\cal A}_{2}^* {\cal D}_{1}+{\cal A}_{1}^* {\cal D}_{2}-{\cal B}_{2}^*{\cal E}_{1}
-{\cal B}_{1}^* {\cal E}_{2}\Big)\nn\\
&-& \sqrt{r} m_\ell Re\Big({\cal A}_{1}^* {\cal G}_{3}+{\cal B}_{1}^* {\cal H}_{3}+{\cal A}_{2}^*{\cal G}_{1}
+{\cal B}_{2}^* {\cal H}_{1}\Big)\Bigg\} \nn\\
&+&32 m_{B_Q}^6 m_\ell \sqrt{\lambda} v \hat s^2 Re\Big({\cal A}_{2}^* {\cal G}_{3}+{\cal B}_{2}^* {\cal H}_{3}\Big),\nn\\
\eea\
and
\bea {\cal T}_{2}^{2HDM}(\hat s) \es - 8 m_{B_Q}^4 v^2 \lambda \Big(\vel {\cal A}_{1} \ver^2 + \vel {\cal B}_{1} \ver^2 
+ \vel {\cal D}_{1} \ver^2 + \vel {\cal E}_{1} \ver^2 \Big) \nnb \\
\ar 8 m_{B_Q}^6 \hat s v^2 \lambda \Big( \vel {\cal A}_{2} \ver^2 + \vel
{\cal B}_{2} \ver^2 + \vel {\cal D}_{2} \ver^2 + \vel {\cal E}_{2} \ver^2 \Big) ~.
\eea\
Integrating Eq.\eqref{DDR} over $z$ in the interval $[-1,1]$, we obtain the differential decay width only in terms of $ \hat s$ as 
\begin{eqnarray}
\frac{d\Gamma_{2HDM}}{d \hat s} (\hat s)= \frac{G_F^2\alpha^2_{em} m_{B_Q}}{8192
\pi^5}| V_{tb}V_{ts}^*|^2 v \sqrt{\lambda} \, \Bigg[{{\cal T}_0^{2HDM}(\hat s)
+\frac{1}{3} {\cal T}_2^{2HDM}(\hat s)}\Bigg]~. \label{decayrate} 
\end{eqnarray}
\subsection{The differential branching ratio}
In this subsection, we present the differential branching ratio as a function of $q^2$ for the $\Lambda_b \rightarrow \Lambda \ell^+ \ell^-$ and $\Xi_b \rightarrow \Xi \ell^+ \ell^-$ decays within the 2HDM framework. 
For the $\Sigma_b \rightarrow \Sigma \ell^+ \ell^-$ channel, we instead compute the differential decay width $\mathrm{d}\Gamma/\mathrm{d}q^2$. To illustrate the dependence of the differential branching ratio on $q^2$, we need the values  in Table~\ref{tab:T1} as input   parameters, together with the form factors \cite{Azizi:2011mw,Aliev:2010uy,Azizi:2011if} that constitute the main inputs. It is worth noting that for the $\Lambda_b \to \Lambda$ transition, more recent alternative parameterizations are available in the literature, computed via modern Light-Cone Sum Rules (LCSR) at next-to-leading order \cite{Wang:2015ndk} and lattice QCD simulations \cite{Detmold:2016pkz}. Both of these frameworks successfully evaluate the complete set of 12 independent helicity form factors. However, since corresponding lattice QCD data are currently unavailable for the $\Sigma_b$ and $\Xi_b$ channels, sum rules remain the only viable option for these modes. Therefore, in order to maintain strict theoretical homogeneity, structural consistency, and a unified systematic error framework across all three decaying channels ($\Lambda_b \to \Lambda$, $\Sigma_b \to \Sigma$, and $\Xi_b \to \Xi$), we adopt the comprehensive full-theory QCD sum rules framework derived in \cite{Aliev:2010uy}, which consistently provides the complete set of 12 transition and tensor form factors for all modes under consideration. The $\Sigma_b$ baryon is a recently studied heavy baryon whose internal structure may involve admixture effects; therefore, its lifetime has not yet been precisely determined either experimentally or theoretically. Due to this uncertainty, it is not appropriate to present a well-defined branching ratio for the decay channel under consideration. Instead, our analysis focuses on the differential decay rate and the partial decay width associated with the relevant transition. The total decay width is taken from the available values reported in the literature (as summarized in the Table~\ref{tab:T1}) and is used as an external input parameter.
\begin{table}[ht]
\centering
\rowcolors{1}{lightgray}{white}
\begin{tabular}{cc}
\hline \hline
   Some Input Parameters  &  Values    \\
\hline \hline
$ m_\mu $            &   $ 0.105658 $ GeV \\
$ m_\tau $           &   $ 1.77686 $ GeV \\
$ m_c $              &   $ 1.275 \pm 0.025 $ GeV \\
$ m_b $              &   $ 4.18  \pm 0.03 $ GeV \\
$ m_t $              &   $ 162.5 $ GeV \\
$ m_W $              &   $ 80.379 $ GeV \\
$ m_{\Lambda_b} $    &   $ 5.6195 \pm 0.0017 $ GeV \\
$ m_{\Lambda} $      &   $ 1.11568\pm 0.000006 $ GeV \\
$ \tau_{\Lambda_b} $ &   $ (1.471 \pm 0.009) \times 10^{-12} \,\text{s} $ \\
$ m_{\Sigma_b} $    &   $ 5.810 \pm 0.00025 $ GeV \\
$ m_{\Sigma} $      &   $ 1.19264 \pm 0.000024$ GeV \\
$ m_{\Xi_b} $       &   $ 5.791 \pm 0.0005 $ GeV \\
$ m_{\Xi} $         &   $ 1.314 \pm 0.0002 $ GeV \\
$ \tau_{\Xi_b} $    &   $ (1.480\pm 0.030)\times 10^{-12} \,\text{s} $ \\
$ \hbar  $           &   $ 6.582\times 10^{-25}\,\text{GeV}\,\text{s} $\\
$ G_{F} $            &   $ 1.166\times 10^{-5}\,\text{GeV}^{-2} $ \\
$ \alpha_{em} $      &   $ \tfrac{1}{137} $ \\
$ |V_{tb}V_{ts}^*|$  &   $ 0.040 $ \\
\hline \hline
\end{tabular}
\caption{The values of some input parameters used in our calculations, taken from PDG~\cite{ParticleDataGroup:2024cfk}.}
\label{tab:T1}
\end{table}
%% 2HDM parametreleri verilecek
In the general framework of the 2HDM, the Yukawa interactions of the charged Higgs boson with fermions are parametrized in terms of model-dependent coefficients X and Y, which correspond to the couplings to down-type and up-type quarks, respectively. These parameters enter the effective Hamiltonian through combinations such as $|Y|^2$  and $ XY^*$, representing pure up-type contributions and mixed up--down type interactions, respectively~\cite{Aliev:1999ap,Ciuchini:1997xe}. In our case, they can be obtained from Ref. \cite{Ciuchini:1997xe} with the following substitutions:

\bea
\vel Y \ver^2  \rar \vel \lambda_{tt} \ver^2 ~~~~~ \mbox{\rm and} ~~~~~ 
X Y^*  \rar \vel \lambda_{tt} \lambda_{bb} \ver e^{i\phi}~. 
\eea

In the type-III 2HDM, the mass of the charged Higgs boson $ m_{H^\pm}$ and the Yukawa couplings $ \lambda_{tt}$ and $\lambda_{bb}$ are free parameters of the model. Unlike type-I and type-II realizations, these Yukawa couplings are, in general, complex quantities. In particular, their product can be written as
\begin{equation}
\lambda_{tt}\lambda_{bb} = |\lambda_{tt}\lambda_{bb}| e^{i\phi},
\end{equation}
where \( \phi \) denotes the CP-violating phase associated with the scalar sector.

This complex structure allows the charged Higgs boson contributions to interfere either constructively or destructively with the SM amplitudes, depending on the values of the coupling parameters and the CP phase. As a result, observable quantities such as branching ratios and asymmetries can be significantly enhanced or suppressed compared to their SM predictions.

It is worth emphasizing that the type-I and type-II versions of the 2HDM can be recovered from the type-III model by imposing specific constraints on the Yukawa sector. These constraints reduce the number of free parameters and eliminate tree-level flavor-changing neutral currents (FCNCs), leading to more restrictive but phenomenologically well-controlled scenarios. 

We investigate the behavior of the differential branching ratio $d\mathcal{B}/dq^2$ for the $\Lambda_b \rightarrow \Lambda \ell^+ \ell^-$ transition within both the SM and the 2HDM frameworks. For the neutral Higgs boson sector, which governs the scalar and pseudoscalar Wilson coefficients $C_{Q_1}$ and $C_{Q_2}$, the numerical analysis is performed by adopting the representative mass values $m_{h^0} = 125\text{ GeV}$, $m_{H^0} = \text{500 GeV}$, and $m_{A^0} = \text{500 GeV}$, along with the CP-even Higgs mixing angle configured as $\sin^2\alpha = \text{0.5}$ (or derived through the alignment limit). It is essential to emphasize that although the contributions of $C_{Q_1,Q_2}$ are structurally suppressed by lepton masses, they undergo significant enhancement in the large $\tan\beta$ regime of the 2HDM. Consequently, these neutral Higgs exchanges are fully integrated into our numerical computations, as they generate notable interference effects that modify the fine structure of the differential decay width and angular asymmetries relative to the pure $C_{9,10}$ driven baselines. In particular, we analyze the impact of long-distance contributions on the $q^2$ spectrum by considering different Higgs mass values with $\lambda_{tt}=0.05$, $0.15$, and $0.30$. The plots are generated for different values of the charged Higgs boson mass, specifically $m_{H^\pm} = 175, 250, 500, 750,$ and $1000~\text{GeV}$, while keeping all other model parameters fixed. 

To ensure the phenomenological viability of our parameter space, we explicitly account for the constraints from the rare radiative decay $b \to s\gamma$. The current experimental world average, $\mathcal{B}(B \to X_s \gamma){\text{exp}} = (3.49 \pm 0.19) \times 10^{-4}$, strongly agrees with the NNLO SM prediction of $(3.40 \pm 0.22) \times 10^{-4}$ \cite{Misiak:2015xwa}. While conventional Type-II models impose a strict lower bound of $m{H^\pm} \gtrsim 800\text{ GeV}$, the Type-III 2HDM framework adopted here introduces arbitrary complex Yukawa matrices ($\chi_{ij}$ or $\xi_{ij}$) rather than rigid $\tan\beta$ dependencies. This additional parametric freedom allows for a destructive interference between the SM $W$-boson and the charged Higgs $H^\pm$ loops within the decay amplitude \cite{Crivellin:2012ye, Mahmoudi:2009zx}. By appropriately configuring the flavor-violating couplings governing the $t-b-H^\pm$ and $s-b-H^\pm$ vertices, the theoretical branching fraction safely accommodates a light charged Higgs boson ($m_{H^\pm} = 175\text{ GeV}$) alongside mass-degenerate heavy neutral scalars ($m_{H^0} = m_{A^0} = 500\text{ GeV}$) within the experimental $1\sigma$ error bands.

\begin{widetext}

\begin{figure}[h!]
\begin{center}
\includegraphics[totalheight=12cm,width=16cm]{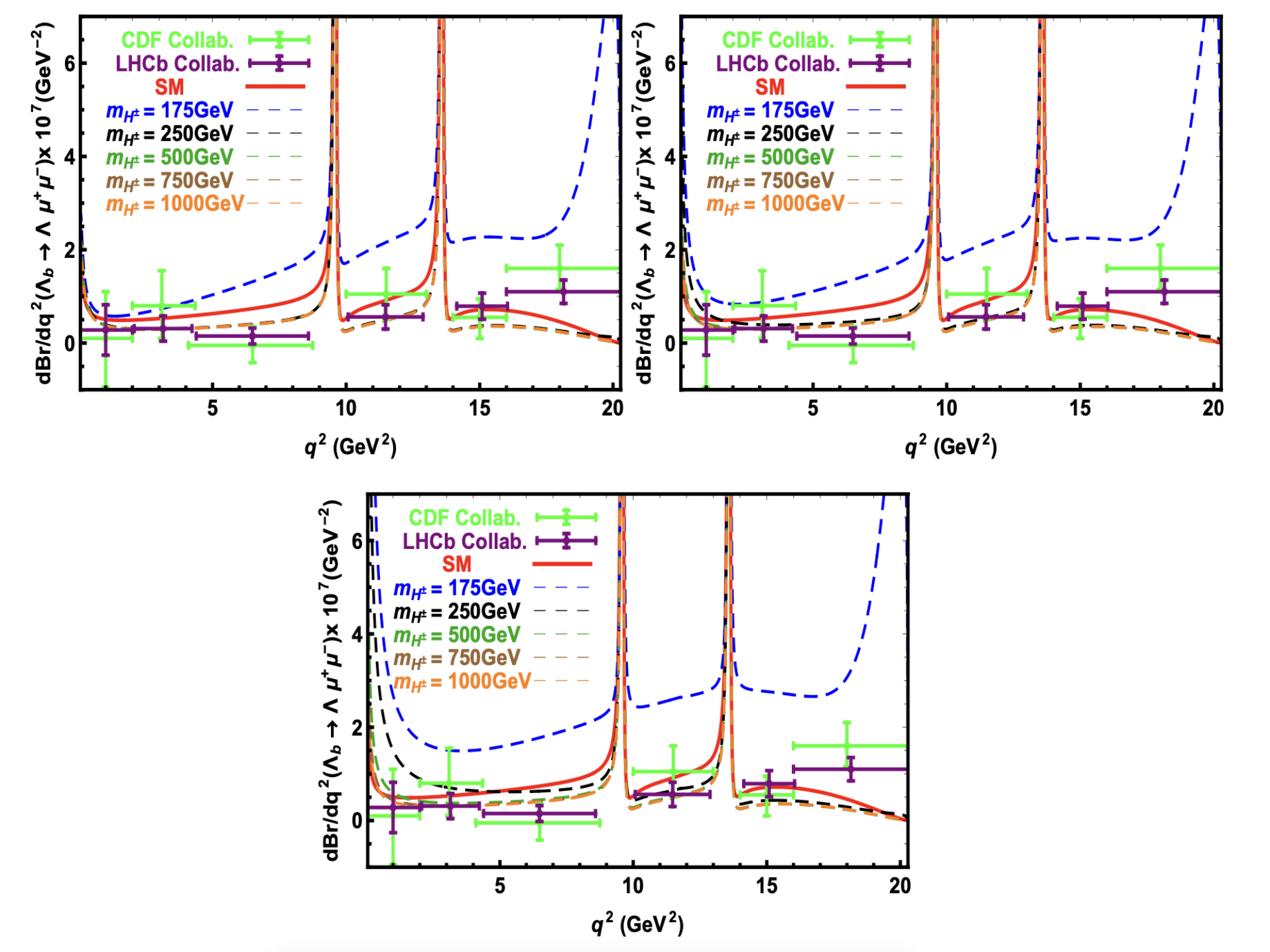}
\end{center}
\caption{The dependence of the $ dBR/dq^2$ on $q^2$  for the $ \Lambda_b \rightarrow \Lambda \mu^+  \mu^-$ transition in  the SM and 2HDM  with long-distance contributions plotted against different Higgs masses for $\lambda_{tt}=0.05$, $\lambda_{tt}=0.15$ and $\lambda_{tt}=0.30$, respectively. The experimental data were obtained from the LHCb and the CDF Collaboration, as cited in Ref.\cite{Aaltonen:2011qs,LHCb:2013uqx} }
\label{fig:F1}
\end{figure}

\end{widetext}

\begin{widetext}

\begin{figure}[h!]
\begin{center}
\includegraphics[totalheight=12cm,width=16cm]{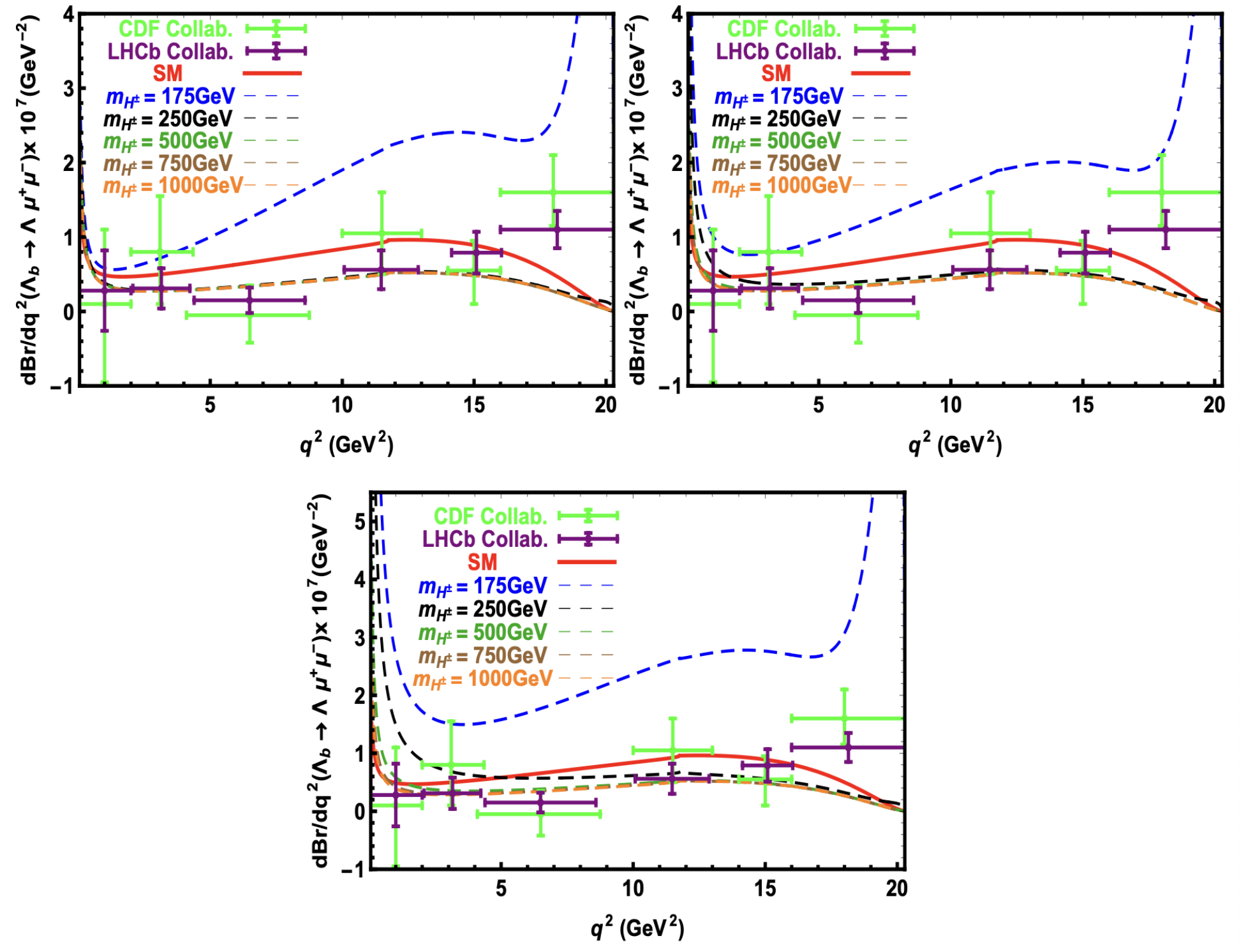}

\end{center}
\caption{The dependence of the $ dBR/dq^2$  on   $q^2$  for the $ \Lambda_b \rightarrow \Lambda \mu^+  \mu^-$ transition in  the SM and 2HDM   without long-distance contributions plotted against different Higgs masses for $\lambda_{tt}=0.05$, $\lambda_{tt}=0.15$ and $\lambda_{tt}=0.30$, respectively. The experimental data were obtained from the LHCb and the CDF Collaboration, as cited in Ref.\cite{Aaltonen:2011qs,LHCb:2013uqx}}
\label{fig:F2}
\end{figure}

\end{widetext}

The theoretical predictions are subsequently compared with the available experimental measurements reported by the LHCb and CDF Collaborations, as documented in Refs.~\cite{CDF:2011buy,LHCb:2013uqx}. It should be noted that these comparisons are restricted to the $ \Lambda_b \rightarrow \Lambda \mu^+  \mu^-$ decay channel, for which experimental data are available. Fig.~\ref{fig:F1} illustrates the results including the long-distance contributions, while Fig.~\ref{fig:F2} displays the corresponding predictions when the long-distance effects are excluded. In the following discussion, we provide a comparative analysis of these scenarios in order to highlight the significance of long-distance contributions.

In Fig.~\ref{fig:F1}, the differential branching ratio exhibits good agreement with the experimental data in the low to intermediate $q^2$ region, specifically for $q^2 = 0$–$16~\text{GeV}^2$. The agreement improves notably for Higgs boson masses exceeding $250~\text{GeV}$, demonstrating closer consistency with the SM predictions. Furthermore, when the parameter $\lambda_{tt}$ is considered, the correspondence with the SM becomes more pronounced. At higher $q^2$ values, beyond approximately $16~\text{GeV}^2$, deviations between the theoretical predictions and the experimental data become more significant. Comparable behavior is observed in Fig.~\ref{fig:F2}, where the long-distance contributions are omitted, indicating that the general trends are robust against the exclusion of these effects. These observations suggest that both the Higgs mass and the parameter $\lambda_{tt}$ play a crucial role in determining the level of agreement between the 2HDM predictions and experimental results.
% for lambda

\begin{widetext}

\begin{figure}[h!]
\begin{center}

\includegraphics[totalheight=12cm,width=16cm]{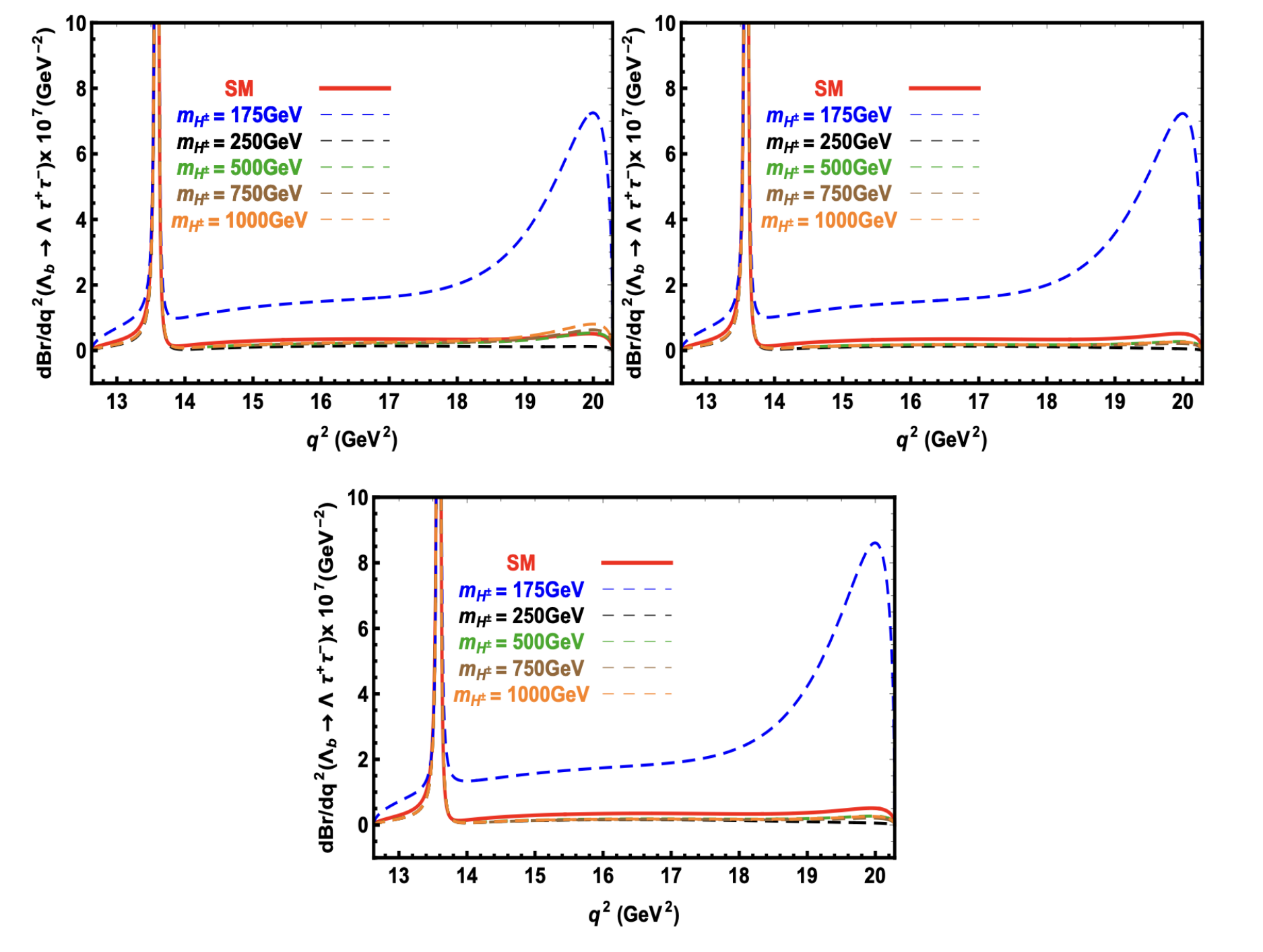}

\end{center}
\caption{The dependence of the $ dBR/dq^2$ on  $q^2$  for the $ \Lambda_b \rightarrow \Lambda \tau^+  \tau^-$ transition in  the SM and 2HDM  with long-distance contributions plotted against different Higgs masses for $\lambda_{tt}=0.05$, $\lambda_{tt}=0.15$ and $\lambda_{tt}=0.30$, respectively.  }
\label{fig:F3}
\end{figure}

\end{widetext}

\begin{widetext}

\begin{figure}[h!]
\begin{center}
\includegraphics[totalheight=12cm,width=16cm]{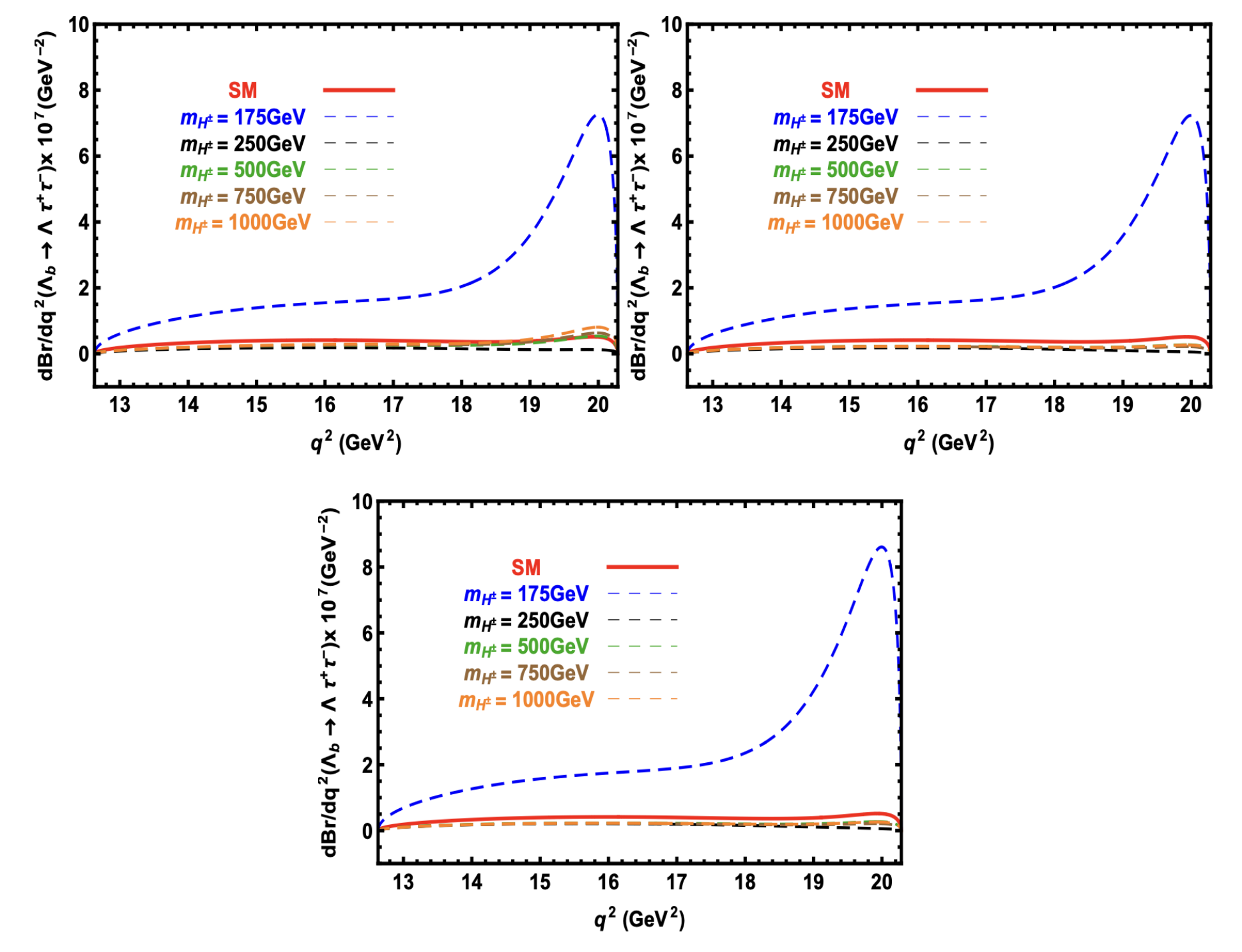}

\end{center}
\caption{The dependence of the $ dBR/dq^2$ on  $q^2$  for the $ \Lambda_b \rightarrow \Lambda \tau^+  \tau^-$ transition in  the SM and 2HDM  without long-distance contributions plotted against different Higgs masses for $\lambda_{tt}=0.05$, $\lambda_{tt}=0.15$ and $\lambda_{tt}=0.30$, respectively.  }
\label{fig:F4}
\end{figure}

\end{widetext}

Figs.~\ref{fig:F3} and ~\ref{fig:F4} illustrate the differential branching ratio for the $\Lambda_b \rightarrow \Lambda \tau^+ \tau^-$ decay channel in the presence of long-distance (LD) contributions and without them, respectively. The plots display the results for both cases sequentially, allowing a direct comparison of the LD effects on the $\tau$ channel. When examining both the LD-included and LD-excluded cases, it is observed that for all values of $q^2$, the scenarios with $m_{H^\pm} = 175$~GeV excluded remain consistent with the SM predictions. Only for $\lambda_{tt} = 0.05$, minor deviations are noticeable at the higher $q^2$ region.

\begin{widetext}

\begin{figure}[h!]
\begin{center}
\includegraphics[totalheight=6cm,width=16cm]{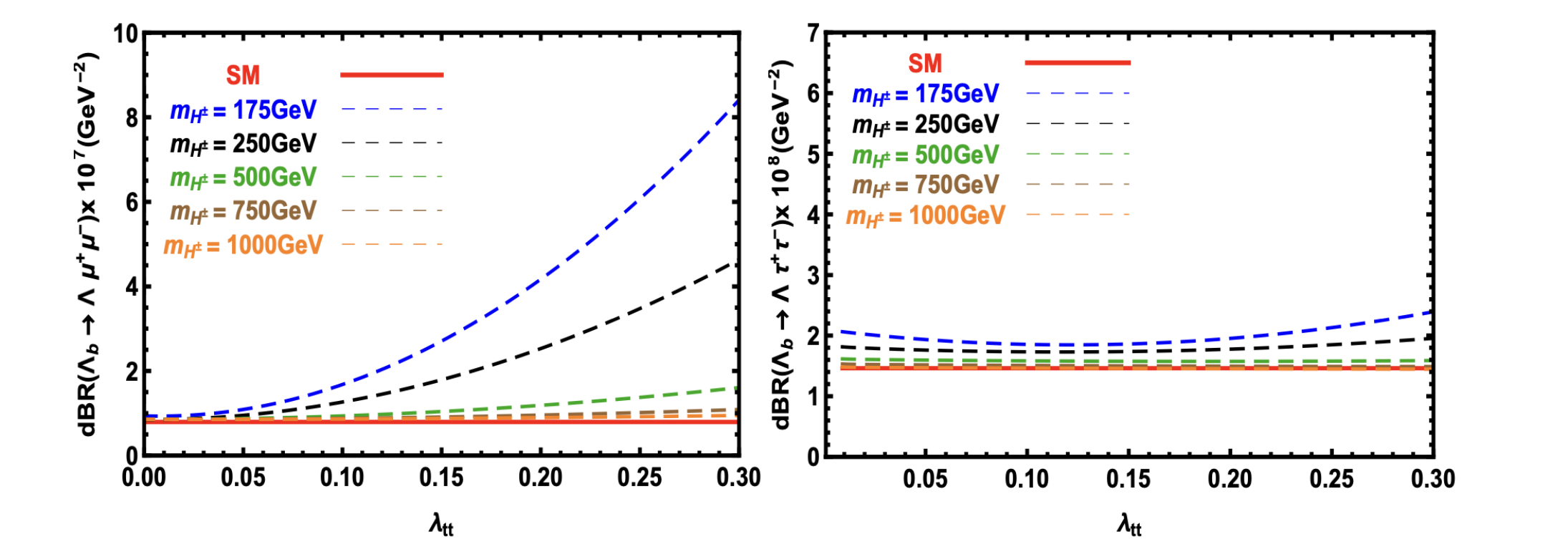}

\end{center}

\caption{The dependence of the $ dBR$  on  $\lambda_{tt}$  for the $ \Lambda_b \rightarrow \Lambda \ell^+  \ell^-$ transition in  the SM and 2HDM.  }
\label{fig:F5}
\end{figure}

\end{widetext}

Fig.~\ref{fig:F5} illustrates the dependence of $dBR$ on $\lambda_{tt}$ for the $ \Lambda_b \rightarrow \Lambda \ell^+ \ell^-$ transition, shown for both $\mu$ and $\tau$ channels. In the $\mu$ channel, the largest deviations are observed at high $q^2$ values, particularly for $m_{H^\pm} = 175$~GeV and $250$~GeV. For other parameter values, the results remain consistent with the SM predictions. In contrast, the $\tau$ channel exhibits behavior fully compatible with the SM across all $q^2$ values.

% for sigma 

\begin{widetext}
\begin{figure}[h!]
\begin{center}
\includegraphics[totalheight=12cm,width=16cm]{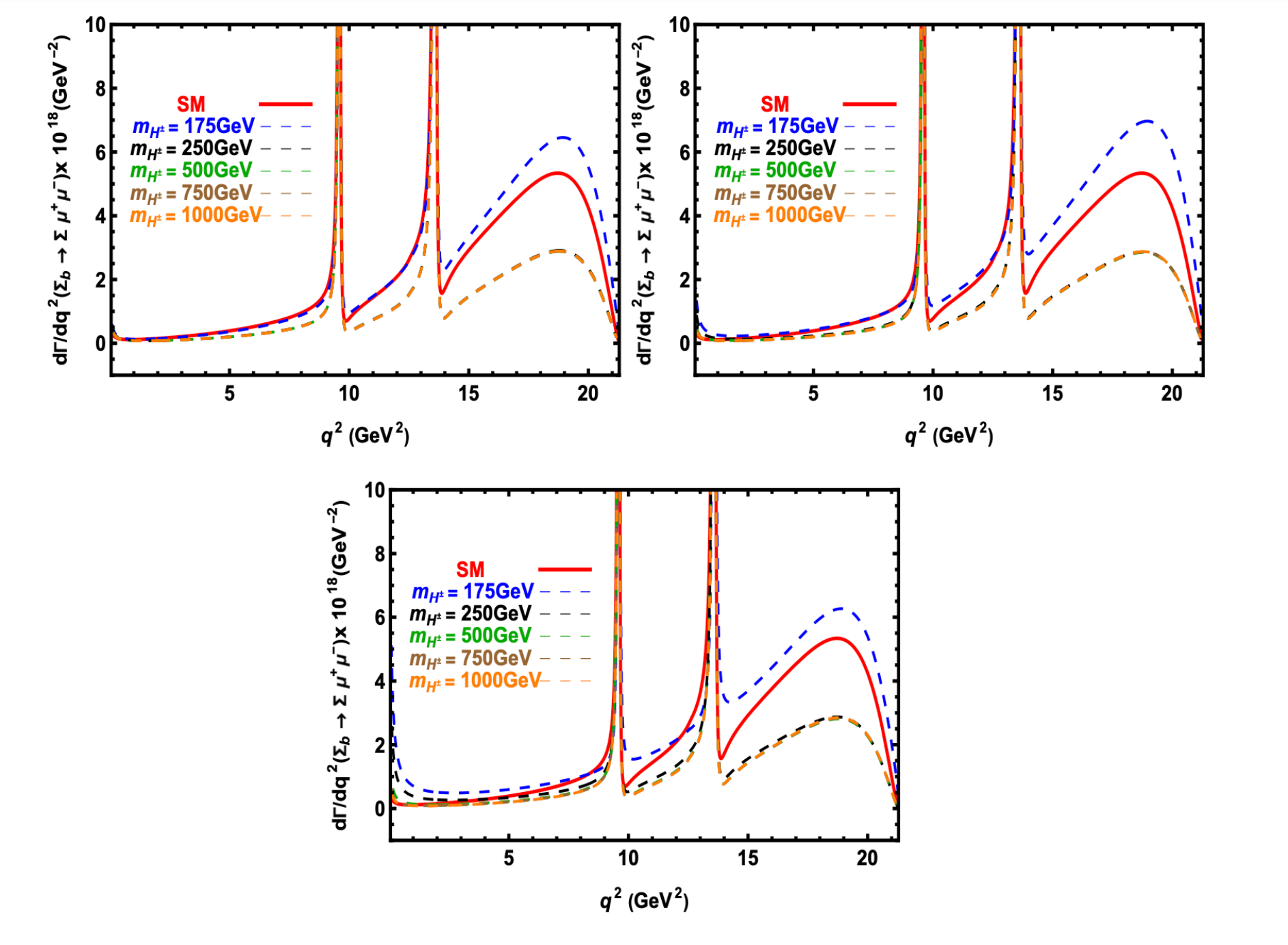}

\end{center}
\caption{The dependence of the $d\Gamma/dq^2$ on  $q^2$  for the $ \Sigma_b \rightarrow \Sigma \mu^+  \mu^-$ transition in the SM and 2HDM  with long-distance contributions plotted against different Higgs masses for $\lambda_{tt}=0.05$, $\lambda_{tt}=0.15$ and $\lambda_{tt}=0.30$, respectively. }
\label{fig:F6}
\end{figure}

\end{widetext}
\begin{widetext}
\begin{figure}[h!]
\begin{center}

\includegraphics[totalheight=12cm,width=16cm]{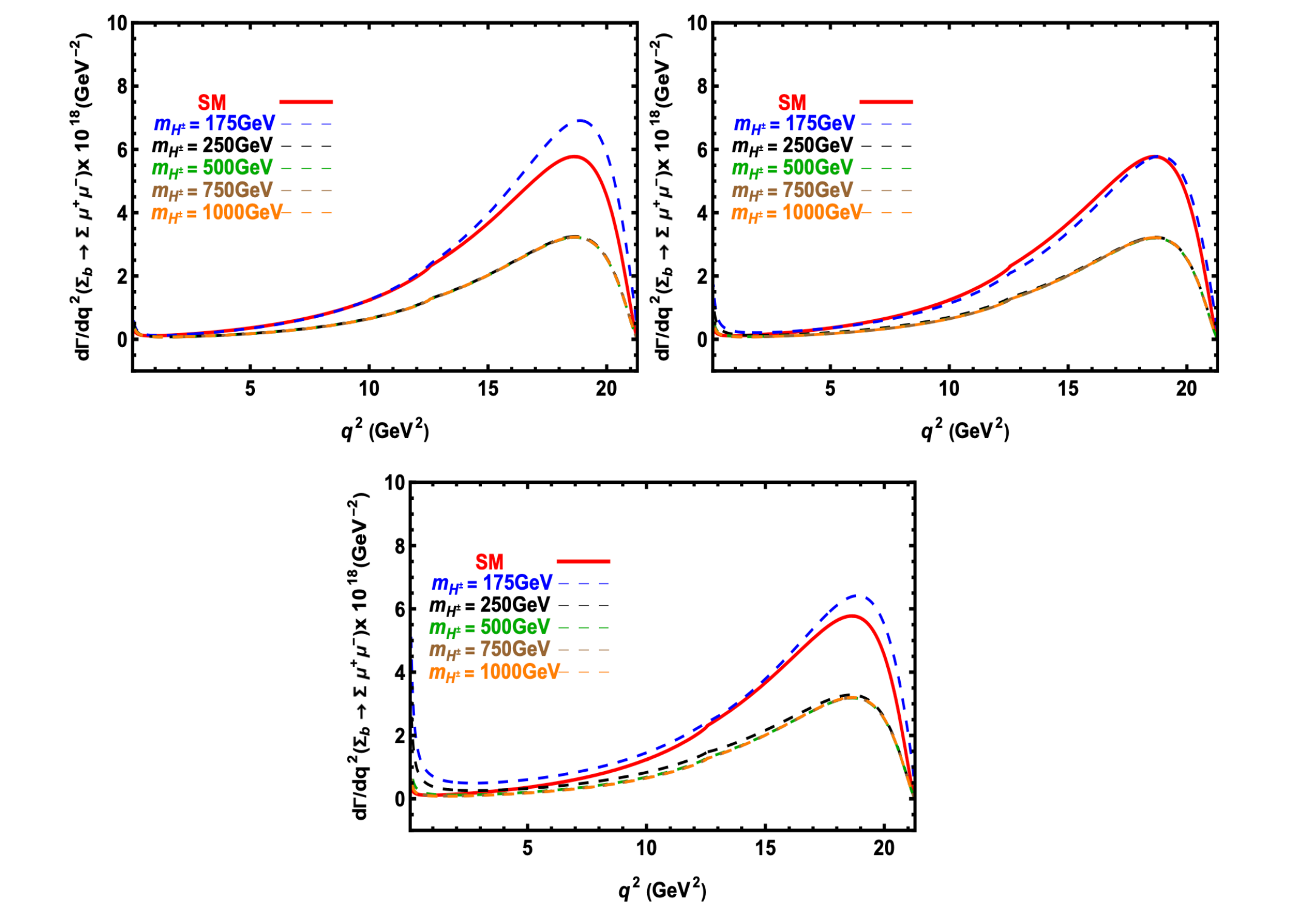}

\end{center}
\caption{The dependence of the $d\Gamma/dq^2$ on   $q^2$  for the $ \Sigma_b \rightarrow \Sigma \mu^+  \mu^-$ transition in  the SM and 2HDM   without long-distance contributions plotted against different Higgs masses for $\lambda_{tt}=0.05$, $\lambda_{tt}=0.15$ and $\lambda_{tt}=0.30$, respectively.  }
\label{fig:F7}
\end{figure}

\end{widetext}

We extend the analysis to the $\Sigma_b \rightarrow \Sigma ,\ell^+\ell^-$ channel and investigate the differential decay width  $d\mathcal{\Gamma}/dq^2$ within both the SM and the 2HDM frameworks. The corresponding spectra, computed under the same benchmark settings as in the $\Lambda_b$ channel, are shown with long-distance contributions in Fig.~\ref{fig:F6} and without long-distance effects in Fig.~\ref{fig:F7}. This comparative approach enables a direct examination of baryonic-channel dependence and highlights the role of the hadronic effects across $\Lambda_b$ and $\Sigma_b$ transitions. 

\begin{widetext}

\begin{figure}[h!]
\begin{center}
\includegraphics[totalheight=12cm,width=16cm]{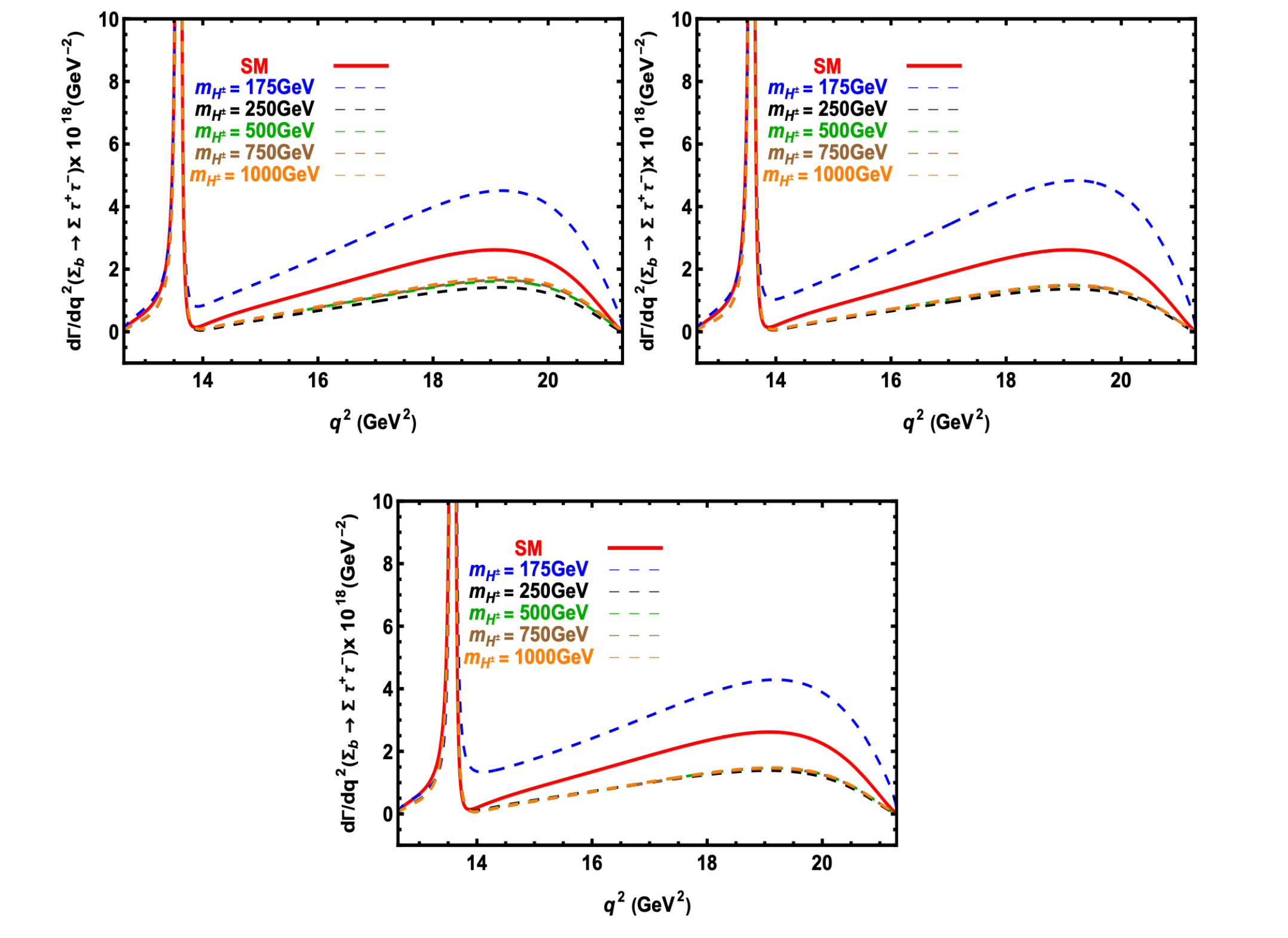}
\

\end{center}
\caption{The dependence of the $d\Gamma/dq^2$ on  $q^2$  for the $\Sigma_b \rightarrow \Sigma  \tau^+  \tau^-$ transition in  the SM and 2HDM  with long-distance contributions plotted against different Higgs masses for $\lambda_{tt}=0.05$, $\lambda_{tt}=0.15$ and $\lambda_{tt}=0.30$, respectively. }
\label{fig:F8}
\end{figure}

\end{widetext}

\begin{widetext}
\begin{figure}[h!]
\begin{center}

\includegraphics[totalheight=12cm,width=16cm]{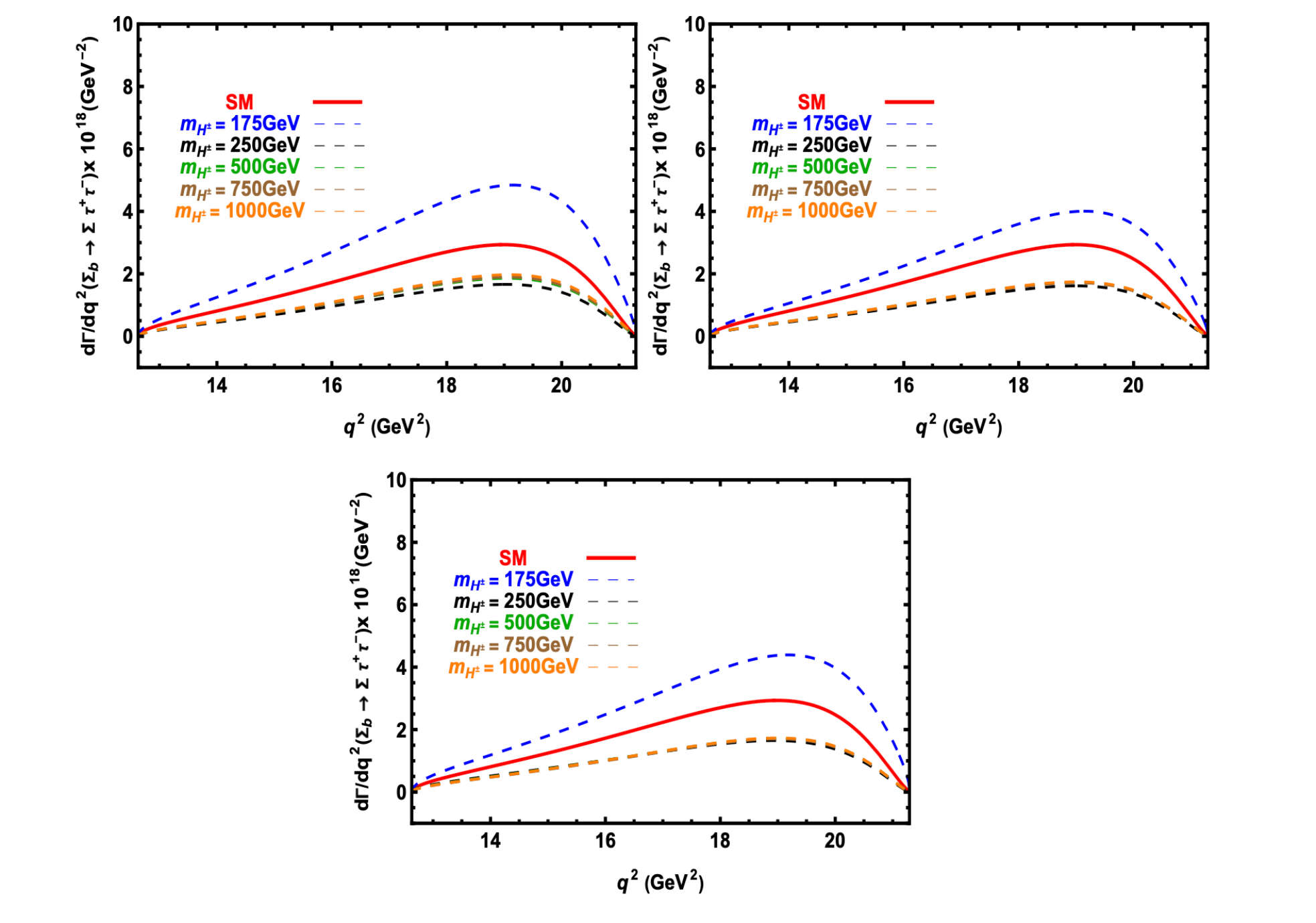}

\end{center}
\caption{The dependence of the $d\Gamma/dq^2$ on $q^2$  for the $ \Sigma_b \rightarrow \Sigma \tau^+  \tau^-$ transition in the SM and 2HDM  without long-distance contributions plotted against different Higgs masses for $\lambda_{tt}=0.05$, $\lambda_{tt}=0.15$ and $\lambda_{tt}=0.30$, respectively. }
\label{fig:F9}
\end{figure}

\end{widetext}

The Figs.~\ref{fig:F8} and ~\ref{fig:F9} show the differential decay width $d\mathcal{\Gamma}/dq^2$ for the $\Sigma_b \rightarrow \Sigma  \tau^+  \tau^-$ transition with and without long-distance contributions. In Fig.~\ref{fig:F8}, the inclusion of LD effects leads to a pronounced resonance peak in the low-$q^2$ region due to intermediate states such as $J/\psi$ making this region less reliable for new physics analysis. In contrast, Fig.~\ref{fig:F9} excludes long-distance contributions, resulting in a smoother distribution governed by short-distance effects, which provides a cleaner framework for comparison with the SM and 2HDM predictions. The 2HDM effects are most prominent at intermediate $q^2$ values ($15 - 18 GeV^2$ )  , where lower Higgs masses and larger $\lambda_{tt}$ couplings lead to significant deviations from the SM. As the Higgs mass increases toward $1000 \text{ GeV}$, the curves converge to the SM prediction, demonstrating decoupling behavior.

\begin{widetext}

\begin{figure}[h!]
\begin{center}
\includegraphics[totalheight=6cm,width=16cm]{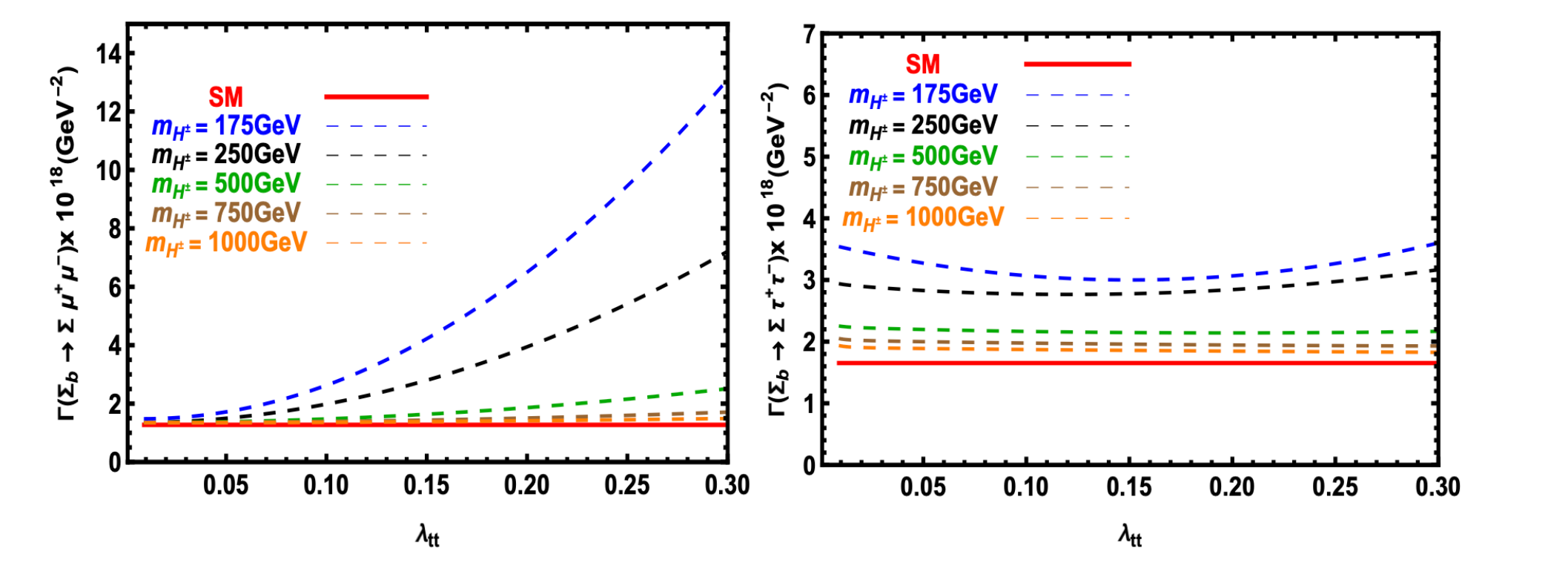}

\end{center}

\caption{The dependence of the $\Gamma$  on  $\lambda_{tt}$  for the $ \Sigma_b \rightarrow \Sigma \ell^+  \ell^-$ transition in the  SM and 2HDM. }
\label{fig:F10}
\end{figure}

\end{widetext}

In  Fig.~\ref{fig:F10}, the total decay width for the $ \Sigma_b \rightarrow \Sigma \ell^+  \ell^-$ transition shows no significant dependence on $\lambda_{tt}$ within the SM, remaining nearly constant. In contrast, the 2HDM predictions exhibit a clear sensitivity to $\lambda_{tt}$, particularly in the $\mu^+  \mu^-$ channel, where the total decay width increases notably for lower charged Higgs masses and approaches the SM limit as the mass increases. The $\tau^+  \tau^-$ channel, however, displays a much weaker dependence on $\lambda_{tt}$ with only minor deviations from the SM. Overall, the $\mu^+  \mu^-$ channel emerges as a more sensitive probe of NP, while the $\tau^+  \tau^-$ channel provides a complementary but less sensitive test.

\begin{widetext}

\begin{figure}[h!]
\begin{center}
\includegraphics[totalheight=12cm,width=16cm]{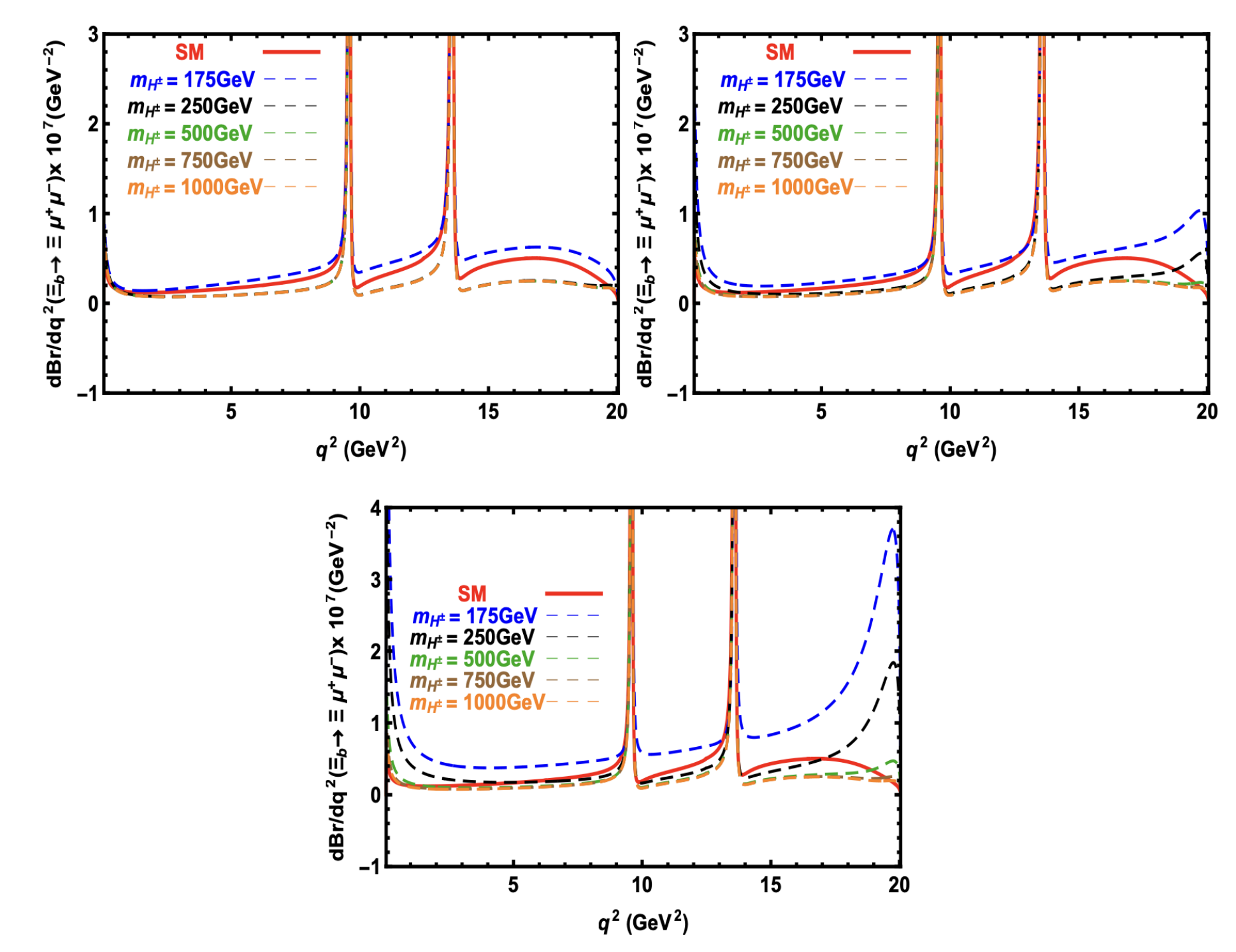}

\end{center}
\caption{The dependence of the $ dBR/dq^2$ on  $q^2$  for the $ \Xi_b \rightarrow \Xi \mu^+  \mu^-$ transition in  the SM and 2HDM  with long-distance contributions plotted against different Higgs masses for $\lambda_{tt}=0.05$, $\lambda_{tt}=0.15$ and $\lambda_{tt}=0.30$, respectively. }
\label{fig:F11}
\end{figure}

\end{widetext}

\begin{widetext}

\begin{figure}[h!]
\begin{center}
\includegraphics[totalheight=12cm,width=16cm]{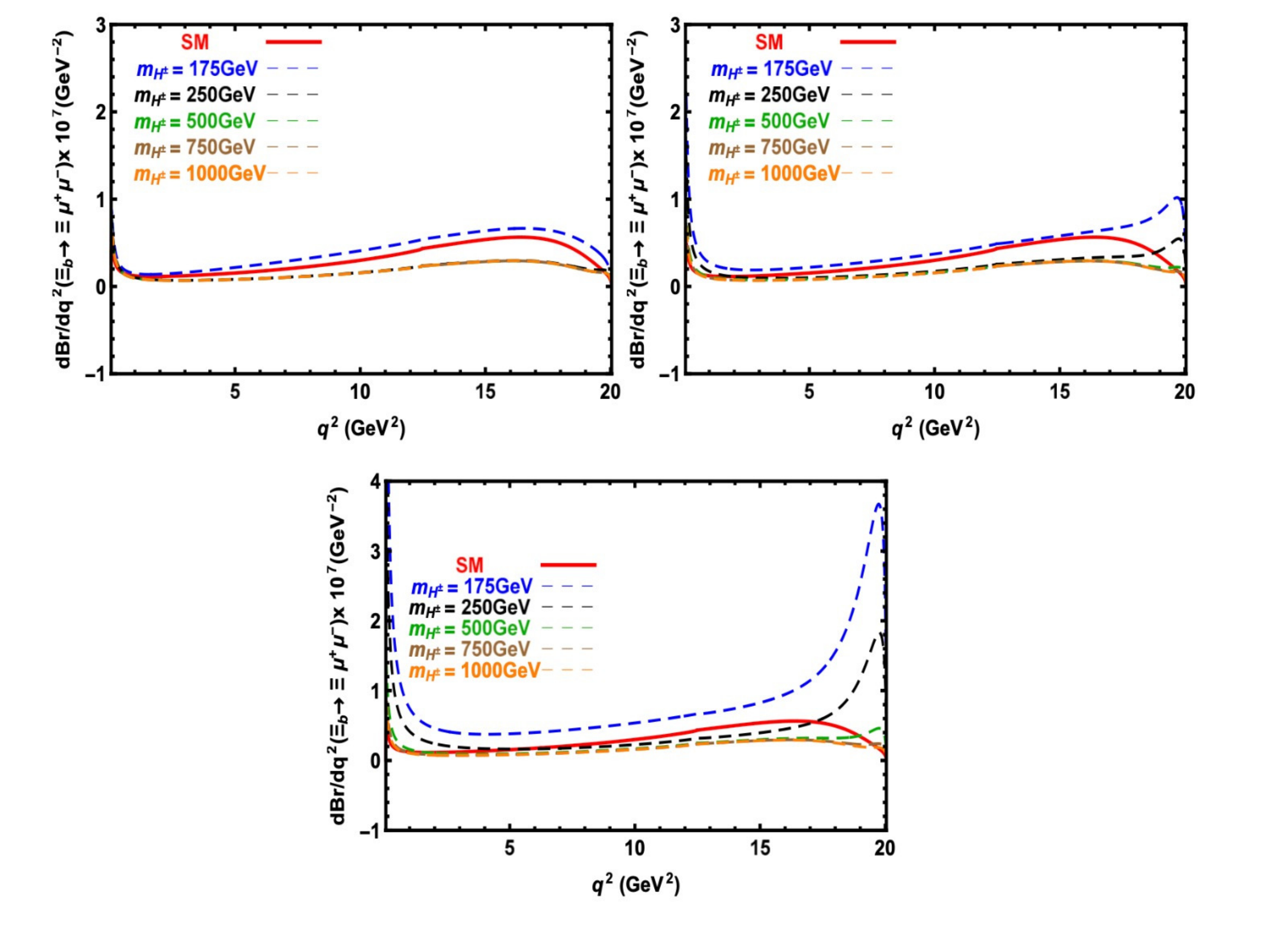}

\end{center}
\caption{The dependence of the $ dBR/dq^2$ on  $q^2$  for the $  \Xi_b \rightarrow \Xi \mu^+  \mu^-$ transition in the SM and 2HDM without long-distance contributions plotted against different Higgs masses for $\lambda_{tt}=0.05$, $\lambda_{tt}=0.15$ and $\lambda_{tt}=0.30$, respectively.  }
\label{fig:F12}
\end{figure}

\end{widetext}

Fig.~\ref{fig:F11} illustrates the $dBr/dq^2$ for the $\Xi_b \to \Xi \mu^+ \mu^-$ transition, comparing the SM with 2HDM predictions across varying charged Higgs masses and coupling constants ($\lambda_{tt}$). The $\Xi_b$ results display patterns similar to those found for the $\Lambda_b$ and $\Sigma_b$ channels: inclusion of LD terms produces pronounced resonance structures in the charmonium regions, while the LD–subtracted curves exhibit a smooth short–distance-dominated behavior. In the muon channel, the largest departures from the SM baseline again appear at high $q^2$, with the effect most noticeable for the lighter charged–Higgs benchmark points; overall, the qualitative dependence on model parameters and on $\lambda_{tt}$ mirrors the trends previously discussed for the other baryonic channels. While the sharp peaks at $q^2 \approx 9.5$ and ${13.5 \text{ GeV}^2}$ represent expected charmonium resonances, the most significant deviations from the SM occur at high $q^2$ values, particularly for lighter Higgs masses like $175 \text{ GeV}$. As the $\lambda_{tt}$ parameter increases across the panels, the enhancement of the branching ratio becomes more pronounced, indicating that this specific decay channel serves as a highly sensitive probe for NP effects BSM. In the absence of long-distance contributions, similar effects are also observed in the Fig.~\ref{fig:F12}. The spectra follow comparable trends to those with long-distance terms included, thereby confirming the robustness of the overall behavior across different scenarios.

% for Ksi

\begin{widetext}

\begin{figure}[h!]
\begin{center}
\includegraphics[totalheight=12cm,width=16cm]{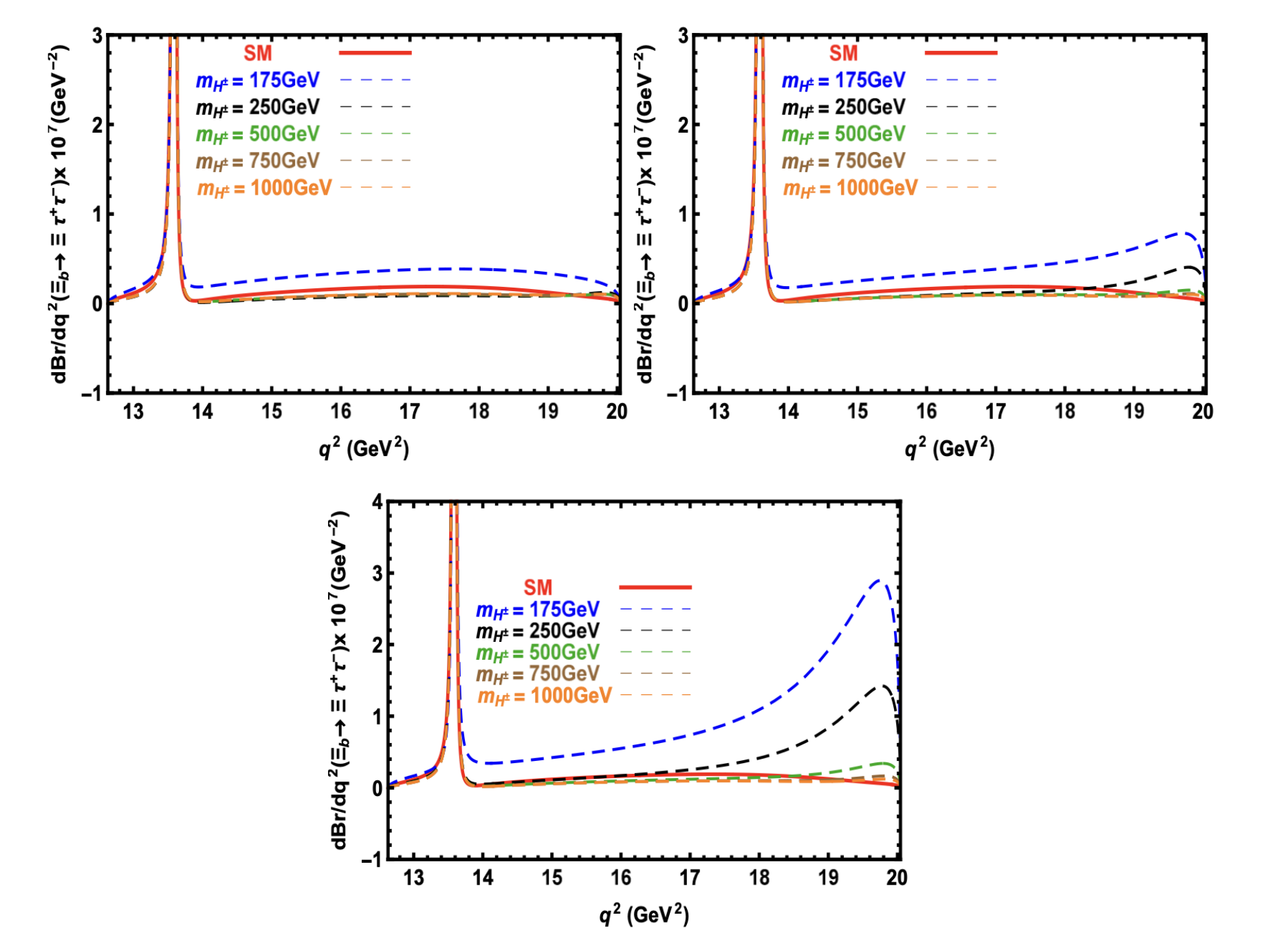}

\end{center}
\caption{The dependence of the $ dBR/dq^2$ on  $q^2$  for the $  \Xi_b \rightarrow \Xi \tau^+  \tau^-$ transition in the SM and 2HDM  with long-distance contributions plotted against different Higgs masses for $\lambda_{tt}=0.05$, $\lambda_{tt}=0.15$ and $\lambda_{tt}=0.30$, respectively. }
\label{fig:F13}
\end{figure}

\end{widetext}

\begin{widetext}

\begin{figure}[h!]
\begin{center}
\includegraphics[totalheight=12cm,width=16cm]{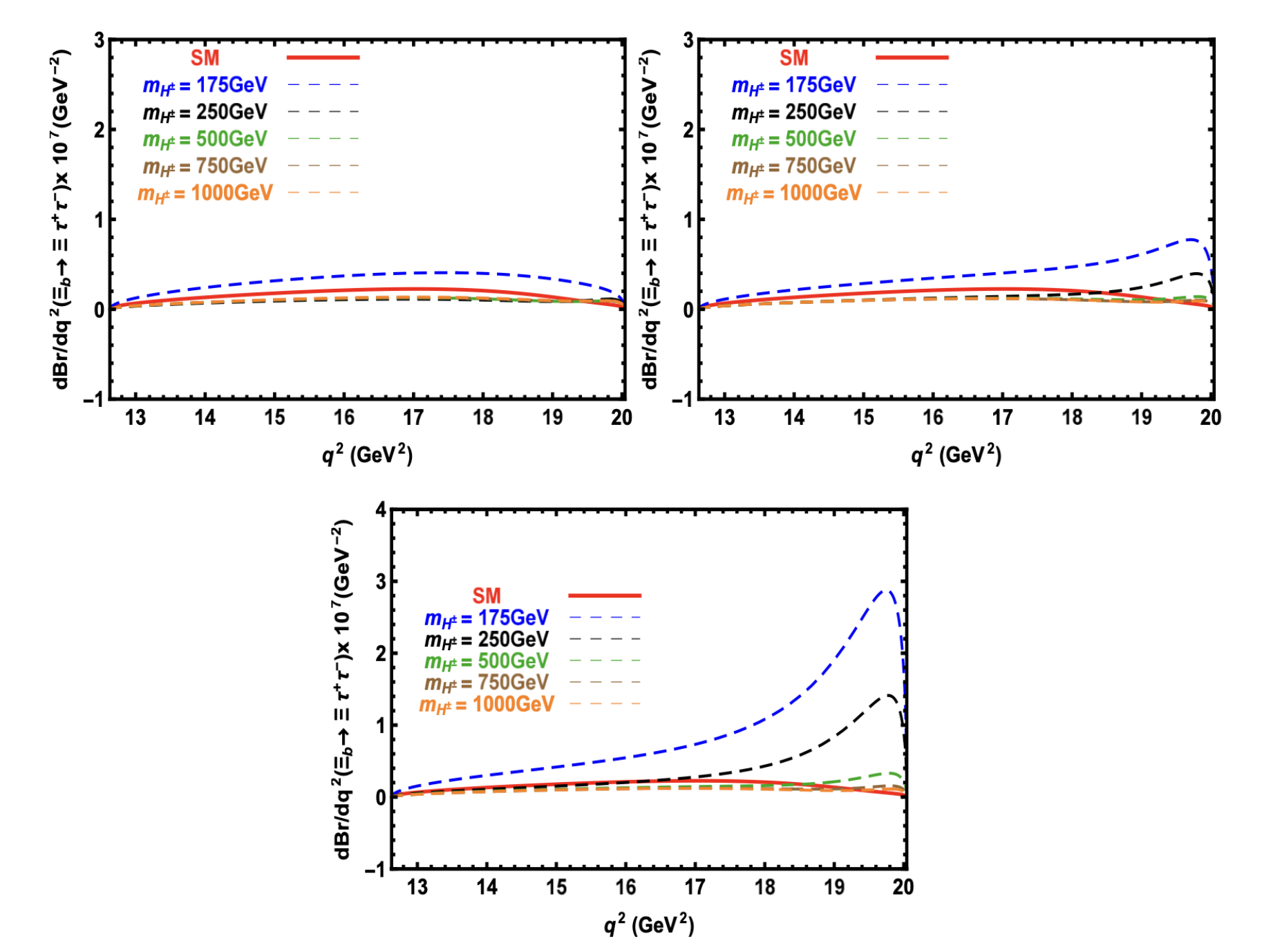}

\end{center}
\caption{The dependence of the $ dBR/dq^2$ on  $q^2$  for the $  \Xi_b \rightarrow \Xi  \tau^+  \tau^-$ transition in the SM and 2HDM  without long-distance contributions plotted against different Higgs masses for $\lambda_{tt}=0.05$, $\lambda_{tt}=0.15$ and $\lambda_{tt}=0.30$, respectively. }
\label{fig:F14}
\end{figure}

\end{widetext}

\begin{widetext}

\begin{figure}[h!]
\begin{center}
\includegraphics[totalheight=6cm,width=16cm]{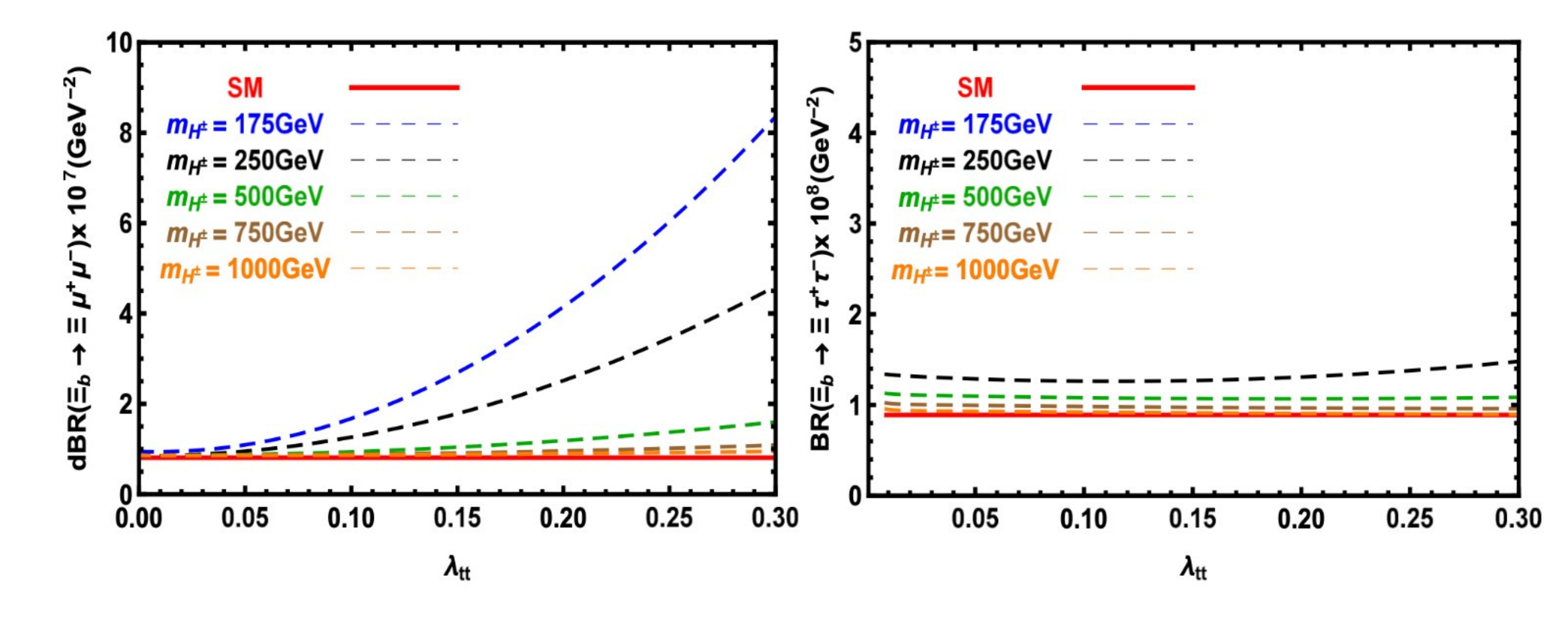}

\end{center}

\caption{The dependence of the $ dBR$ on  $\lambda_{tt}$  for the $  \Xi_b \rightarrow \Xi \ell^+  \ell^-$ transition in  the SM and 2HDM. }
\label{fig:F15}
\end{figure}

\end{widetext}

Figs.~\ref{fig:F13} and ~\ref{fig:F14} present the analysis of the $\tau$ channel for the $q^2$-dependent differential branching ratio. The results exhibit similar trends to those observed in the muon channel, confirming the consistency of the underlying dynamics across different lepton flavors. In particular, the spectra obtained with and without long-distance contributions reproduce the characteristic features discussed in the previous analysis.Similar to the $\Lambda_b$ and $\Sigma_b$ channels, Fig.~\ref{fig:F15} illustrates the dependence of the differential branching ratio on $\lambda_{tt}$. The observed trends are consistent with those reported for the previous channels, confirming that the $\lambda_{tt}$-driven effects manifest similarly across different baryonic transitions.

\subsection{The branching ratio}

Within the SM, the branching ratio predictions including the associated form factor uncertainties are evaluated for the $  \Lambda_b \rightarrow \Lambda \ell^+  \ell^-$ decays, considering both the muon and tau channels, and are presented in Table~\ref{tab:T2}. These SM estimates serve as theoretical benchmarks against possible deviations induced by new physics scenarios may be probed. For comparison, the available experimental measurements from the LHCb and CDF Collaborations are presented; however, such data currently exist only for the $\Lambda_{b}\rightarrow \Lambda \mu^{+}  \mu^{-} $ channel. The results indicate that the SM predictions are compatible with the experimental values within the present uncertainties in the muon channel, whereas no experimental measurements are available for the tau channel, rendering it an open sector for future high-precision studies. Table~\ref{tab:T2} also includes our SM predictions for other considered channels, namely  $\Xi_b \rightarrow \Xi \mu^+ \mu^-$ and $ \Xi_{b}\rightarrow \Xi \tau^{+}  \tau^{-} $. For the $\Sigma_b \rightarrow \Sigma \mu^+ \mu^-$,$ \Sigma_{b}\rightarrow \Sigma \tau^{+}  \tau^{-} $ channels in the Table~\ref{tab:TSigma}, as stated above, we present the total decay width rather than the branching ratio due to the lack of reliable experimental information on its total width and the absence of a precise lifetime determination. Our predictions in these baryonic and leptonic channels can guide the related experiments aiming to search for these modes and measure branching ratios or total decay widths of them.  
% CDF kaynak arXiv:1107.3753 [hep-ex]].
% LHCb kaynak arXiv:1306.2577 [hep-ex]]
\begin{widetext}

\begin{table}
\begin{tabular}{|c|c|c|c|}
\hline\hline
         Decay Modes       & {Present Work ($BR\times10^{6}$)}& {CDF Collab. ($BR\times10^{6}$)\cite{Aaltonen:2011qs})}& {LHCb Collab.($BR\times10^{6}$) \cite{LHCb:2013uqx}} \\ \hline\hline
    {$\Lambda_{b}\rightarrow \Lambda \mu^{+}  \mu^{-} $ }  & {  $1.294^{+1.21}_{-0.73}$  } & {$1.73 \pm 0.42 \pm 0.55$} & {$0.96 \pm 0.16 \pm 0.13 \pm 0.21$} \\
    \hline
{$ \Lambda_{b}\rightarrow \Lambda \tau^{+}  \tau^{-} $}  &{  $ 0.187^{+0.331}_{-0.06}$} &  - & - \\
 \hline
{$ \Xi_{b}\rightarrow \Xi \mu^{+}  \mu^{-} $  } &  { $0.187^{+0.518}_{-0.117}$} &  - & -\\
 \hline
{$ \Xi_{b}\rightarrow \Xi \tau^{+}  \tau^{-} $  } &  { $0.066^{+0.183}_{-0.065}$} &  - & - \\
 \hline \hline
\end{tabular}%
\caption{The branching ratio values within the SM for the $  \Lambda_b \rightarrow \Lambda \ell^+  \ell^-$ and  $\Xi_b \rightarrow \Xi \ell^+ \ell^-$ channels are presented, including the associated form factor uncertainties. For reference, the experimental results reported by the LHCb and CDF Collaborations are included only for the $\Lambda_b \rightarrow \Lambda \mu^+ \mu^-$ channel.}
\label{tab:T2}
\end{table}

\end{widetext}

\begin{widetext}

\begin{table}
\begin{tabular}{|c|c|c|c|}
\hline\hline
         Decay Modes       & {Present Work ($\Gamma_{total}\times10^{18}$)} \\ \hline\hline
    
{$ \Sigma_{b}\rightarrow \Sigma \mu^{+}  \mu^{-}$}   & { $3.944^{+5.788}_{-2.475}$ }    \\
 \hline
{$ \Sigma_{b}\rightarrow \Sigma \tau^{+}  \tau^{-} $}  & {  $1.310^{+2.025}_{-0.814}$ } \\
 \hline \hline
\end{tabular}%
\caption{The total decay width values within the SM for the $ \Sigma_{b}\rightarrow \Sigma \ell^{+}  \ell^{-}$ are presented, including the associated form factor uncertainties.}
\label{tab:TSigma}
\end{table}

\end{widetext}
The branching ratio predictions including the associated form factor uncertainties within the 2HDM type-III framework for the $\Lambda_b \rightarrow \Lambda \ell^+ \ell^-$ and $\Xi_b \rightarrow \Xi \ell^+ \ell^-$ decay channels are calculated for different values of the charged Higgs boson mass $m_{H^\pm} = 175, 250, 500, 750,$ and $1000~\text{GeV}$, as well as for the parameter values $\lambda_{tt} = 0.05, 0.15,$ and $0.30$, where for the $\Sigma_b \rightarrow \Sigma \ell^+ \ell^-$ channel the total decay width is presented instead of the branching ratio. These results are sequentially presented in Table~\ref{tab:T3}, Table~\ref{tab:T4} and Table~\ref{tab:T5}, respectively, providing a comprehensive comparison of the branching ratios or total decay width across the three baryonic channels under consideration.

\begin{widetext}

\begin{table}
\begin{tabular}{|c|c|c|c|}
\hline\hline
          Decay Modes       & {$\lambda_{tt}=0.05$ ($BR\times10^{6}$)}& {$\lambda_{tt}=0.15$ ($BR\times10^{6}$)}& {$\lambda_{tt}=0.30$ ($BR\times10^{6}$)} \\ 
\hline\hline
{$ \Lambda_{b}\rightarrow \Lambda \mu^{+}  \mu^{-}$} ($m_{H^{\pm}}=175 GeV$)  & { $3.018^{+4.38}_{-1.27}$ }   &  { $3.651^{+2.98}_{-2.1}$ } & { $4.613^{+5.96}_{-2.13}$ } \\
 \hline
{$ \Lambda_{b}\rightarrow \Lambda \tau^{+}  \tau^{-} $} ($m_{H^{\pm}}=175 GeV$)  & {  $0.855^{+2.2}_{-0.12}$ } &  $0.714^{+1.86}_{-0.1}$ & $1.056^{+2.49}_{-0.197}$\\
\hline\hline
{$ \Lambda_{b}\rightarrow \Lambda \mu^{+}  \mu^{-}$ ($m_{H^{\pm}}=250 GeV$) } &  { $0.810^{+0.68}_{-0.47}$} &  $0.954^{+0.8}_{-0.55}$ & $1.463^{+1.22}_{-0.855}$\\
 \hline
{$ \Lambda_{b}\rightarrow \Lambda \tau^{+}  \tau^{-}$ ($m_{H^{\pm}}=250 GeV$) } &  { $0.079^{+0.13}_{-0.03}$} &  $0.078^{+0.15}_{-0.03}$& $0.091^{+0.13}_{-0.04}$\\
\hline\hline
{$ \Lambda_{b}\rightarrow \Lambda \mu^{+}  \mu^{-}$ ($m_{H^{\pm}}=500 GeV$) } &  { $0.768^{+0.648}_{-0.452}$} &  $0.752^{+0.709}_{-0.425}$ & $0.890^{+0.75}_{-0.522}$\\
 \hline
{$ \Lambda_{b}\rightarrow \Lambda \tau^{+}  \tau^{-}$ ($m_{H^{\pm}}=500 GeV$) } &  { $0.114^{+0.254}_{-0.03}$} & $0.101^{+0.184}_{-0.03}$  &$0.102^{+0.185}_{-0.03}$ \\
 \hline \hline
{$ \Lambda_{b}\rightarrow \Lambda \mu^{+}  \mu^{-}$ ($m_{H^{\pm}}=750 GeV$) } &  { $0.764^{+0.645}_{-0.449}$} &  $0.768^{+0.649}_{-0.452}$ & $0.757^{+0.713}_{-0.428}$\\
 \hline
{$ \Lambda_{b}\rightarrow \Lambda \tau^{+}  \tau^{-}$ ($m_{H^{\pm}}=750 GeV$) } &  { $0.115^{+0.274}_{-0.02}$} &  $0.095^{+0.166}_{-0.03}$ & $0.094^{+0.165}_{-0.03}$\\
\hline\hline
{$ \Lambda_{b}\rightarrow \Lambda \mu^{+}  \mu^{-}$ ($m_{H^{\pm}}=1000 GeV$) } &  { $0.767^{+0.648}_{-0.451}$} &  $0.726^{+0.686}_{-0.411}$ & $0.776^{+0.656}_{-0.456}$\\
 \hline
{$ \Lambda_{b}\rightarrow \Lambda \tau^{+}  \tau^{-}$ ($m_{H^{\pm}}=1000 GeV$) } &  { $0.126^{+0.322}_{-0.02}$} & $0.100^{+0.183}_{-0.03}$  & $0.099^{+0.181}_{-0.03}$\\
\hline\hline
\end{tabular}%
\caption{Branching ratio predictions, including the associated form factor uncertainties for the $\Lambda_b \rightarrow \Lambda \ell^+ \ell^-$ decay channel within the 2HDM type-III framework, calculated for different values of the charged Higgs mass $m_{H^\pm}$ and the parameter $\lambda_{tt}$.}
\label{tab:T3}
\end{table}

\end{widetext}

Table~\ref{tab:T3} presents the branching ratio predictions including the associated form factor uncertainties for the $\Lambda_b \rightarrow \Lambda \ell^+ \ell^-$ decay within the 2HDM type-III framework. The results indicate a pronounced sensitivity to the parameter $\lambda_{tt}$ at lower charged Higgs masses, particularly in the muon channel, where the branching ratio reaches $4.613 \times 10^{-6}$ for $m_{H^\pm}=175$~GeV and $\lambda_{tt}=0.30$. In contrast, the tau channel exhibits a more moderate dependence on $\lambda_{tt}$ with consistently smaller BR values. As the charged Higgs mass increases, the branching ratios in both channels decrease and tend to converge, reflecting a gradual approach to SM–like behavior for higher Higgs masses.

The SM predictions, Table~\ref{tab:T2}, for the muon channel ($\Lambda_b \to \Lambda \mu^+ \mu^-$) show a branching ratio of $ 1.294^{+1.21}_{-0.73} \times 10^{-6} $, which is consistent with experimental data from CDF and LHCb within the large presented uncertainties. In contrast, the 2HDM predictions in Table~\ref{tab:T3} demonstrate that New Physics effects, particularly at a lower Higgs mass ($m{H^\pm} = 175 \text{ GeV}$) and higher coupling ($\lambda_{tt} = 0.30$), can significantly enhance the $ BR $ up to $4.613^{+5.96}_{-2.13} \times 10^{-6}$. While the tau channel ($\Lambda_b \to \Lambda \tau^+ \tau^-$) exhibits a more moderate dependence on these parameters, both channels tend to converge toward SM-like values as the charged Higgs mass increases, illustrating that the current theoretical uncertainties still allow for significant BSM contributions in the low-mass regime.

\begin{widetext}

\begin{table}
\begin{tabular}{|c|c|c|c|}
\hline\hline
          Decay Modes       & {$\lambda_{tt}=0.05$ ($\Gamma_{total}\times10^{18}$)}& {$\lambda_{tt}=0.15$ ($\Gamma_{total}\times10^{18}$)}& {$\lambda_{tt}=0.30$ ($\Gamma_{total}\times10^{18}$)} \\ 
\hline\hline
{$ \Sigma_b \rightarrow \Sigma \mu^{+}  \mu^{-}$} ($m_{H^{\pm}}=175 GeV$)  & { $4.400^{+7.440}_{-2.416}$ }   &  { $4.599^{+6.427}_{-2.111}$ } & { $4.774^{+7.985}_{-2.394}$ } \\
{$ \Sigma_b \rightarrow \Sigma \tau^{+}  \tau^{-} $} ($m_{H^{\pm}}=175 GeV$)  & {  $2.096^{+3.911}_{-1.083}$ } &  $1.993^{+3.259}_{-0.898}$ & $2.221^{+3.782}_{-0.917}$\\
\hline\hline
{$\Sigma_b \rightarrow \Sigma \mu^{+}  \mu^{-}$ ($m_{H^{\pm}}=250 GeV$) } &  { $2.195^{+3.235}_{-1.373}$} &  {$2.235^{+3.319}_{-1.400}$} & {$2.446^{+3.712}_{-1.539} $}\\
{$ \Sigma_b \rightarrow \Sigma \tau^{+}  \tau^{-}$ ($m_{H^{\pm}}=250 GeV$) } &  { $0.737^{+1.122}_{-0.467}$} &  $0.721^{+1.096}_{-0.460}$& $0.751^{+1.150}_{-0.481}$\\
\hline\hline
{$ \Sigma_b \rightarrow \Sigma \mu^{+}  \mu^{-}$ ($m_{H^{\pm}}=500 GeV$) } &  { $2.172^{+3.190}_{-1.363}$} &  $2.235^{+3.192}_{-1.363}$ & $2.185^{+3.252}_{-1.383}$\\
{$ \Sigma_b \rightarrow \Sigma \tau^{+}  \tau^{-}$ ($m_{H^{\pm}}=500 GeV$) } &  { $0.820^{+1.285}_{-0.504}$} & $0.773^{+1.194}_{-0.483}$  &$0.768^{+1.192}_{-0.483}$ \\
 \hline \hline
{$ \Sigma_b \rightarrow \Sigma \mu^{+}  \mu^{-}$ ($m_{H^{\pm}}=750 GeV$) } &  { $2.174^{+3.192}_{-1.364}$} &  $2.165^{+3.183}_{-1.361}$ & $2.159^{+3.193}_{-1.363}$\\
{$ \Sigma_b \rightarrow \Sigma \tau^{+}  \tau^{-}$ ($m_{H^{\pm}}=750 GeV$) } &  { $0.831^{+1.310}_{-0.510}$} &  $0.765^{+1.175}_{-0.480}$ & $0.758^{+1.167}_{-0.477}$\\
\hline\hline
{$ \Sigma_b \rightarrow \Sigma \mu^{+}  \mu^{-}$ ($m_{H^{\pm}}=1000 GeV$) } &  { $2.177^{+3.197}_{-1.366}$} &  $2.177^{+3.187}_{-1.362}$ & $2.159^{+3.185}_{-1.361}$\\
{$ \Sigma_b \rightarrow \Sigma \tau^{+}  \tau^{-}$ ($m_{H^{\pm}}=1000 GeV$) } &  { $0.861^{+1.370}_{-0.523}$} & $0.861^{+1.199}_{-0.486}$  & $0.77^{+1.191}_{-0.484}$\\
\hline\hline
\end{tabular}%
\caption{The total decay width predictions, including the associated form factor uncertainties, for the $\Sigma_b \rightarrow \Sigma \ell^+ \ell^-$ decay channel within the 2HDM type-III framework, calculated for different values of the charged Higgs mass $m_{H^\pm}$ and the parameter $\lambda_{tt}$.}
\label{tab:T4}
\end{table}

\end{widetext}
Table~\ref{tab:T4} presents the total decay width predictions including the associated form factor uncertainties for the $\Sigma_b \rightarrow \Sigma \ell^+ \ell^-$ decay within the 2HDM type-III framework. The results show that the muon channel exhibits moderate sensitivity to $\lambda_{tt}$, with the highest total decay width reaching $3.944^{+5.788}_{-2.475} \times 10^{-18}$ at $m_{H^\pm}=175$~GeV and $\lambda_{tt}=0.30$. The tau channel displays smaller total decay width values and a more subdued dependence on $\lambda_{tt}$. As the charged Higgs mass increases, the total decay widths in both channels gradually converge, approaching a pattern reminiscent of SM predictions, indicating a diminishing influence of the 2HDM contributions at higher Higgs masses.

The comparison for the $\Sigma_b \to \Sigma \mu^+ \mu^-$ channel shows that while the SM predicts a baseline total decay width of $3.944^{+5.788}_{-2.475} \times 10^{-18}$ in Table~\ref{tab:TSigma}, the 2HDM (Type-III) framework allows for a notable enhancement at lower charged Higgs masses ($m_{H^\pm} = 175 \text{ GeV}$), reaching up to $4.774^{+7.985}_{-2.394} \times 10^{-18}$ in Table~\ref{tab:T4}. However, as the Higgs mass increases toward $1000 \text{ GeV}$, these BSM predictions tend to decrease and even fall slightly below the SM central value. Ultimately, due to the large overlapping uncertainty bands in both models, the $\Sigma$ channel remains a key area where high-precision experimental measurements are needed to identify potential NP.

\begin{widetext}

\begin{table}
\begin{tabular}{|c|c|c|c|}
\hline\hline
          Decay Modes       & {$\lambda_{tt}=0.05$ ($BR\times10^{7}$)}& {$\lambda_{tt}=0.15$ ($BR\times10^{7}$)}& {$\lambda_{tt}=0.30$ ($BR\times10^{7}$)} \\ 
\hline\hline
{$ \Xi_b \rightarrow \Xi \mu^{+}  \mu^{-}$} ($m_{H^{\pm}}=175 GeV$)  & { $8.550^{+30.671}_{-7.389}$ }   &  { $7.145^{+25.844}_{-6.224}$ } & { $10.55^{+35.525}_{-8.591}$ } \\
{$ \Xi_b \rightarrow \Xi \tau^{+}  \tau^{-} $} ($m_{H^{\pm}}=175 GeV$)  & {  $1.255^{+3.749}_{-1.179}$ } &  $1.127^{+3.331}_{-1.337}$ & $1.709^{+11.933}_{-2.836}$\\
\hline\hline
{$ \Xi_b \rightarrow \Xi \mu^{+}  \mu^{-}$ ($m_{H^{\pm}}=250 GeV$) } &  { $0.795^{+2.098}_{-0.458}$} &  $0.779^{+1.923}_{-0.413}$ & $0.911^{+2.219}_{-0.480}$\\
{$ \Xi_b \rightarrow \Xi \tau^{+}  \tau^{-}$ ($m_{H^{\pm}}=250 GeV$) } &  { $0.318^{+1.057}_{-0.355}$} &  $0.092^{+1.889}_{-0.521}$& $0.566^{+5.074}_{-1.187}$\\
\hline\hline
{$ \Xi_b \rightarrow \Xi \mu^{+}  \mu^{-}$ ($m_{H^{\pm}}=500 GeV$) } &  { $1.143^{+3.681}_{-0.843}$} &  $1.011^{+2.856}_{-0.640}$ & $1.020^{+2.875}_{-0.643}$\\
{$ \Xi_b \rightarrow \Xi \tau^{+}  \tau^{-}$ ($m_{H^{\pm}}=500 GeV$) } &  { $0.363^{+1.142}_{-0.389}$} & $0.308^{+1.209}_{-0.393}$  &$1.039^{+1.779}_{-0.509}$ \\
 \hline \hline
{$ \Xi_b \rightarrow \Xi \mu^{+}  \mu^{-}$ ($m_{H^{\pm}}=750 GeV$) } &  { $1.151^{+3.891}_{-0.895}$} &  $0.952^{+2.615}_{-0.582}$ & $0.946^{+2.594}_{-0.576}$\\
{$ \Xi_b \rightarrow \Xi \tau^{+}  \tau^{-}$ ($m_{H^{\pm}}=750 GeV$) } &  { $0.717^{+1.150}_{-0.392}$} &  $0.671^{+1.079}_{-0.366}$ & $0.345^{+1.23}_{-0.395}$\\
\hline\hline
{$ \Xi_b \rightarrow \Xi \mu^{+}  \mu^{-}$ ($m_{H^{\pm}}=1000 GeV$) } &  { $1.265^{+4.481}_{-1.040}$} &  $1.007^{+2.830}_{-0.634}$ & $0.999^{+2.801}_{-0.626}$\\
{$ \Xi_b \rightarrow \Xi \tau^{+}  \tau^{-}$ ($m_{H^{\pm}}=1000 GeV$) } &  { $0.747^{+1.20}_{-0.408}$} & $0.674^{+1.080}_{-0.370}$  & $0.696^{+1.125}_{-0.377}$\\
\hline\hline
\end{tabular}%
\caption{Branching ratio predictions, including the associated form factor uncertainties, for the $\Xi_b \rightarrow \Xi \ell^+ \ell^-$ decay channel within the 2HDM type-III framework, calculated for different values of the charged Higgs mass $m_{H^\pm}$ and the parameter $\lambda_{tt}$.}
\label{tab:T5}
\end{table}

\end{widetext}
Table~\ref{tab:T5} summarizes the branching ratio predictions including the associated form factor uncertainties for the $\Xi_b \rightarrow \Xi \ell^+ \ell^-$ decays in the 2HDM type-III framework. The muon channel shows a pronounced sensitivity to both $\lambda_{tt}$ and $m_{H^\pm}$, with BR values decreasing as the Higgs mass increases, while the tau channel exhibits smaller BRs and milder dependence on the model parameters, reflecting a more stable behavior. 

A closer inspection of Table~\ref{tab:T5} reveals certain non-monotonic trajectories and numerical variations across the configured 2HDM with Type-III parameter space, which may potentially point toward alternative physics scenarios. Specifically, for the $\Xi_{b} \rightarrow \Xi \tau^{+} \tau^{-}$ channel at $m_{H^{\pm}} = 250\text{ GeV}$, the branching ratio exhibits a visible variation around $\lambda_{tt} = 0.15$ before rising at $\lambda_{tt} = 0.30$. This behavior could be interpreted as a reflection of the underlying competition between the linear interference terms and the quadratic 2HDM contributions within this general Yukawa framework. Furthermore, the numerical transitions observed at $m_{H^{\pm}} = 500\text{ GeV}$ indicate that such non-linear responses could be more pronounced in tauonic modes than in muon channels. This situation might be attributed to the lack of helicity suppression for the heavier $\tau$-leptons, allowing the scalar amplitudes to scale with $m_\ell$, combined with possible kinematical threshold effects near the high-$q^2$ region. Rather than being anomalous, these localized modifications might accommodate NP contributions where the scalar sector potentially modulates the baryon decay distributions.

The SM predicts a relatively low branching ratio of $0.187^{+0.518}_{-0.117} \times 10^{-6}$ in Table~\ref{tab:T2}, which is the smallest among the muon-transition baryonic decays listed. In contrast, the 2HDM framework (from the first graph and general trends in the data) shows that for lower charged Higgs masses like $m_{H^\pm} = 175 \text{ GeV}$, this value can be significantly enhanced, specifically in the high-$q^2$ region in the Table~\ref{tab:T5}. While the SM provides a baseline with considerable uncertainty, the 2HDM predictions suggest that even small coupling parameters ($\lambda_{tt}$) can notably increase the decay rate, making the $\Xi$ channel a highly sensitive, albeit rare, indicator for NP. 

In summary, the comprehensive numerical analysis of both the SM baseline and the various 2HDM Type-III scenarios reveals important methodological and phenomenological features governed by the LCSR framework. The noticeably large and asymmetric uncertainty bands observed across the predicted branching ratios and decay widths (as tabulated in Tables II, III, IV and V) are intrinsic signatures of the LCSR approach rather than calculation artifacts. Because the mathematical dependence of the baryon transition form factors on the primary scale and threshold parameters—such as the continuum thresholds ($s_0$) and Borel tuning parameters ($M^2$)—is inherently non-linear, even independent and symmetric input variations naturally project onto highly asymmetric, non-Gaussian final error distributions. Consequently, while these broad uncertainty margins reflect the current precision limits of hadronic inputs and lead to overlapping error intervals between the SM and specific 2HDM scenarios (particularly at heavier charged Higgs masses $m_{H^{\pm}} \geq 500\text{ GeV}$), they underscore the necessity of shifting the phenomenological focus toward the dynamic behavior and relative shape variations of the observables across the $q^2$ spectrum. By prioritizing the evaluation of functional trajectories, angular asymmetries, and kinematic trends rather than static central values, systematic normalization errors significantly cancel out, making it entirely possible to isolate genuine physical signals and trace the explicit modifications induced by the new physics parameters.

\subsection{The lepton forward backward asymmetry}

\begin{widetext}

\begin{figure}[h!]
\begin{center}
\includegraphics[totalheight=12cm,width=16cm]{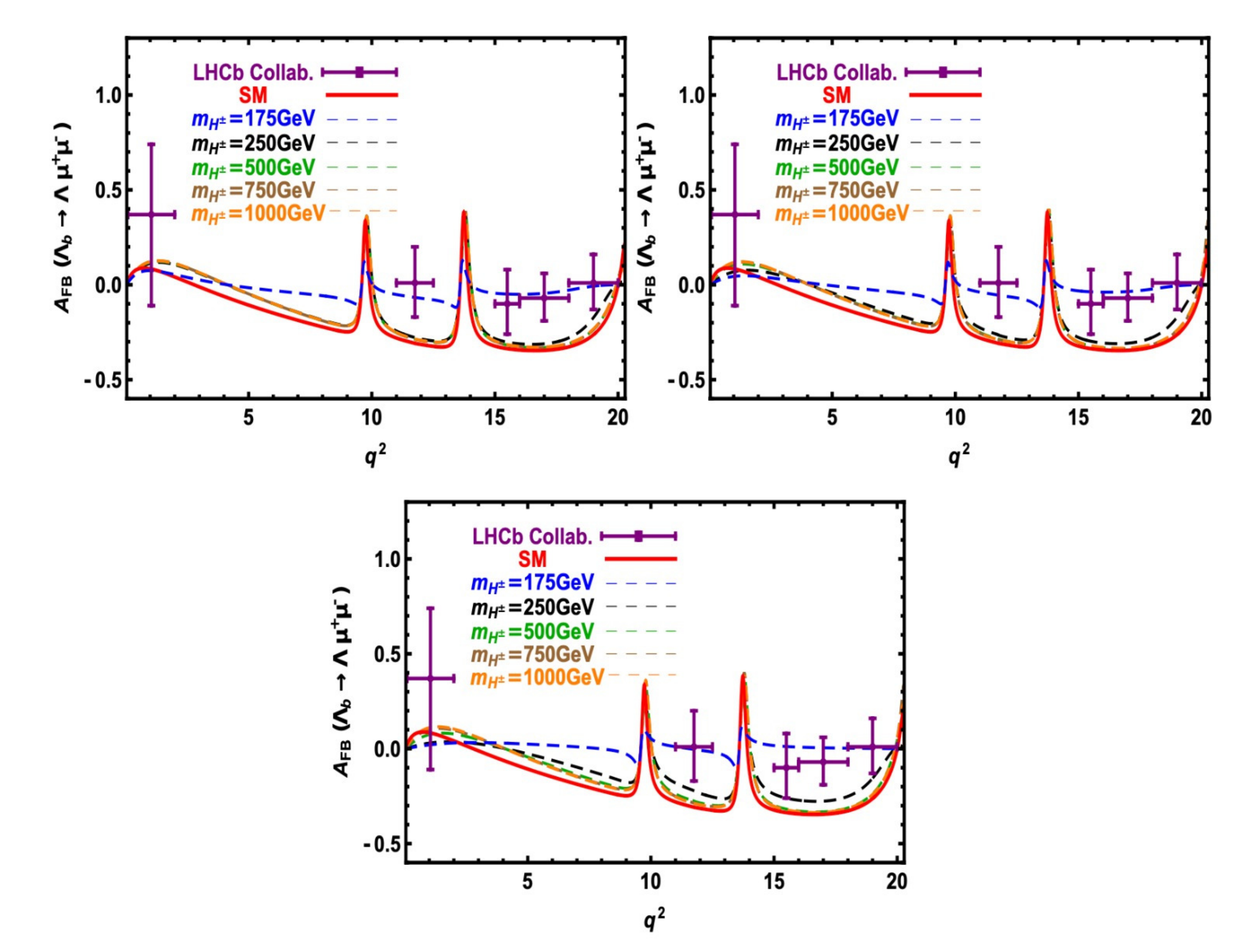}

\end{center}
\caption{The representation of the $A_{FB}$ with long-distance contributions as a function of $q^2$ for the $\Lambda_b \rightarrow \Lambda \mu^+ \mu^-$ transition within the SM and 2HDM frameworks, plotted against different charged Higgs masses for $\lambda_{tt} = 0.05$, $\lambda_{tt} = 0.15$, and $\lambda_{tt} = 0.30$, respectively, and compared with the available experimental data reported by the LHCb Collaboration. }
\label{fig:F16}
\end{figure}

\end{widetext}

\begin{widetext}

\begin{figure}[h!]
\begin{center}
\includegraphics[totalheight=12cm,width=16cm]{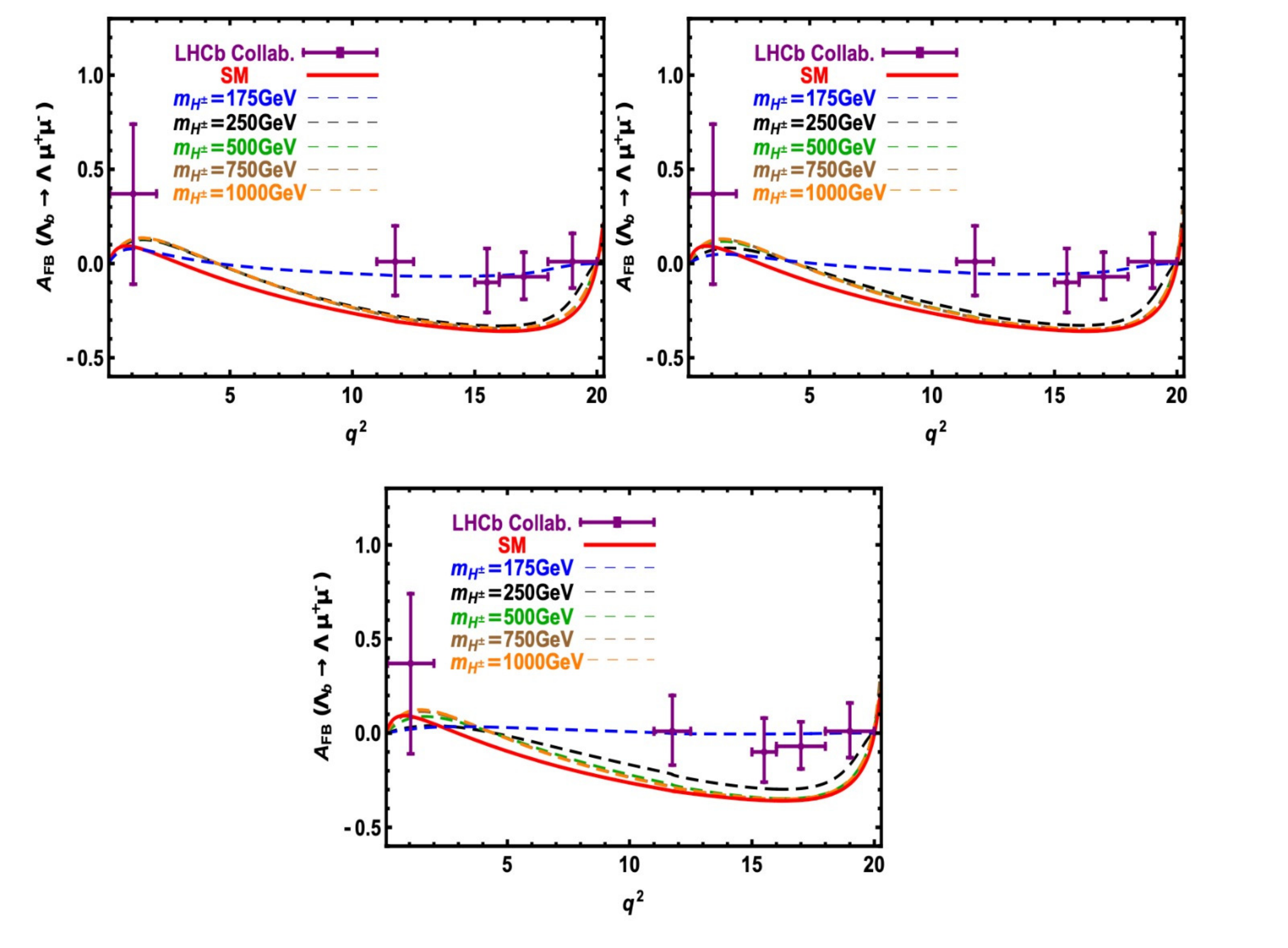}

\end{center}
\caption{The representation of the $A_{FB}$ without long-distance contributions as a function of $q^2$ for the $\Lambda_b \rightarrow \Lambda \mu^+ \mu^-$ transition  within the SM and 2HDM frameworks, plotted against different charged Higgs masses for $\lambda_{tt} = 0.05$, $\lambda_{tt} = 0.15$, and $\lambda_{tt} = 0.30$, respectively, and compared with the available experimental data reported by the LHCb Collaboration. }
\label{fig:F17}
\end{figure}

\end{widetext}

\begin{widetext}

\begin{figure}[h!]
\begin{center}
\includegraphics[totalheight=12cm,width=16cm]{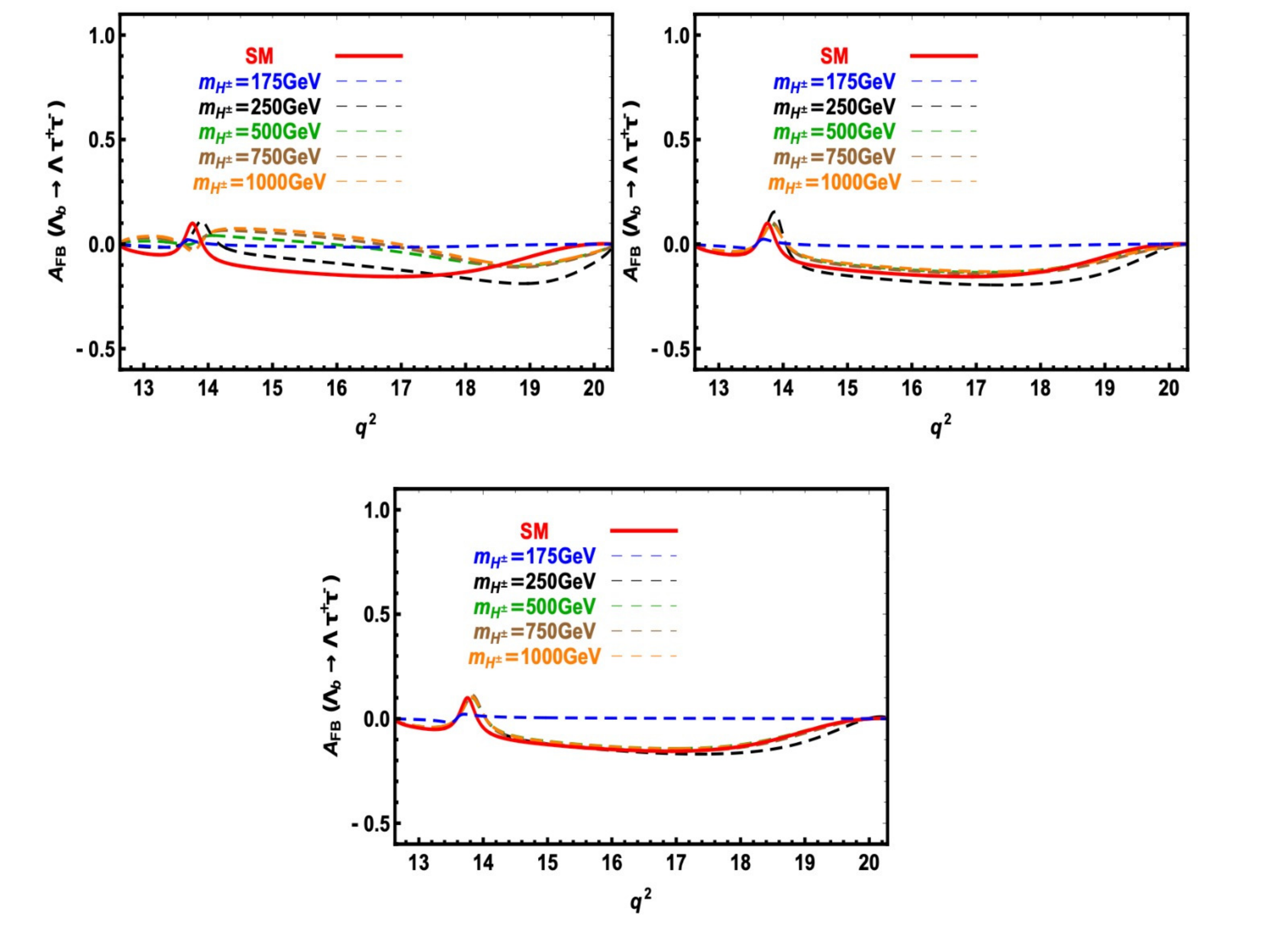}
\

\end{center}
\caption{The representation of the $ A_{FB}$ as a function of $q^2$   for the $ \Lambda_b \rightarrow \Lambda \tau^+  \tau^-$  transition in the SM and 2HDM with long-distance contributions plotted against different Higgs masses for $\lambda_{tt}=0.05$, $\lambda_{tt}=0.15$ and $\lambda_{tt}=0.30$, respectively. }
\label{fig:F18}
\end{figure}

\end{widetext}

\begin{widetext}

\begin{figure}[h!]
\begin{center}
\includegraphics[totalheight=12cm,width=16cm]{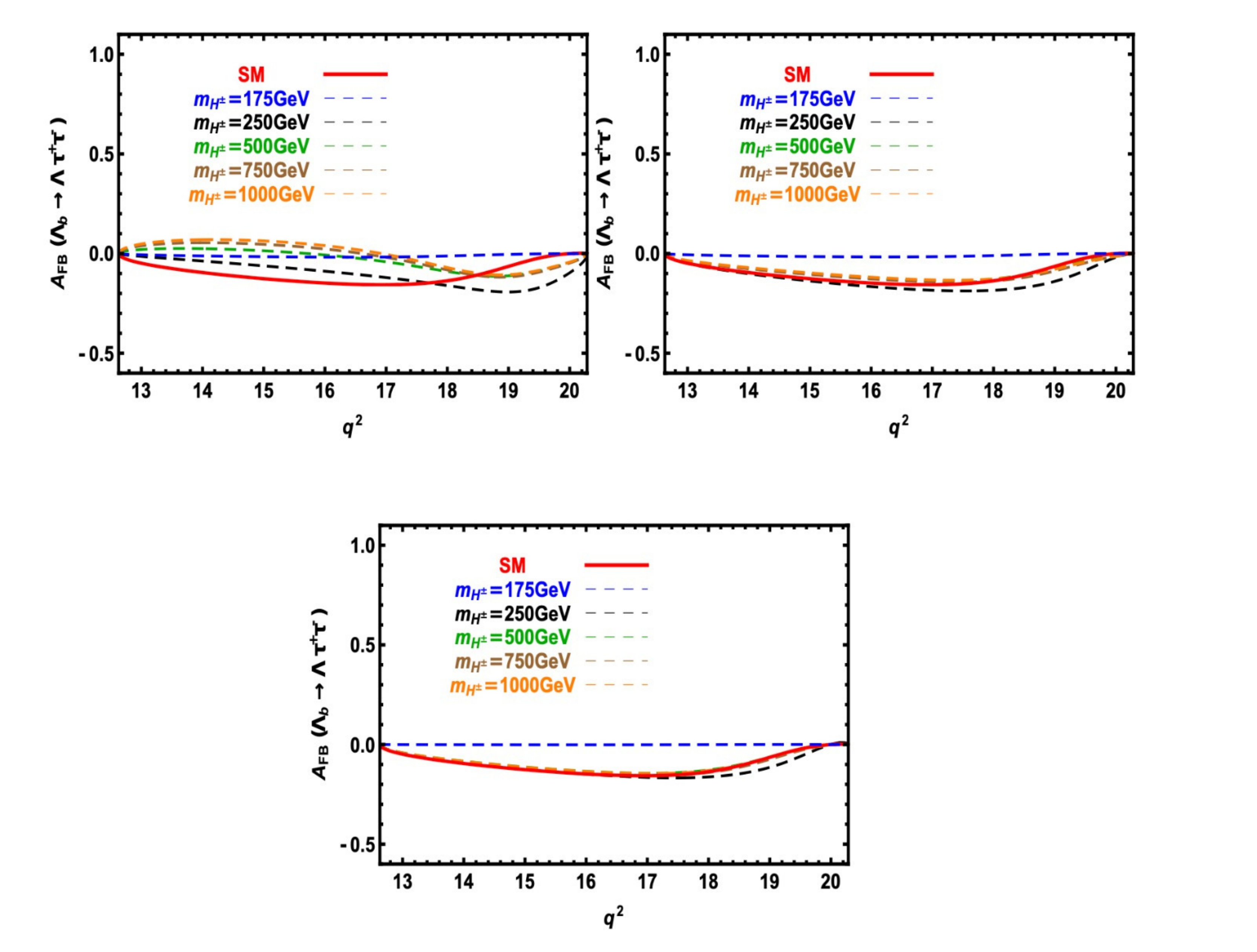}

\end{center}
\caption{The representation of the $ A_{FB}$ as a function of $q^2$   for the $ \Lambda_b \rightarrow \Lambda \tau^+  \tau^-$  transition in the SM and 2HDM  without long-distance contributions plotted against different Higgs masses for $\lambda_{tt}=0.05$, $\lambda_{tt}=0.15$ and $\lambda_{tt}=0.30$, respectively. }
\label{fig:F19}
\end{figure}

\end{widetext}

This subsection addresses the lepton forward–backward asymmetry ($A_{FB}$), a key observable well known for its sensitivity to new physics contributions, defined as
\begin{widetext}
\begin{eqnarray}
A_{FB}(q^2)= \dfrac{\int_{0}^{1} \dfrac{d\Gamma}{d q^2 d cos\Theta_l } d cos\Theta_l - \int_{-1}^{0} \dfrac{d\Gamma}{d q^2 d cos\Theta_l } d cos\Theta_l }{\int_{0}^{1} \dfrac{d\Gamma}{d q^2 d cos\Theta_l } d cos\Theta_l +\int_{-1}^{0} \dfrac{d\Gamma}{d q^2 d cos\Theta_l } d cos\Theta_l }.
\end{eqnarray}
\end{widetext}
In  Figs.~\ref{fig:F16} and ~\ref{fig:F17}, the $q^2$ dependence of the $A_{FB}$ is plotted for the $\mu$ channel within the SM and 2HDM scenarios with Type III, with results shown both including and excluding the LD contributions. Furthermore, the obtained predictions are compared with the available experimental data from LHCb, and results are presented for different values of the charged Higgs boson mass, $m_{H^\pm} = 175, 250, 500, 750,$ and $1000~\text{GeV}$, as well as for the parameter values $\lambda_{tt} = 0.05, 0.15,$ and $0.30$. Corresponding analyses for the $\tau$ channel are shown in  Fig.~\ref{fig:F18} and ~\ref{fig:F19}, where similar trends are depicted.

From  Figs.~\ref{fig:F16}-- \ref{fig:F19}, it becomes evident that the experimental data display a particularly strong agreement with the predictions corresponding to $m_{H^\pm} = 175~\text{GeV}$ in the $\mu$ channel, while the other charged Higgs mass values remain in closer alignment with the SM expectations. In contrast, the $\tau$ channel is not experimentally accessible with sufficient precision, preventing a meaningful comparison. This indicates that the consistency between theoretical predictions and experimental observations can only be reliably assessed in the $\mu$ channel and depends sensitively on the chosen charged Higgs mass scenario.

% Sigma AFB

\begin{widetext}

\begin{figure}[h!]
\begin{center}
\includegraphics[totalheight=12cm,width=16cm]{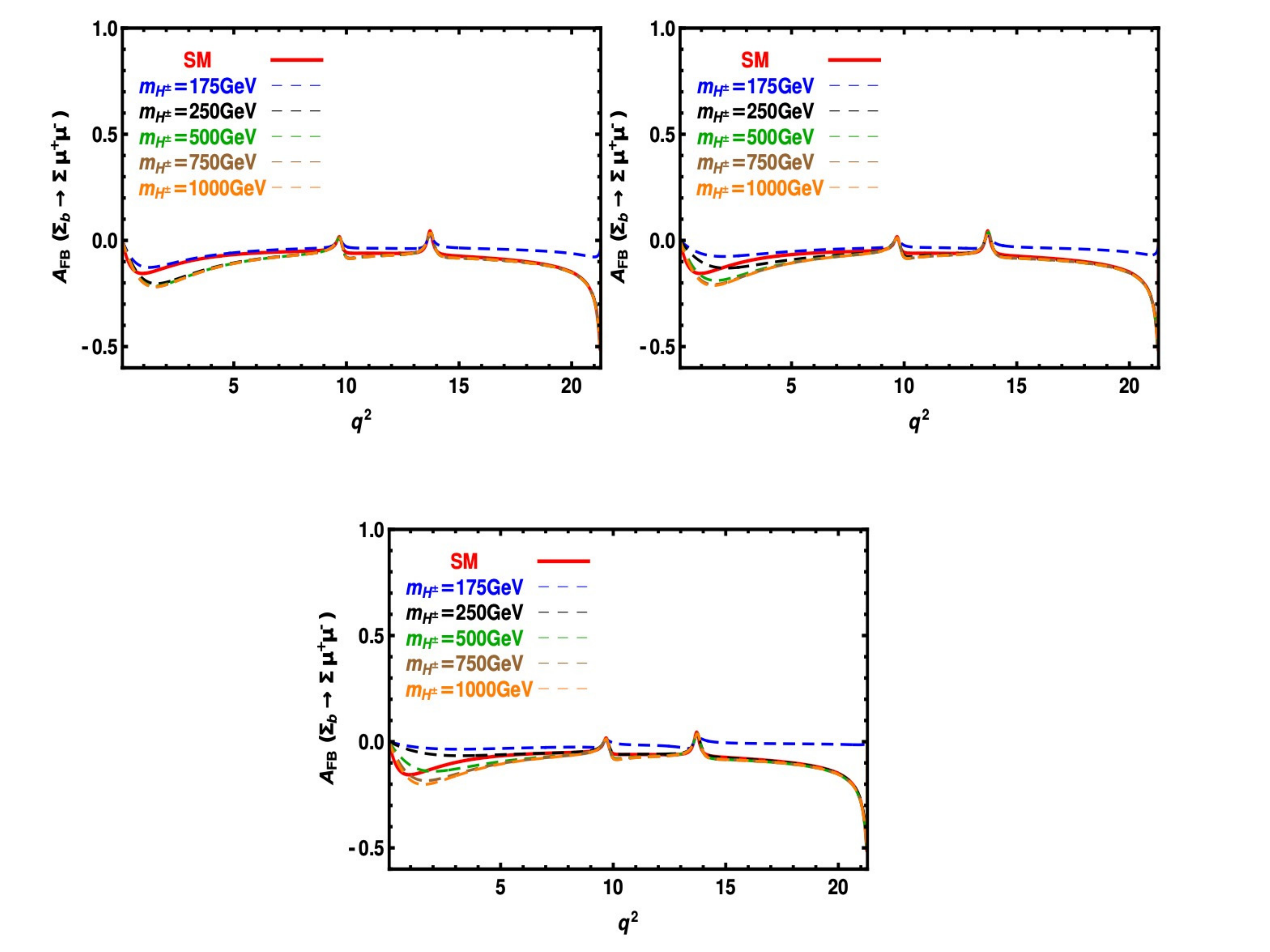}

\end{center}
\caption{The representation of the $ A_{FB}$ as a function of $q^2$   for the $ \Sigma_b \rightarrow \Sigma \mu^+  \mu^-$  transition in the SM and 2HDM with long-distance contributions plotted against different Higgs masses for $\lambda_{tt}=0.05$, $\lambda_{tt}=0.15$ and $\lambda_{tt}=0.30$, respectively. }
\label{fig:F20}
\end{figure}

%$\Sigma_b \rightarrow \Sigma  \tau^+  \tau^-$
\end{widetext}

\begin{widetext}

\begin{figure}[h!]
\begin{center}
\includegraphics[totalheight=12cm,width=16cm]{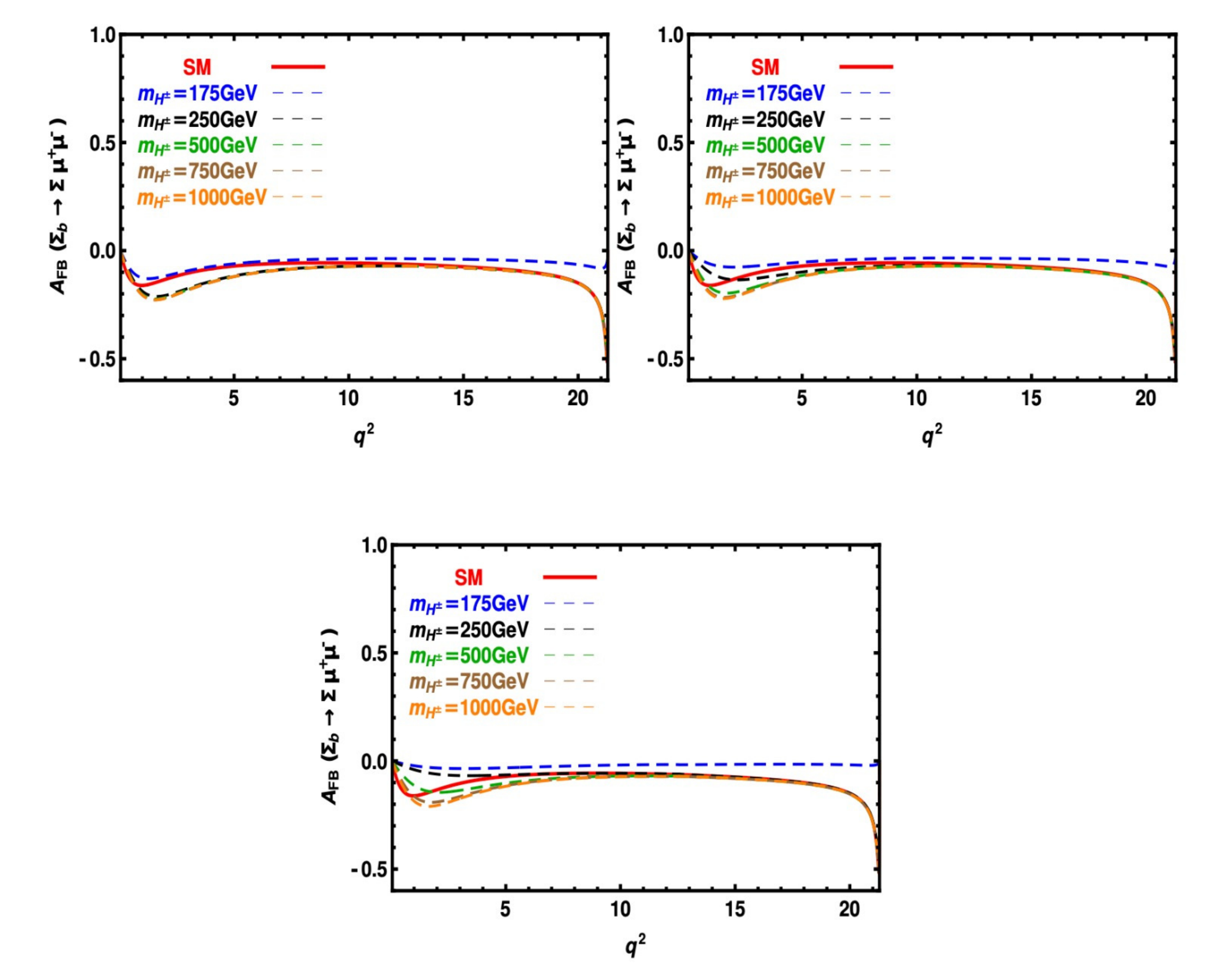}

\end{center}
\caption{The representation of the $ A_{FB}$ as a function of $q^2$ for the $ \Sigma_b \rightarrow \Sigma \mu^+  \mu^-$  transition in the SM and 2HDM without long-distance contributions plotted against different Higgs masses for $\lambda_{tt}=0.05$, $\lambda_{tt}=0.15$ and $\lambda_{tt}=0.30$, respectively. }
\label{fig:F21}
\end{figure}

\end{widetext}

\begin{widetext}

\begin{figure}[h!]
\begin{center}
\includegraphics[totalheight=12cm,width=16cm]{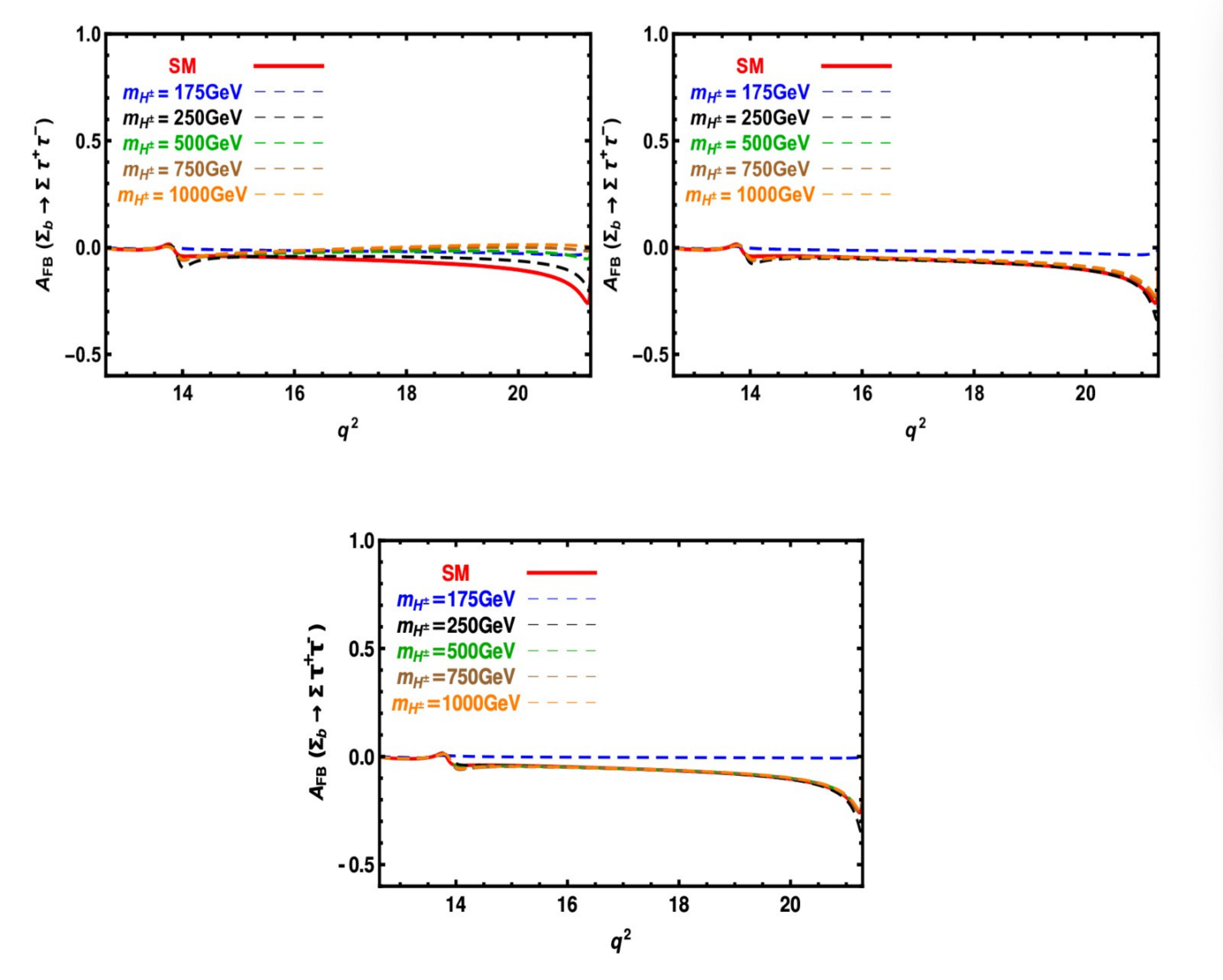}

\end{center}
\caption{The representation of the $ A_{FB}$ as a function of $q^2$   for the $ \Sigma_b \rightarrow \Sigma \tau^+  \tau^-$  transition in the SM and 2HDM with long-distance contributions plotted against different Higgs masses for $\lambda_{tt}=0.05$, $\lambda_{tt}=0.15$ and $\lambda_{tt}=0.30$, respectively. }
\label{fig:F22}
\end{figure}

\end{widetext}

\begin{widetext}

\begin{figure}[h!]
\begin{center}
\includegraphics[totalheight=12cm,width=16cm]{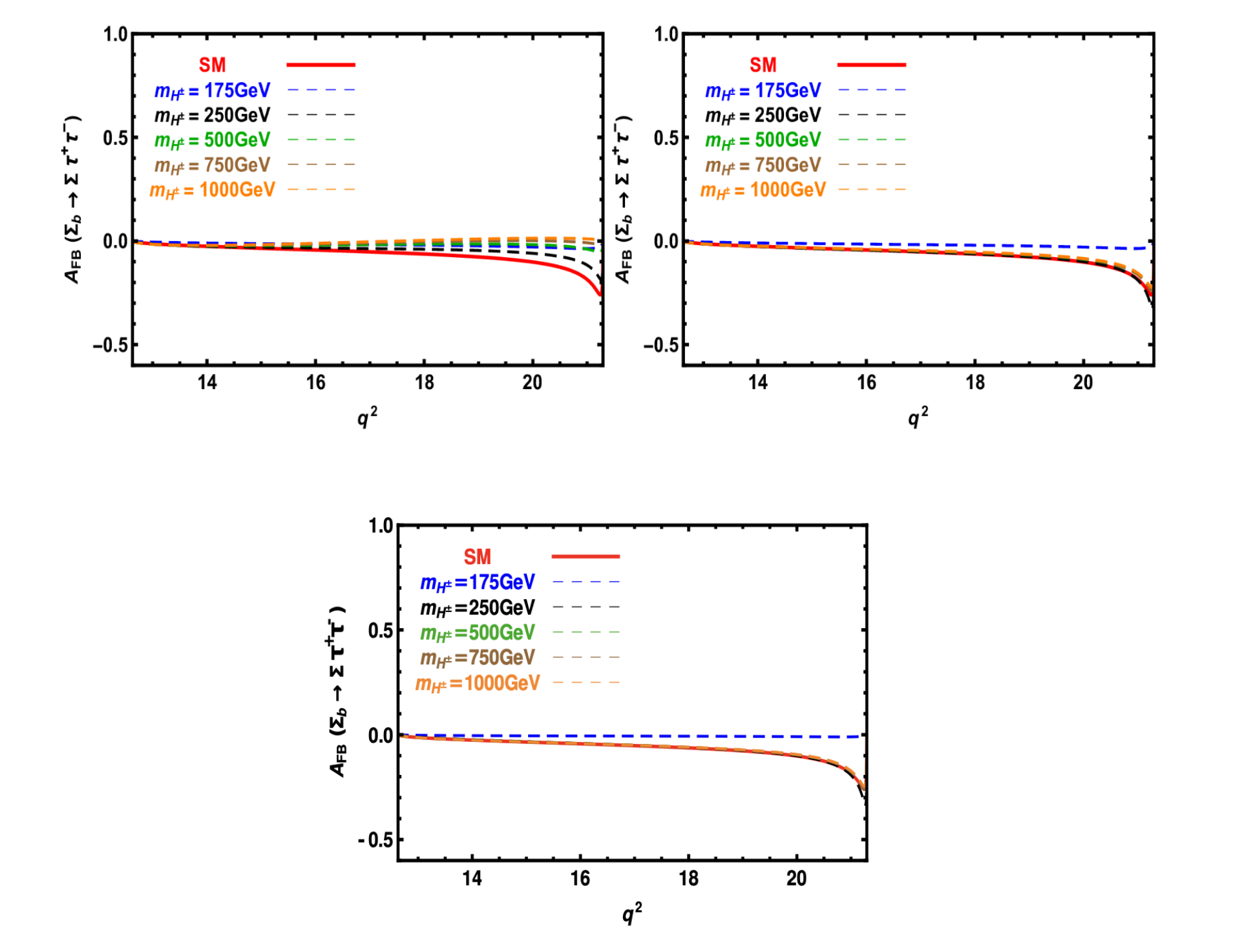}

\end{center}
\caption{The representation of the $ A_{FB}$ as a function of $q^2$   for the $ \Sigma_b \rightarrow \Sigma \tau^+  \tau^-$  transition in the SM and 2HDM without long-distance contributions plotted against different Higgs masses for $\lambda_{tt}=0.05$, $\lambda_{tt}=0.15$ and $\lambda_{tt}=0.30$, respectively. }
\label{fig:F23}
\end{figure}

\end{widetext}

% Ksi

\begin{widetext}

\begin{figure}[h!]
\begin{center}
\includegraphics[totalheight=12cm,width=16cm]{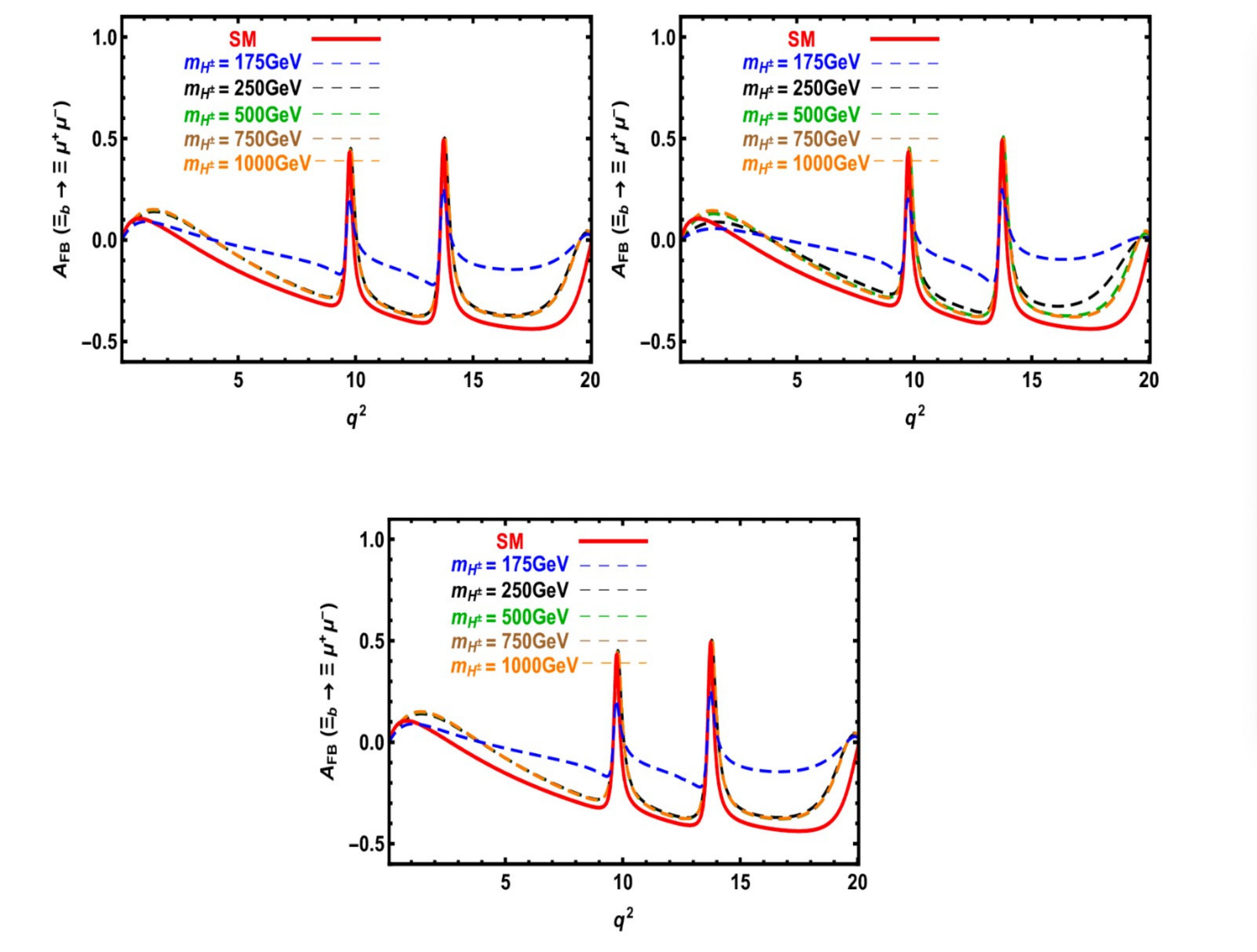}

\end{center}
\caption{The representation of the $ A_{FB}$ as a function of $q^2$   for the $\Xi_b \rightarrow \Xi \mu^+  \mu^-$  transition in the SM and 2HDM  with long-distance contributions plotted against different Higgs masses for $\lambda_{tt}=0.05$, $\lambda_{tt}=0.15$ and $\lambda_{tt}=0.30$, respectively. }
\label{fig:F24}
\end{figure}

\end{widetext}
\begin{widetext}

\begin{figure}[h!]
\begin{center}
\includegraphics[totalheight=12cm,width=16cm]{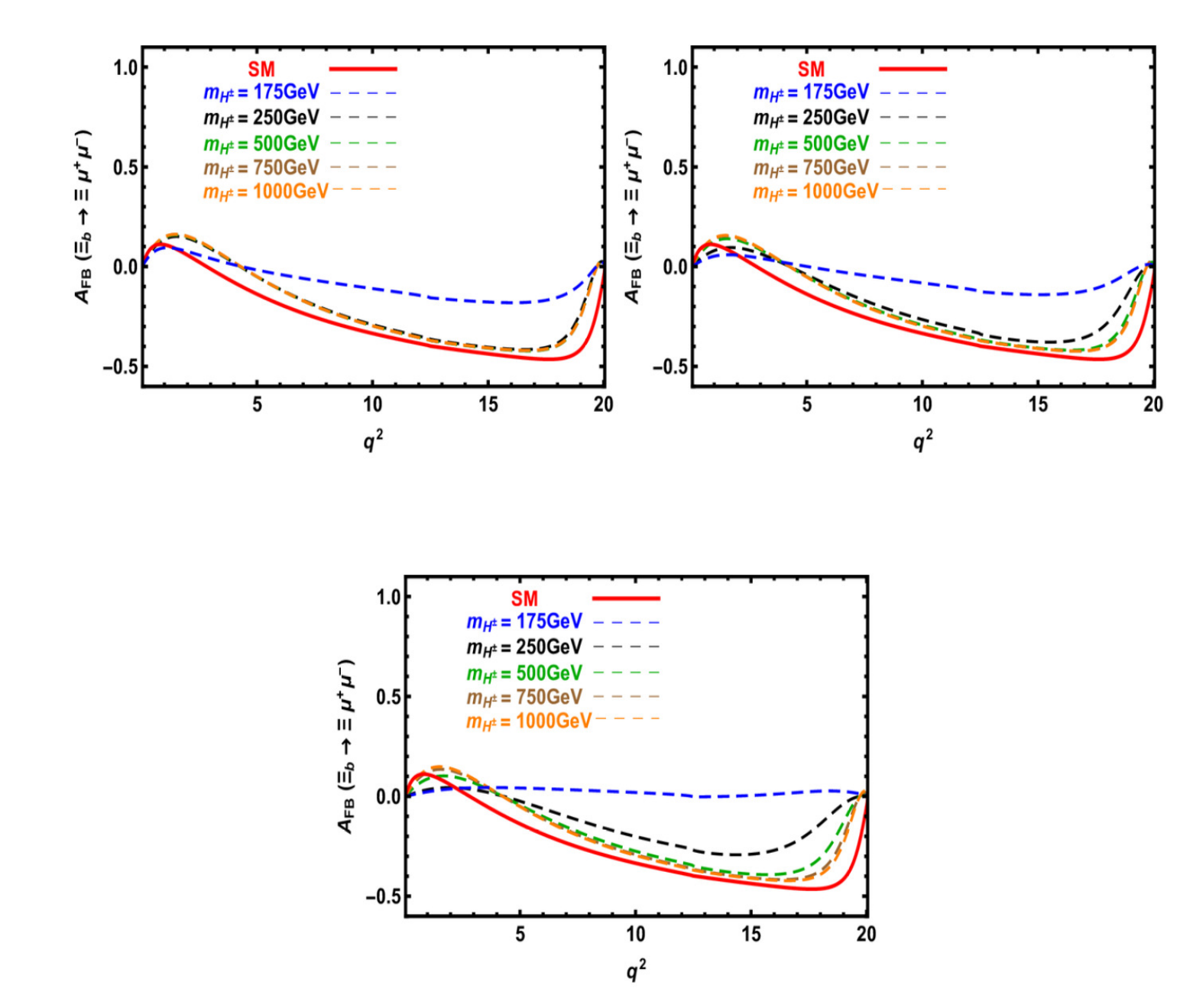}

\end{center}
\caption{The representation of the $ A_{FB}$ as a function of $q^2$   for the $ \Xi_b \rightarrow \Xi \mu^+  \mu^-$  transition in the SM and 2HDM  without long-distance contributions plotted against different Higgs masses for $\lambda_{tt}=0.05$, $\lambda_{tt}=0.15$ and $\lambda_{tt}=0.30$, respectively. }
\label{fig:F25}
\end{figure}

\end{widetext}

\begin{widetext}

\begin{figure}[h!]
\begin{center}
\includegraphics[totalheight=12cm,width=16cm]{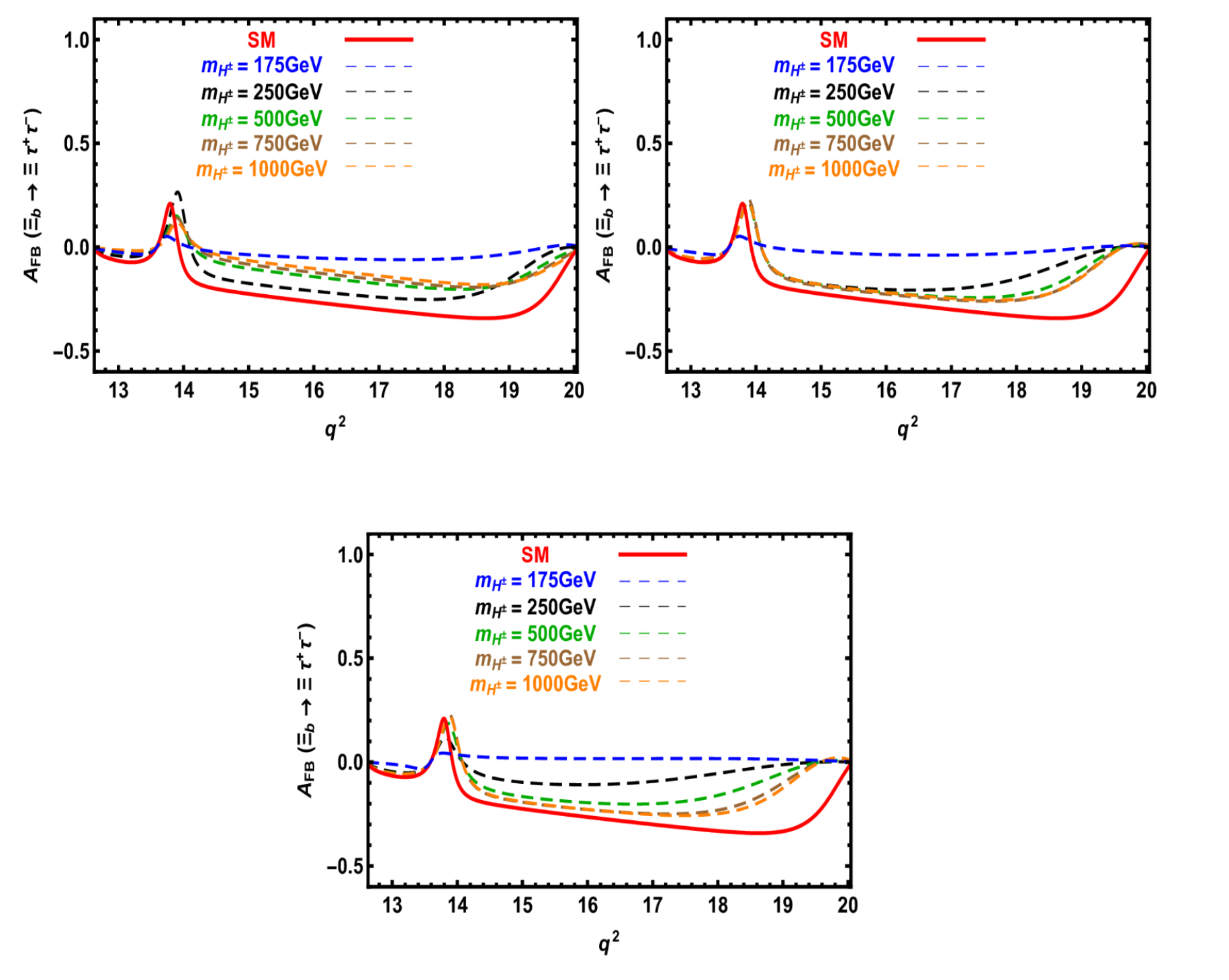}

\end{center}
\caption{The representation of the $ A_{FB}$ as a function of $q^2$   for the $ \Xi_b \rightarrow \Xi \tau^+  \tau^-$  transition in the SM and 2HDM  with long-distance contributions plotted against different Higgs masses for $\lambda_{tt}=0.05$, $\lambda_{tt}=0.15$ and $\lambda_{tt}=0.30$, respectively. }
\label{fig:F26}
\end{figure}

\end{widetext}

\begin{widetext}

\begin{figure}[h!]
\begin{center}
\includegraphics[totalheight=12cm,width=16cm]{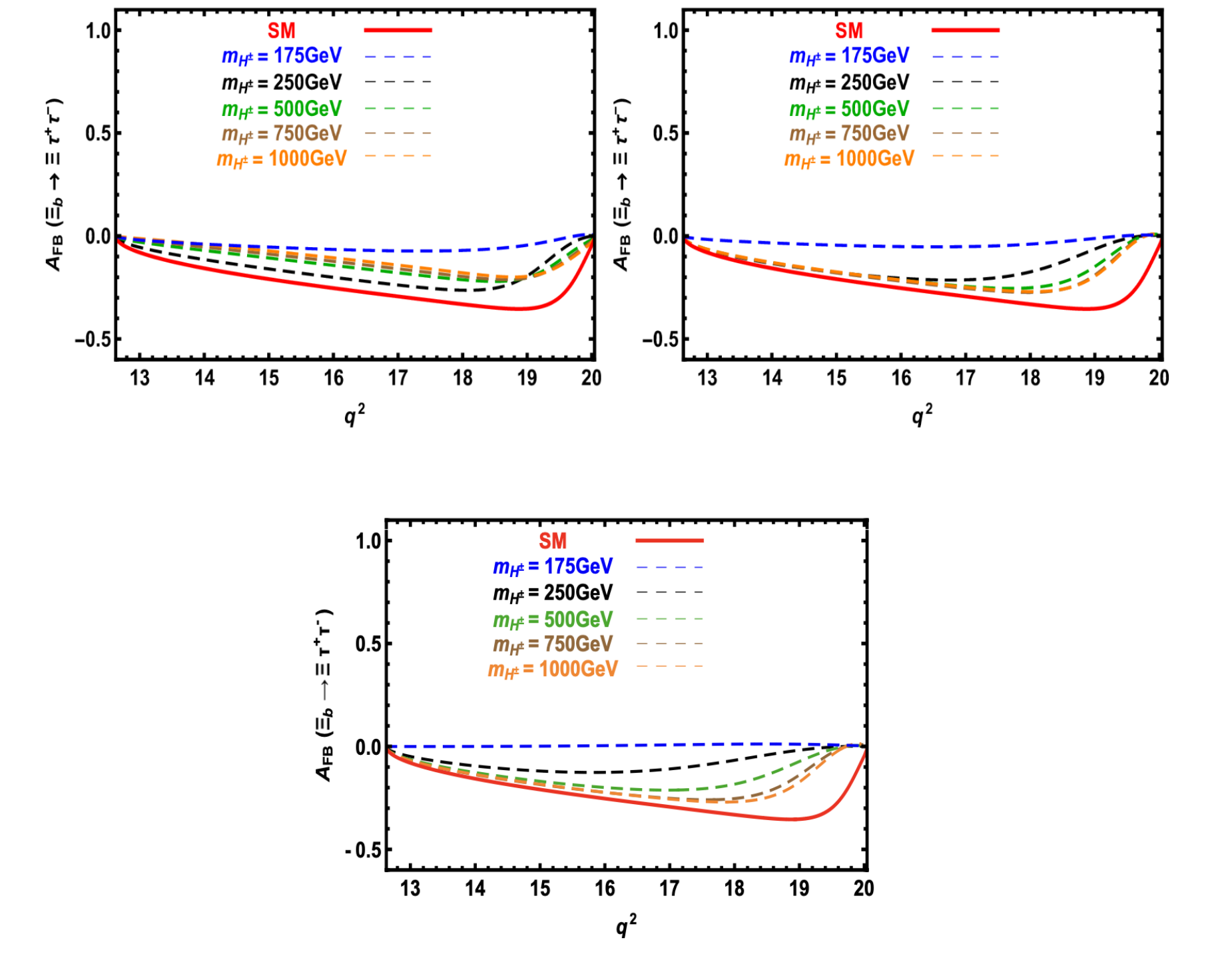}

\end{center}
\caption{The representation of the $ A_{FB}$ as a function of $q^2$   for the $ \Xi_b \rightarrow \Xi  \tau^+  \tau^-$  transition in the SM and 2HDM  without long-distance contributions plotted against different Higgs masses for $\lambda_{tt}=0.05$, $\lambda_{tt}=0.15$ and $\lambda_{tt}=0.30$, respectively. }
\label{fig:F27}
\end{figure}

\end{widetext}
Similar $A_{FB}$ plots have been generated for the $\Sigma_b \rightarrow \Sigma \ell^+ \ell^-$ and $\Xi_b \rightarrow \Xi \ell^+ \ell^-$ channels, following the same methodology as for the $\Lambda_b$ transition. These results, presented in Figs.~\ref{fig:F20}--\ref{fig:F27}, systematically explore both the $\mu$ and $\tau$ channels, highlighting the effects of different charged Higgs masses and $\lambda_{tt}$ parameter values on the forward-backward asymmetry. For both the $\mu$ and $\tau$ channels in the $\Sigma_b$ and $\Xi_b$ decays, it is observed that the case with $m_{H^\pm} = 175~\text{GeV}$ exhibits noticeable deviations from the other mass scenarios, while the overall patterns remain largely consistent across the models.

The $A_{FB}$ across the $\Lambda_b$,$\Sigma_b$ and $\Xi_b$  muon decay channels provides a sensitive probe for New Physics, specifically highlighting the divergence between the SM and the 2HDM with Type-III framework. While all channels maintain consistent charmonium resonance peaks at $q^2 \approx 9.5$ and $13.5 \, \text{GeV}^2$, the most significant model deviations occur at the lowest charged Higgs mass ($m_{H^\pm} = 175 \, \text{GeV}$), where the 2HDM tends to flatten the asymmetry or shift the zero-crossing points toward the positive region, contrary to the deeper negative values predicted by the SM. As the Higgs mass increases toward $1000 \, \text{GeV}$, these BSM signatures converge with the SM baseline, suggesting that low-mass Higgs scenarios leave unique ``fingerprints'' in high-$q^2$ observables that could be distinguished through high-precision experimental measurements.

In the $\tau$ decay channels, the $A_{FB}$ is generally more suppressed than in $\mu$ channels due to the greater tau lepton mass, yet it remains a valuable indicator of New Physics at high $q^2$ values. While the SM predicts a consistent trend toward negative asymmetry in the $\Lambda_b$,$\Sigma_b$ and $\Xi_b$ sectors, the 2HDM (particularly at $m_{H^\pm} = 175 \, \text{GeV}$) significantly dampens this effect, keeping the asymmetry near zero. This ``flattening'' of the $A_{FB}$ curve serves as a primary signature of the 2HDM Type-III framework in tauonic transitions, with the divergence from the SM becoming more pronounced as the coupling parameter $\lambda_{tt}$ increases.

Future determinations of $\textit{A}_{FB}$ in different lepton channels, and their confrontation with our predictions, will be crucial for probing baryonic structures and constraining BSM parameters.

\section{Summary and Concluding Remarks}
\label{sec:Disc}
%%%%%%%%%%%%%%%%%%%%%%%%%%%%%%%%%%%%%%%%%%%%%%%%%%%%%%%%%%%
In this study, the rare dileptonic decays $ \Lambda_b \rightarrow \Lambda \ell^+  \ell^-$, $\Sigma_b \rightarrow \Sigma \ell^+ \ell^-$ and $\Xi_b \rightarrow \Xi \ell^+ \ell^-$ have been investigated within the frameworks of the SM and the general Type-III 2HDM. The main motivation of this work stems from the inability of the SM to fully address certain fundamental questions, highlighting the need to explore beyond the SM physics. The 2HDM is regarded as one of the simplest extensions of the SM Higgs sector. In the Type-III 2HDM, flavor non-diagonal couplings in the fermion sector are allowed, giving rise to tree-level FCNC effects, which in turn affect the $b \rightarrow s \ell^{+}\ell^{-}$ transitions. By employing the decay amplitudes and transition matrix elements, the effects of the Type-III 2HDM on various observables such as the differential decay width, the differential branching ratio, the total decay width, the total branching ratio, and lepton forward–backward asymmetry have been systematically evaluated.
	
For the $ \Lambda_b \rightarrow \Lambda \mu^+  \mu^-$ transition, the differential branching ratio shows good agreement with experimental data (LHCb and CDF Collaborations) in the low to intermediate $q^2$ region ($0$–$16$ GeV$^2$). When the charged Higgs boson mass ($m_{H^\pm}$) exceeds $250$ GeV and the $\lambda_{tt}$ parameter is taken into account, the agreement becomes more pronounced, exhibiting closer consistency with the SM predictions. However, at higher $q^2$ values (beyond approximately $16$ GeV$^2$), deviations between theoretical predictions and experimental data become more significant.

The inclusion of LD contributions generates prominent resonance structures in the charmonium regions, while LD-subtracted curves display a smooth, short-distance dominated behavior. The overall trends remain robust even when LD contributions are excluded. In the $ \Lambda_b \rightarrow \Lambda \tau^+  \tau^-$ channel, it is observed that, except for the $m_{H^\pm} = 175 GeV$ scenario, the results are consistent with SM predictions across all $q^2$ values. Overall, the $\tau$ channel exhibits behavior fully compatible with the SM over the entire $q^2$ range. The $\Sigma_b \rightarrow \Sigma \ell^+ \ell^-$ and $\Xi_b \rightarrow \Xi \ell^+ \ell^-$  channels display similar patterns to the $ \Lambda_b$ channel, with the largest deviations again appearing at high $q^2$ values and for lighter $m_{H^\pm}$ (particularly $175 GeV$ and $250 GeV$). Specifically, for the $\Xi_b \to \Xi \mu^+ \mu^-$ transition, the 2HDM predictions—particularly at $m_{H^\pm} = 175 \, \text{GeV}$—exhibit a significant enhancement in the branching ratio at high $q^2$ values compared to the SM baseline, suggesting that this region is uniquely sensitive to the scalar and pseudoscalar exchange mediated by the charged Higgs.

The $ A_{FB}$ is an important observable, known for its sensitivity to new physics contributions. In the $ \Lambda_b \rightarrow \Lambda \mu^+  \mu^-$ channel, it has been observed that the experimental data exhibit particularly strong agreement with predictions for $m_{H^\pm} = 175 GeV$, while other charged Higgs mass values remain closer to the SM expectations. Since the $\tau$ channel is not experimentally accessible and currently lacks precise measurements, no definitive conclusion can be drawn from this mode. Therefore, the assessment of the consistency between theoretical predictions and experimental data relies primarily on the $\mu$ channel and remains highly sensitive to the chosen charged Higgs mass scenario. In the $\Sigma_b$ and $\Xi_b$ decays, the $m_{H^\pm} = 175\,\mathrm{GeV}$ case similarly exhibits noticeable deviations compared to the other mass scenarios.

Overall, the studied baryonic channels ($ \Lambda_b, \Xi_b$) display comparable dependencies on the model parameters in terms of both differential and total branching ratios as well as the forward–backward asymmetry. For the $\Sigma_b \rightarrow \Sigma \ell^+ \ell^-$ channel, we instead analyze the differential decay width and present the total decay width rather than the branching ratio, due to the lack of reliable experimental information on its lifetime as well as the forward–backward asymmetry.

Our results indicate that low charged Higgs masses ($m_{H^\pm} = 175 GeV$) and large $\lambda_{tt}$ values maximize BSM effects and, in some cases (particularly for $A_{FB}$), provide the best agreement with existing experimental data in $\Lambda_b$ and $\mu$ channel. While the SM consistently predicts deeper negative $A_{FB}$ values in the high $q^2$ region across all channels, the 2HDM (Type-III) results in a distinct flattening of the asymmetry curves; this damping effect is most prominent in the $\Sigma_b$ and $\tau$ sectors, where the asymmetry is pulled significantly closer to the zero-axis compared to the SM expectations.

These findings demonstrate that the Type-III 2HDM has the potential to significantly impact dileptonic baryon decays, especially in the low charged Higgs mass region. The model parameters, $m_{H^\pm}$ and $\lambda_{tt}$, play a crucial role in determining the extent to which $\mu$-channel observables ($dBR$,$d\Gamma$,$\Gamma$, $BR$, $A_{FB}$) deviate from SM predictions. Future experimental studies at updated LHCb and/or Belle II detectors offer excellent opportunities to probe these rare decay channels. In particular, precise measurements of $A_{FB}$ in different hadronic and leptonic channels will be essential for a deeper understanding of baryonic structures and for constraining beyond the SM parameters.

The results obtained in this study thus serve as a milestone in evaluating the feasibility of the Type-III 2HDM in the ongoing search for physics beyond the SM in fundamental particle physics.

\end{document}